\documentclass[amsmath,amssymb,aps,prd,10pt,twocolumn]{revtex4}
\usepackage{graphicx}
\usepackage{mathtools}
\usepackage{makecell}
\usepackage{siunitx}
\DeclareMathOperator\erfc{erfc}
\DeclareMathOperator\cov{cov}
\begin{document}

\title{Complex Analysis of Askaryan Radiation: A Fully Analytic Model in the Time-Domain}

\author{Jordan C. Hanson}
\email{jhanson2@whittier.edu}
\affiliation{Department of Physics and Astronomy, Whittier College}
\author{Raymond Hartig}
\affiliation{Department of Physics and Astronomy, Whittier College}
\date{\today}

\begin{abstract}
The detection of ultra-high energy (UHE, $\geq$10 PeV) neutrinos via detectors designed to utilize the Askaryan effect has been a long-time goal of the astroparticle physics community.  The Askaryan effect describes radio-frequency (RF) radiation from high-energy cascades.  When a UHE neutrino initiates a cascade, cascade properties are imprinted on the radiation.  Thus, observed radiation properties must be used to reconstruct the UHE neutrino event.  Analytic Askaryan models have three advantages when used for UHE neutrino reconstruction.  First, cascade properties may be derived from the match between analytic function and observed data.  Second, analytic models minimize computational intensity in simulation packages.  Third, analytic models can be embedded in firmware to enhance the real-time sensitivity of detectors.  We present a fully analytic Askaryan model in the time-domain for UHE neutrino-induced cascades in dense media that builds upon prior models in the genre.  We then show that our model matches semi-analytic parameterizations used in Monte Carlo simulations for the design of IceCube-Gen2.  We find correlation coefficients greater than 0.95 and fractional power differences $<5$\% between the the fully analytic and semi-analytic approaches.
\end{abstract}


\maketitle

\section{Introduction}

The extrasolar flux of neutrinos with energies between [0.01-1] PeV has been measured by the IceCube collaboration \cite{10.1126/science.1242856}.  Previous analyses have shown that the discovery of UHE neutrinos (UHE-$\nu$) will require an expansion in detector volume because the flux is expected to decrease with energy \cite{10.1016/j.astropartphys.2010.06.003,10.1088/1475-7516/2010/10/013,10.1103/physrevd.98.062003,10.1088/1475-7516/2020/03/053,10.1103/physrevd.102.043021}.  The UHE-$\nu$ flux could potentially explain the origin of UHE cosmic rays (UHECR), and provides the opportunity to study electroweak interactions at record-breaking energies \cite{Ackermann:201946d,Ackermann:20195ec}.  Utilizing the Askaryan effect expands the effective volume of UHE-$\nu$ detector designs, because this effect offers a way to detect UHE-$\nu$ with radio pulses that travel more than 1 km in sufficiently RF-transparent media such as Antarctic and Greenlandic ice \cite{10.3189/2015jog14j214, 10.3189/2015jog15j057, 10.1016/j.astropartphys.2011.11.010}.

The Askaryan effect occurs within a dense medium with an index of refraction $n$.  A relativistic particle with $v > c/n$ initiates a high-energy cascade with negative total charge.  The charge radiates energy in the RF bandwidth, and the radiation may be detected if the medium does not significantly attenuate the signal \cite{askaryan1,zhs}.  The IceCube EHE analysis has constrained the UHE-$\nu$ flux to be $E_\nu^2 \phi_\nu \leq 2 \times 10^{-8}$ GeV cm$^{-2}$ s$^{-1}$ sr$^{-1}$ between $[5\times 10^{15} - 2\times 10^{19}]$ eV \cite{10.1103/physrevd.98.062003}.  Arrays of $\mathcal{O}(100)$ \textit{in situ} detectors encompassing effective areas of $\approx 10^4$ m$^2$ sr per station, spaced by $\mathcal{O}(1)$ RF attenuation length could discover a UHE-$\nu$ flux beyond the EHE limits.  The most suitable ice formations exist in Antarctica and Greenland, and a group of prototype Askaryan-class detectors has been deployed.  These detectors seek to probe unexplored UHE-$\nu$ flux parameter-space from astrophysical and cosmogenic sources \cite{10.1103/PhysRevD.85.062004,10.1088/1475-7516/2020/03/053,10.1103/physrevd.102.043021,10.1103/physrevd.99.122001}.

Askaryan radiation was first measured in the laboratory in silica sand, and later ice \cite{saltzberg,10.1103/PhysRevD.74.043002,ask_ice}.  Cascade properties affect the amplitude and phase of the radiation.  At RF wavelengths, cascade particles radiate coherently, and the radiation amplitude scales with the total track length of the excess negative charge.  The RF pulse shape is influenced by the \textit{longitudinal length} of the cascade, and the pulse is strongest when the viewing angle is close to the Cherenkov angle, $\theta_{\rm C}$.  The \textit{excess charge profile} describes the excesse negative charge versus longitudinal position on the cascade axis.  Radiation wavelengths shorter than the \textit{lateral width} of the cascade, perpendicular to the cascade axis, are attenuated.  At energies far above 10 PeV in ice, however, excess charge profiles generated by \textit{electromagnetic} cascades experience the LPM effect and can have multiple peaks \cite{10.1016/j.astropartphys.2009.06.005,10.1103/physrevd.82.074017}.  This theoretical foundation has been constructed from a variety of experimental and simulation results.

The field of Askaryan-class detectors requires this foundation for at least two reasons.  First, the theoretical form of the Askaryan RF pulse is used to optimize RF detector designs.  Askaryan models are incorporated into simulations \cite{dookayka2011characterizing,testbed,10.1140/epjc/s10052-020-7612-8} in order to calculate expected signals and aid in detector design.  For example, reconstruction tools for the radio component of IceCube-Gen2 combine machine learning and insights from Askaryan radiation physics \cite{10.1140/epjc/s10052-019-6971-5,10.1088/1748-0221/15/09/p09039,IFT}.  Second, Askaryan models are used as templates to search large data sets for signal candidates \cite{10.1088/1475-7516/2020/03/053,10.1016/j.astropartphys.2014.09.002}.  The signal-to-noise ratios (SNRs) at RF channels are expected to be small (SNR $\approx 3$), because the amplitude of the radiated field decreases with the vertex distance ($1/r$), and the signal is attenuated by the ice \cite{10.3189/2015jog14j214,Barwick:2018497,ALLISON201963}.  Low SNR signals reqire correspondingly low RF trigger thresholds, but signals must be sampled for a bandwidth of [0.1-1] GHz.  Thus, RF channels are triggered at high rates by thermal noise.  UHE-$\nu$ signals will be hidden within millions of thermal triggers.  Template-waveform matching between models and data is a powerful technique for isolating RF signals from high-energy particles \cite{10.1016/j.astropartphys.2015.04.002,10.1016/j.astropartphys.2014.09.002}.

Askaryan models fall into three categories: full Monte Carlo (MC), semi-analytic, and fully analytic.  The original work by E. Zas, F. Halzen, and T. Stanev (ZHS) \cite{zhs} was a full MC model.  The properties of cascades with total energy $\leq 1$ PeV were examined.  A parameterization for the Askaryan field below 1 GHz was offered, attenuating modes above 1 GHz via a frequency-dependent form factor tied to the lateral cascade width.  The semi-analytic approach was introduced by J. Alvarez-Mu\~{n}iz \textit{et al} (ARVZ) \cite{10.1103/physrevd.84.103003}.  This approach accounts for fluctuations in the charge excess profile, and provides an analytic vector potential observed at the Cherenkov angle.  The vector potential at the Cherenkov angle is labeled the form factor, and observed fields are derived from the derivative of the vector potential once convolved with a charge excess profile from MC.  Recent work also accounts for differences in fit parameters from electromagnetic and hadronic cascades, and other interaction channels, while matching full MC simulations \cite{PhysRevD.101.083005}.

Finally, fully analytic models of Askaryan radiation from first principles have been introduced.  J. Ralston and R. Buniy (RB) gave a fully analytic model valid for observations of cascades in the near and far-field, with the transition encapsulated by a parameter $\eta$ \cite{10.1103/physrevd.65.016003}.  The result was complex frequency-domain model.  Recently, a model and software implementation was given by J. C. Hanson and A. Connolly (JCH+AC) that built upon RB by providing an analytic form factor derived from GEANT4 simulations, and accounted for LPM elongation \cite{10.1016/j.astropartphys.2017.03.008}.  This work connected the location of poles in the complex frequency plane to $\eta$ and the form factor.  The poles combine to form a low-pass filter for the Askaryan radiation.  The JCH+AC results match the ZHS results while demonstrating the physical origins of model parameters.  The RB and JCH+AC results are given in the Fourier domain, but most UHE-$\nu$ searches (like template matching) have taken place in the time-domain.  \textit{The goals of this work are to produce a fully analytic time-domain model accounting for complex poles, valid for all viewing angles $\theta$ and $\eta < 1$, and to demonstrate that it matches semi-analytic models.}

In Section \ref{sec:unit}, the cascade geometry, units, and vocabulary are defined.  In Section \ref{sec:ff}, we describe how the JCH+AC form factor fits into the current model \cite{10.1016/j.astropartphys.2017.03.008}.  In Section \ref{sec:onc}, the analytic Askaryan field, observed at $\theta = \theta_{\rm C}$ (\textit{on-cone}), is presented.  In Section \ref{sec:ofc}, the analytic Askaryan field observed for $\theta \neq \theta_{\rm C}$ (\textit{off-cone}) is presented.  In Section \ref{sec:fit}, fully analytic fields are matched to semi-analytic fields generated with NuRadioMC \cite{10.1140/epjc/s10052-020-7612-8} at 10 PeV (electromagnetic cascades) and 100 PeV (hadronic cascades).  Though the LPM effect is activated in NuRadioMC, it has a negligible influence on the waveform comparison at these energies.  In Section \ref{sec:conc}, the results are summarized and potential applications of the model are described.

\section{Units, Definitions, and Conventions}
\label{sec:unit}

\begin{figure}[ht]
\centering
\includegraphics[width=0.5\textwidth]{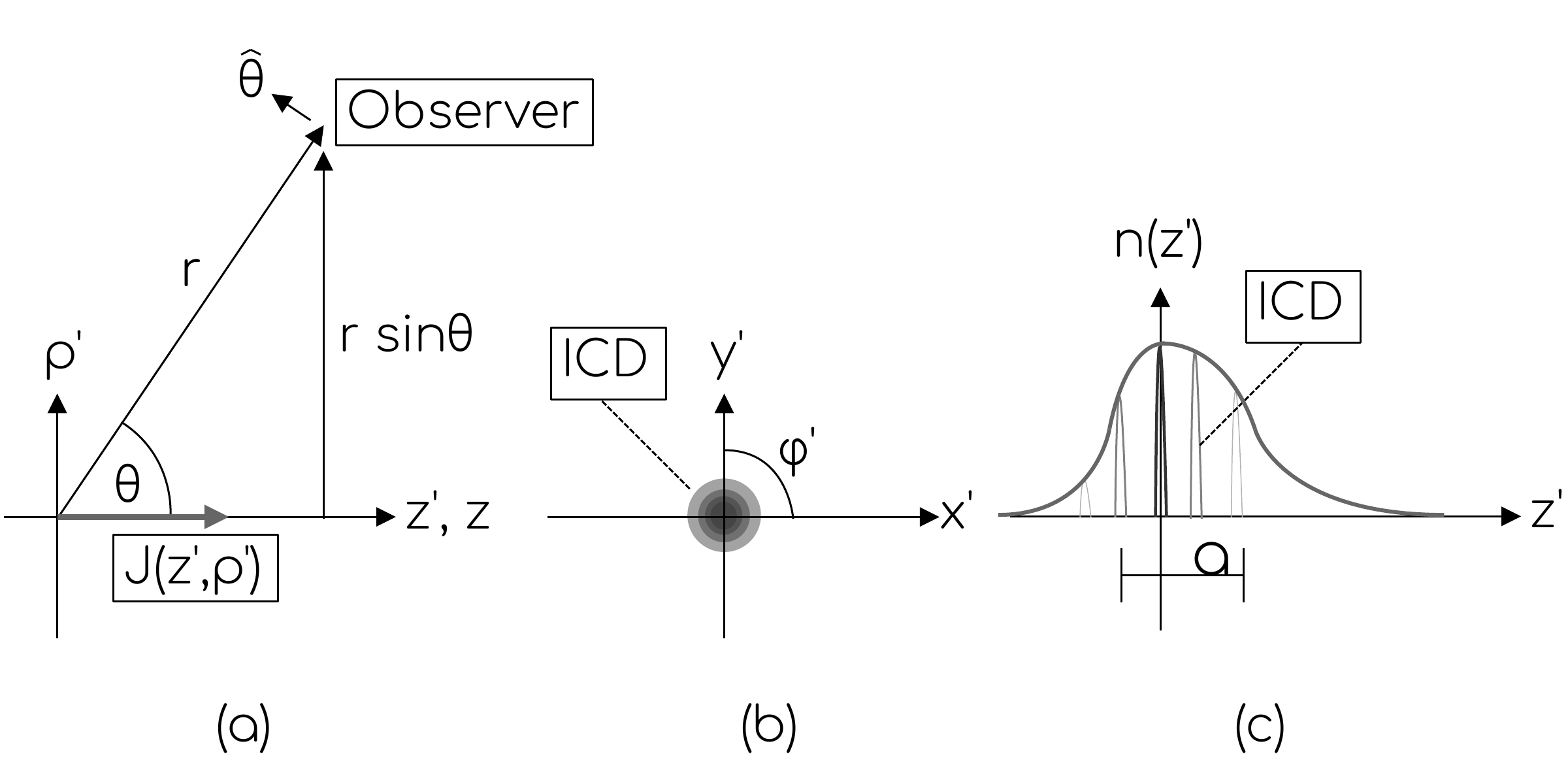}
\caption{\label{fig:geo} (a) Side view of the coordinate systems used in the analysis. Spherical unprimed coordinates refer to the observer.  Primed cylindrical coordinates refer to $\vec{J}(\rho',z')$. (b) Front view of the coordinate system.  The instantaneous charge density (ICD) is assumed to have no $\phi'$-dependence. (c) The function $n(z')$ describes the total cascade excess charge, and it has a characteristic width $a$.  The ICD has an instantaneous width much smaller than $a$ \cite{10.1016/j.astropartphys.2017.03.008}.}
\end{figure}

The coordinate system of the Askaryan radiation from a vector current density $\vec{J}$ is shown in Fig. \ref{fig:geo} (a)-(b).  Primed cylindrical coordinates refer $\vec{J}$, and the unprimed spherical coordinates refer to the observer.  The zenith or \textit{viewing angle} is measured with respect to the \textit{longitudinal axis} ($z'$).  The observer displacement is $r = |\vec{x} - \vec{x}'|$, in the $\hat{r}$ direction.  The origin is located where the cascade has the highest instantaneous charge density (ICD).  The ICD is treated with cylindrical symmetry, so it has no $\phi'$-dependence.  This assumption is based on the large number of cascade particles and momentum conservation.  The lateral extent of the ICD is along the \textit{lateral axis} ($\rho'$).  The viewing angle is $\theta$ in spherical coordinates, and the Cherenkov angle occurs when $\theta$ satisfies $\cos(\theta_{\rm C}) = 1/n_{\rm ice}$ with $n_{\rm ice} = 1.78 \pm 0.003$ \cite{bog}.

In Fig. \ref{fig:geo} (c), an example excess charge profile $n(z')$ is shown with characteristic longitudinal length $a$.  The individual ICDs represent the excess charge density for small windows of time, and $n(z')$ refers to the total excess charge as a function of $z'$.  Approximating the central portion of $n(z')$ as a Gaussian distribution $N(\mu,\sigma)$ corresponds to setting $a = 2 \sigma$.  Askaryan radiation occurs because $n(z')$ represents excess negative charge \cite{zhs,PhysRevD.65.103002,10.1016/j.astropartphys.2017.03.008}.  Cascades may be characterized as \textit{electromagnetic}, initiated by charged outgoing leptons from UHE-$\nu$ interactions, or \textit{hadronic}, initiated by the interaction between the UHE-$\nu$ and the nucleus.  Electromagnetic cascades follow the Greisen distribution and hadronic cascades follow the Gaisser-Hillas distribution.  An example of such an implementation via the ARVZ semi-analytic parameterization is AraSim \cite{10.1016/j.astropartphys.2011.11.010}.

The units of the electromagnetic field in the Fourier domain are V/m/Hz, often converted in the literature to V/m/MHz. To make the distance-dependence explicit, both sides of field equations are multiplied by $r$, as in $r\vec{E} = ... $, making the units V/Hz. Throughout this work, an overall field normalization constant $E_{\rm 0}$ is used.  $E_{\rm 0}$ may be linearly scaled with energy, as in other Askaryan models.  We show that the on-cone field amplitude is proportional to $E_{\rm 0}$ times a characteristic frequency-squared, so the units of $E_{\rm 0}$ are V/Hz$^2$.  For off-cone results, we show that the field amplitude is proportional to $E_{\rm 0}$ times a characteristic frequency divided by a characteristic pulse width, and the units of $E_{\rm 0}$ remain V/Hz$^2$.

In Section \ref{sec:ff2}, we review briefly the energy-dependence of the longitudinal length $a$ in both the electromagnetic and hadronic cases.  For the Greisen distribution with critical energy $E_{\rm crit}$, it can be shown that if $n_{\rm max} = n(z_{\rm max})$, where $z_{\rm max} = \ln{(E_{\rm C}/E_{\rm crit})}$, then $n_{\rm max} a \sim E_{\rm C}/E_{\rm crit}$.  Thus, the area under the curve $n(z')$ scales with the total cascade energy $E_{\rm C}$.  RB demonstrated that the Askaryan radiation amplitude is proportional to $n_{\rm max} a$ and therefore $E_{\rm C}$.  The cascade develops over a length $\approx a$, but the radiation is coherent over a length $\Delta z'_{\rm coh}$ for which the displacement is constant to first order relative to a wavelength.  The $\eta$ parameter is the square of the ratio of $a$ to $\Delta z'_{\rm coh}$:

\begin{equation}
\eta = \left( \frac{a}{\Delta z'_{\rm coh}} \right)^2 = \frac{k}{r} (a\sin\theta)^2
\end{equation}

In far-field, $\eta<1$.  In the first JCH+AC model, a limiting frequency $\omega_{\rm C}$ (Equation \ref{eq:eta1}) was shown to filter the Askaryan radiation \cite{10.1016/j.astropartphys.2017.03.008}:

\begin{equation}
\eta = \frac{\omega}{\omega_{\rm C}}
\label{eq:eta1}
\end{equation}

The effect of $\omega_{\rm C}$ is described in Section \ref{sec:onc}. The Askaryan radiation is primarily polarized in the $\hat{\theta}$-direction, with a small amount along $\hat{r}$ \cite{10.1016/j.astropartphys.2017.03.008,10.1103/physrevd.84.103003}.  The wavevector is $k = (2\pi)/(n\lambda)$, where $n$ is the index of refraction.  A 3D wavevector was defined by RB, equivalent to $\vec{q} = nk(1, \vec{\rho}/R)$.  The vector current density is treated by RB as a charge density times the velocity of the ICD: $\vec{J}(t,\vec{x}') = \rho(z'-vt,\rho') \vec{v}$.  Further, the charge density is factored into $n(z')$ times ICD: $\rho(z'-vt,\rho') = n(z') f(z'-vt,\rho')$.  The form factor $\widetilde{F}$ is the three-dimensional spatial Fourier transform of the ICD \cite{10.1103/physrevd.65.016003}.

The result for $\widetilde{F}$ was derived analytically by JCH+AC \cite{10.1016/j.astropartphys.2017.03.008}, and that derivation is briefly described in Section \ref{sec:ff1}.  JCH+AC define a parameter $\sigma$, and $\widetilde{F}$ is a function of $\sigma$: $\widetilde{F}(\sigma)$.  The variable $\sigma$ is related to the ratio of lateral ICD width to radiated wavelength.  In the derivation of $\widetilde{F}$, it is convenient to set $\sigma$ equal to the ratio of angular frequency to the low-pass cutoff frequency $\omega_{\rm CF}$ of $\widetilde{F}$:

\begin{equation}
\sigma = \frac{\omega}{\omega_{\rm CF}} \label{eq:sigma_0}
\end{equation}

Armed with $\widetilde{F}$, the longitudinal length $a$ and the corresponding energy-dependence on $E_{\rm 0}$, the RB field equations $\vec{\mathcal{E}}$, and the displacement $r$, the Askaryan electromagnetic field may be assembled according to the following form \cite{10.1103/physrevd.65.016003}:

\begin{equation}
r\vec{E}(\omega,\theta) = E_{\rm 0} \left( \frac{\omega}{2\pi} \right) \psi \vec{\mathcal{E}}(\omega,\theta) \widetilde{F}(\omega,\theta) \label{eq:main}
\end{equation}

The factor $E_{\rm 0}$ is proportional to cascade energy.  The factor $\omega$ is the angular frequency.  The variable $\psi$ is $\psi = -i \exp(ikr) \sin\theta$.  The function $\vec{\mathcal{E}}(\omega,\theta)$ contains the vector and complex pole structure of the field (see \cite{10.1103/physrevd.65.016003} and \cite{10.1016/j.astropartphys.2017.03.008}).  The model represented by Equation \ref{eq:main} is an \textit{all-$\theta$, all-$\omega$} model.  That is, Equation \ref{eq:main} is valid at all frequencies and all viewing angles, provided one accepts the approximation of the central portion of $n(z')$ as Gaussian.  The first goal of this work is to build an \textit{all-$\theta$, all-$t$} model in the time-domain, derived from Equation \ref{eq:main}, and the second goal is to compare it to semi-analytic parameterizations.

\section{The Form Factor and Longitudinal Length Parameter}
\label{sec:ff}

To arrive at the main electromagnetic field in the time-domain, the individual pieces of Equation \ref{eq:main} must first be assembled.  The first piece will be the form factor $\widetilde{F}$ that accounts for the 3D ICD, followed by some remarks about the energy-dependence of the longitudinal length parameter $a$.

\subsection{The Form Factor}
\label{sec:ff1}

The form factor is the 3D Fourier transform of the ICD $f(z',\vec{\rho}')$, with $\vec{q} = nk(1, \vec{\rho}/R)$ \cite{10.1103/physrevd.65.016003}:

\begin{equation}
F(\vec{q}) = \int d^3 x' f(z',\rho') e^{-i \vec{q} \cdot \vec{x}'}
\end{equation}

The goal is to evaluate $\widetilde{F}$ in the Fourier domain for an ICD definition informed by cascade simulations.  Simulations of the cascade induced by UHE-$\nu$ indicate a thin wave of charge in $z'$ spread uniformly in $\phi'$, that decreases exponentially in $\rho'$.  Using these observations JCH+AC complete the derivation in \cite{10.1016/j.astropartphys.2017.03.008}.  The final result was a simple analytic formula:

\begin{equation}
\boxed{
\widetilde{F} = \frac{1}{(1+(\omega/\omega_{\rm CF})^2)^{3/2}}
} \label{eq:F_main}
\end{equation}

The form factor acts as a low-pass filter with the cutoff-frequency $\omega_{\rm CF}$:

\begin{equation}
\widetilde{F} \approx \frac{\omega_{\rm 0}^2}{(\omega+i\omega_{\rm 0})(\omega-i\omega_{\rm 0})} \label{eq:two_pole}
\end{equation}

The definition $\omega_{\rm 0} = \sqrt{2/3} ~ \omega_{\rm CF}$ has been used in the final step.  Equation \ref{eq:two_pole} matches the original ZHS parameterization (see Equation 20 of \cite{zhs}).

\begin{figure}
\centering
\includegraphics[width=0.45\textwidth]{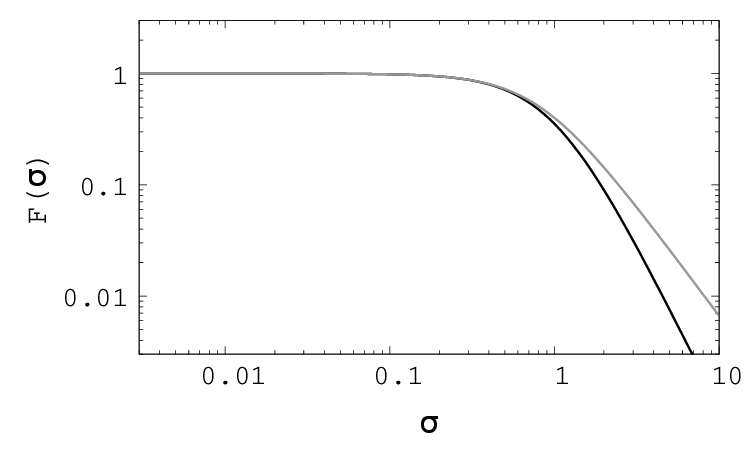}
\caption{\label{fig:F}  (Black) Equation \ref{eq:F_main}, graphed versus $\sigma = \omega/\omega_{\rm CF}$. (Gray) The two-pole approximation.}
\end{figure}

\subsubsection{A Note about the Moli\`{e}re Radius}

In Section \ref{sec:fit:off}, the decay constant $l$ of the lateral component of the ICD is inferred from best-fit values of $\omega_{\rm 0}$.  The connection between the $l$-parameter and $\omega_0$ was described by JCH+AC \cite{10.1016/j.astropartphys.2017.03.008}.  Put simply, the ICD decays by a factor of $1/e$ a lateral distance $l$ from the cascade axis.  Note, however, that the $l$-parameter is not the Moli\`{e}re radius.  The Moli\`{e}re radius is the lateral radius which forms a cylinder containing 90\% of the energy deposition of the cascade.  For ice with a density of $0.917$ g cm$^{-3}$, one can estimate $R_{\rm M} \approx 9.2$ cm using standard formulas.  Although it is tempting to compare $l$ to $R_{\rm M}$, these parameters have different definitions. Knowing that $l$ is related to $\omega_{\rm 0}$, $l$ may be estimated as $\lambda/2$ in ice at the cutoff-frequency.  At 3 GHz in ice, $\lambda/2 \approx 2.8$ cm, and at 1 GHz in ice, $\lambda/2 \approx 8.4$ cm.  Although the results are at the same order of magnitude as $R_{\rm M}$, there are three effects limiting the high-frequency spectrum of the radiation: $\omega_{\rm 0}$, $\omega_{\rm C}$, and the viewing angle.  Thus, $l < R_{\rm M}$ is possible for a radiation spectrum limited to $\lesssim 1$ GHz.  

\subsection{The Longitudinal Length Parameter}
\label{sec:ff2}

The next piece required in the assembly of the main electromagnetic field is the energy-dependence of the overall amplitude, and the energy dependence of the longitudinal length parameter, $a$, which is a part of $\vec{\mathcal{E}}$ in Equation \ref{eq:main} \cite{10.1103/physrevd.65.016003}.  What follows are two separate discussions, one for electromagnetic cascades, and one for hadronic cascades.  Though we share these calculations for convenience, note that a variety of theoretical and experimental results on this topic are available \cite{saltzberg} \cite{ANDRINGA2011360} \cite{10.1088/1742-6596/1879/3/032089}.  

\subsubsection{Electromagnetic Case}

The number of charged particles versus distance in radiation lengths $n(z')$ in an electromagnetic cascade taking place in a dense medium with initial cascade energy $E_C$, \textit{critical energy} $E_{crit}$, normalization parameter $n_{\rm 0}$, and \textit{age} $s$ is \cite{10.1016/j.astropartphys.2017.03.008}

\begin{equation}
n(z') = \frac{n_0}{\sqrt{\ln(E_C/E_{crit})}} \exp \left \lbrace z'\left(1 - \frac{3}{2}\ln(s) \right) \right \rbrace
\end{equation}

To find the energy-dependent width of the Greisen distribution, four steps are necessary: (1) normalization of $n(z')$ as a fraction of the maximum excess charge, (2) conversion of $n(z')$ to $n(s)$, (3) determination of the width of $n(s)$ by approximating the central portion as a Gaussian distribution, and (4) conversion of the width from $s$ units to radiation lengths $z'$, and then converting those results to a distance.  Define the ratio $R = n(z_{\rm max} \pm a/2)/n_{\rm max}$, so the FWHM occurs when $R = 0.5$.  The final result in radiation lengths is

\begin{equation}
\boxed{
a = \sqrt{\ln(E_{\rm C}/E_{\rm crit})} \sqrt{-6\ln(R)}
} \label{eq:main_1}
\end{equation}

Since $R<1$, $\ln(R)<0$ and $a$ is real-valued, and $a$ in Equation \ref{eq:main_1} is in radiation lengths.  In solid ice the density is $\rho_{ice} = 0.917$ g cm$^{-3}$, and the electromagnetic radiation length is $z_0 = 36.08$ g cm$^{-2}$ \cite{10.1016/j.astropartphys.2017.03.008}.  Converting to distance gives 

\begin{equation}
\boxed{
a = \frac{z_0}{\rho_{ice}}\sqrt{\ln(E_{\rm C}/E_{\rm crit})} \sqrt{-6\ln(R)}
}
\end{equation}

Note that $a \propto \sqrt{\ln(E_{\rm C})}$, as shown by RB and others.  The product $n_{\rm max} a$ is proportional to the energy $E_C/E_{\rm crit}$.  For this reason RB took $n_{\rm max} a$ as the field normalization rather than $E_{\rm C}$ \cite{10.1103/physrevd.65.016003}.  As an example, let $R = 0.4$, and $E_{\rm crit} \approx 10^8$ eV, gives $a \approx 4$ meters for $E_{\rm C} = 10^{16}$ eV.  We show in Section \ref{sec:fit} that our fitted $a$-values are close to 4 meters when matched to semi-analytic parameterizations.

\subsubsection{Hadronic Case}

The Gaisser-Hillas distribution describes hadronic cosmic-ray air showers, but has also been applied to hadronic cascades in dense media in codes like AraSim \cite{testbed,10.1016/j.astropartphys.2011.11.010}.  The original function reads

\begin{equation}
n(z') = n_{max} \left( \frac{z'-z_{0}}{z_{max} - z_{0}} \right)^{(z_{max} - z_{0})/\lambda} e^{ \frac{z_{max} - z'}{\lambda}}
\end{equation}

The variables are defined as follows: $n_{max}$ is the instantaneous maximum number of particles in the cascade, $z'$ is the longitudinal distance in radiation lengths, $z_0$ is the initial starting point, $\lambda$ is the interaction length, and $z'_{max}$ is the location of $n_{max}$.  Using the same steps as the electromagnetic case, we find

\begin{equation}
\boxed{
a = \sqrt{\lambda z'_{\rm max}} \sqrt{-8 \ln(R)}
} \label{eq:main_2}
\end{equation}

The $a$ parameter again goes as $\sqrt{z_{max}} \propto \sqrt{\ln(E_{\rm C})}$ which produces similar lengths as the electromagnetic case when scaled by the appropriate interaction length and ice density.

\section{On-Cone Field Equations}
\label{sec:onc}

The $\hat{\theta}$-component of the electromagnetic field at $\theta = \theta_{\rm C}$ will now be built in the time-domain from Equation \ref{eq:main}.  Setting $\theta = \theta_{\rm C}$ in the general RB field equations (Appendix \ref{app:a}), with Equation \ref{eq:F_main} for $\widetilde{F}$, $\sigma = \omega/\omega_{\rm CF}$ and $\eta = \omega/\omega_{\rm CF}$, and letting $E_{\rm 0}$ be proportional to cascade energy $E_{\rm C}$ produces Equation 45 from JCH+AC \cite{10.1016/j.astropartphys.2017.03.008}:

\begin{equation}
r\widetilde{E}(\omega,\theta_{\rm C}) = \frac{(-i\omega) E_0 \sin(\theta_{\rm C}) e^{i\omega r/c}}{(1-i \omega/\omega_{\rm C})^{1/2} (1+(\omega/\omega_{\rm CF})^2)^{3/2}} \label{eq:re_1}
\end{equation}

More detail is provided in Appendix \ref{app:a}. Let the retarded time be $t_{\rm r} = t - r/c$, and let $\omega_0 = \sqrt{\frac{2}{3	}}\omega_{\rm CF}$ and $\hat{E}_{\rm 0} = E_{\rm 0} \sin\theta_{\rm C}$.  Finally, let $\epsilon = \omega_0/\omega_{\rm C}$.  The inverse Fourier transform of Equation \ref{eq:re_1} is

\begin{widetext}
\begin{equation}
r E(t,\theta_{\rm C}) = \frac{\hat{E}_0 i \omega_{\rm C} \omega_{\rm 0}^2}{\pi} \frac{d}{dt_{\rm r}}\int_{-\infty}^{\infty} \frac{ e^{-i\omega t_r}}{ \left(2 i \omega_{\rm C} + \omega \right) (\omega+i\omega_{\rm 0}) (\omega - i\omega_{\rm 0}) } d\omega \label{eq:part1}
\end{equation}
\end{widetext}

In Equation \ref{eq:part1}, the derivative with respect to the retarded time $d/dt_{\rm r}$ is introduced to remove a factor of $(-i\omega)$ from the numerator.  Accounting for the complex poles and the sign of $t_{\rm r}$, complex integration and expansion to first-order in $\epsilon$ yields

\begin{widetext}
\begin{equation}
\boxed{
r E(t,\theta_{\rm C}) = \frac{1}{3} \hat{E}_0 \omega_{\rm CF}^2
\begin{cases}
\left(1 - \frac{1}{2}\epsilon \right)e^{\omega_{\rm 0} t_r} ~~~~~~~~~~~~~~~~~~~~~ t_{\rm r} < 0 \\
\left( 2 e^{-2\omega_{\rm C} t_r} - \left(1+\frac{1}{2}\epsilon \right) e^{-\omega_{\rm 0}t_r} \right) ~~ t_{\rm r} > 0
\end{cases}
} \label{eq:on_cone}
\end{equation}
\end{widetext}

Equation \ref{eq:on_cone} represents the time-domain solution for the on-cone $\hat{\theta}$-component of the Askaryan electric field.  The expansion to first-order in $\epsilon$ is only performed so the final result resembles semi-analytic results for $\vec{E} = -\partial\vec{A}/\partial t_{\rm r}$ \cite{10.1103/physrevd.84.103003,PhysRevD.101.083005}. Table \ref{tab:features_1} summarizes the definitions of the parameters in Equation \ref{eq:on_cone}.  Fit results for the parameters of Table \ref{tab:features_1} are shown in Section \ref{sec:fit}.

\begin{table}
\renewcommand{\arraystretch}{1.5}
\begin{tabular}{| c | c |}
\hline
Parameter & Definition \\ \hline
$\hat{E}_{\rm 0}$ & $E_{\rm 0} \sin(\theta_{\rm C})$ \\
$E_{\rm 0}$ & $\approx n_{\rm max}a$ \\
$\omega_{\rm 0}$ & $\sqrt{\frac{2}{3}} \omega_{\rm CF}$ \\
$\omega_{\rm CF}$ & $(c\sqrt{2\pi}\rho_{\rm 0})/(n\sin\theta)$ (see Eqs. 22,23, and 46 of \cite{10.1016/j.astropartphys.2017.03.008}) \\
$\omega_{\rm C}$ & $(rc)/(na^2\sin^2\theta)$ (see Eq. 39 of \cite{10.1016/j.astropartphys.2017.03.008})\\
$\epsilon$ & $\omega_{\rm 0}/\omega_{\rm C}$ \\
$t_{\rm r}$ & $t - r/c$ \\ \hline
\end{tabular}
\caption{\label{tab:features_1} The parameters used to build Equation \ref{eq:on_cone}.  Fitted values in comparison to semi-analytic parameterizations are shown in Section \ref{sec:fit}.}
\end{table} 

\begin{figure}
\centering
\includegraphics[width=0.45\textwidth,trim=0cm 6cm 0cm 7cm,clip=true]{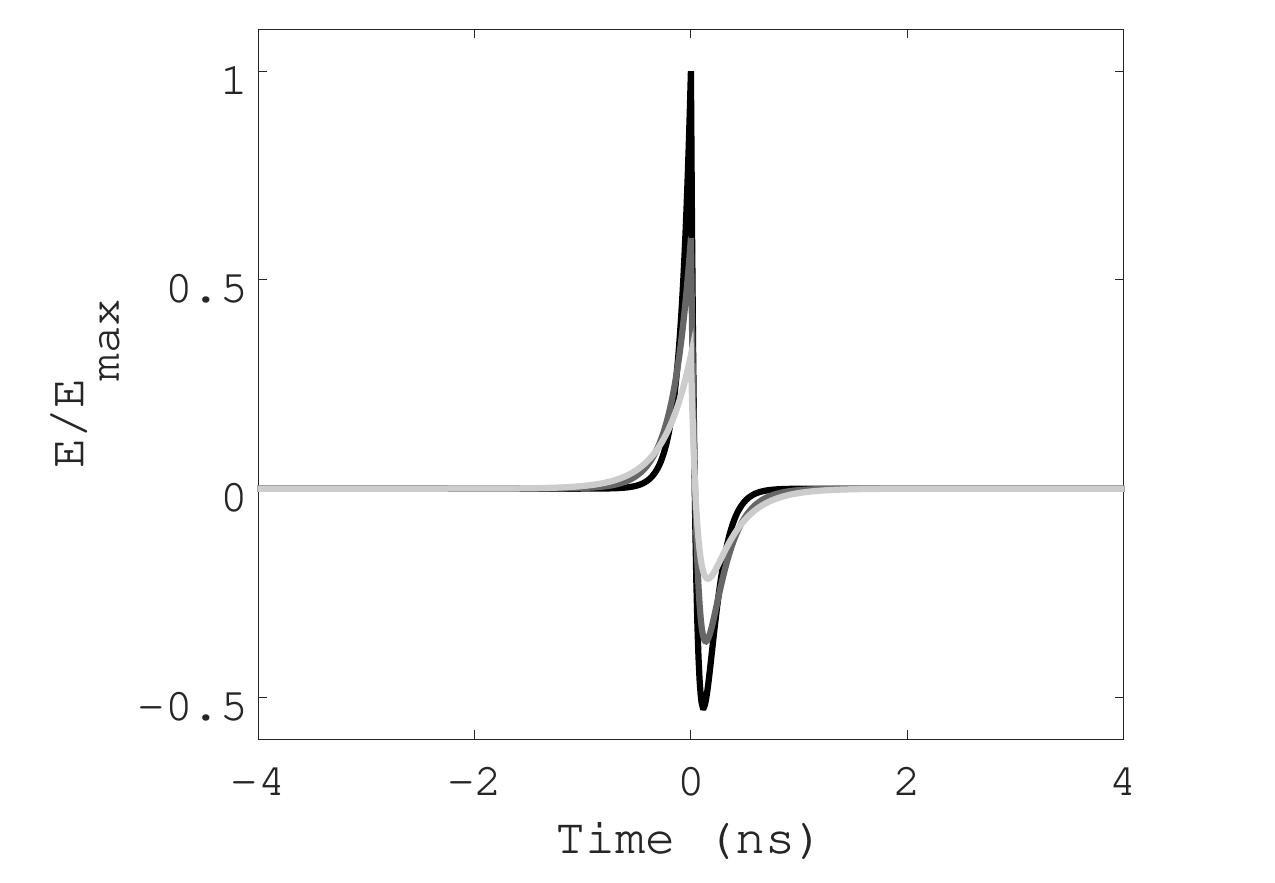}
\includegraphics[width=0.45\textwidth,trim=0cm 6cm 0cm 7cm,clip=true]{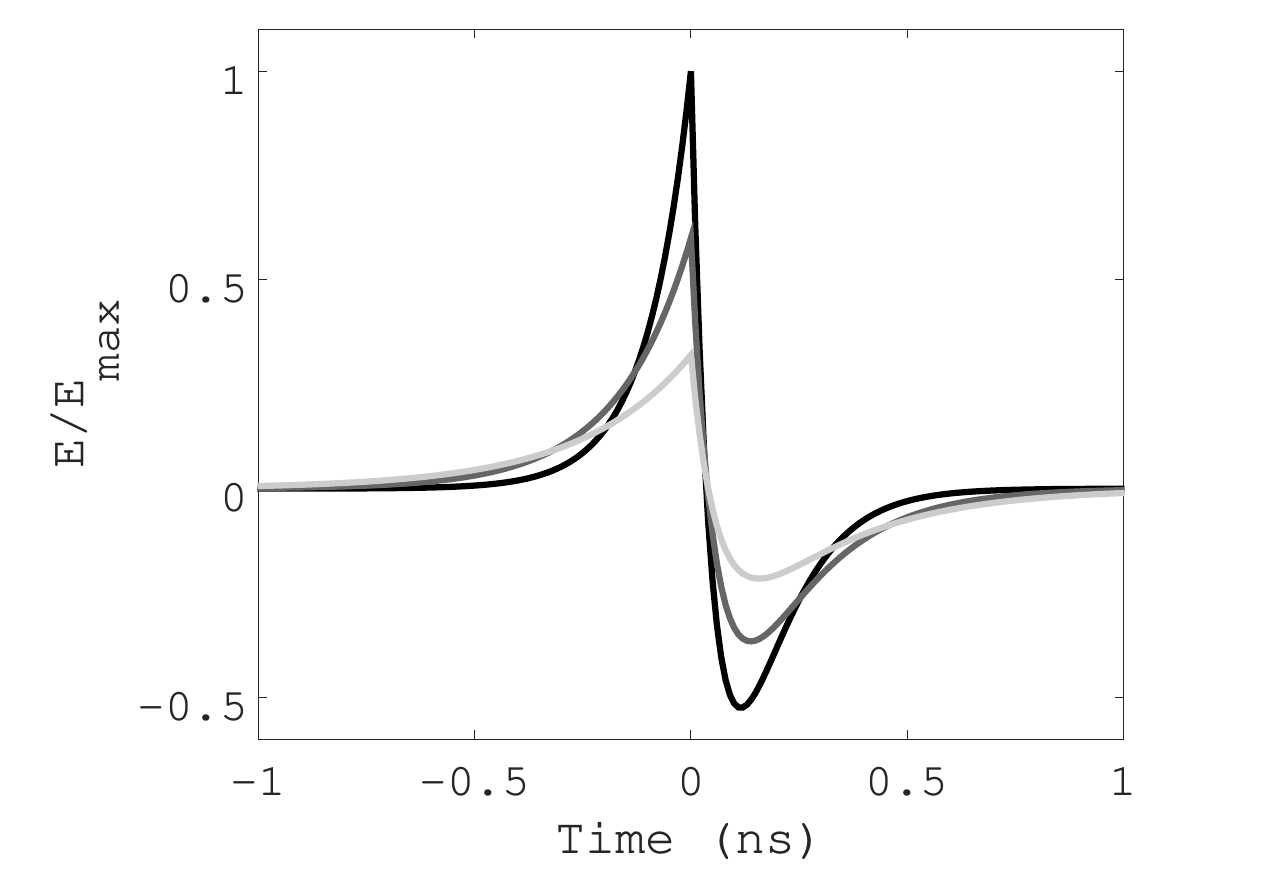}
\caption{\label{fig:on_cone} (Top) Equation \ref{eq:on_cone} from $[-4,4]$ ns, with (black) $\omega_{\rm C} = 2\pi(1.25)$ GHz, $\omega_{\rm 0} = 2\pi(1.56)$ GHz, $\epsilon = 1.25$, (gray) $\omega_{\rm C} = 2\pi(1.25)$ GHz, $\omega_{\rm 0} = 2\pi(0.94)$ GHz, $\epsilon = 0.75$, (light gray) $\omega_{\rm C} = 2\pi(1.25)$ GHz, $\omega_{\rm 0} = 2\pi(0.625)$ GHz, $\epsilon = 0.5$.  The amplitudes of all curves are normalized to the peak of the $\epsilon = 1.25$ (black) data. (Bottom) Same as top panel, plotted between $[-1,1]$ ns.}
\end{figure}

Notice that the amplitude is asymmetric, and the the parameter $\epsilon$ influences the asymmetry.  The $\epsilon$ parameter was studied in JCH+AC in detail.  For example, Fig. 10 of \cite{10.1016/j.astropartphys.2017.03.008} shows that $\epsilon \approx [0.1 - 1]$ for inverse lateral width $l^{-1} = \sqrt{2\pi} \rho_{\rm 0} \approx 20$ m$^{-1}$ and $a \approx 4$ m.  The best-fit results for $\epsilon$ and $a$ are shown in Section. \ref{sec:fit}.  JCH+AC showed that the expression for $\epsilon$ is the product of the ratio of the lateral to longitudinal length, and the ratio of the longitudinal length to the observer displacement, making it a physical parameter connecting the event geometry to the cacscade shape \cite{10.1016/j.astropartphys.2017.03.008}.  Figure \ref{fig:on_cone} displays normalized examples of Equation \ref{eq:on_cone} for different values of $\omega_{\rm 0}$, $\omega_{\rm C}$, and $\epsilon$.

\subsection{Verification of the Uncertainty Principle}
\label{sec:sigma0}

As a check on the procedures used to perform the inverse Fourier transform that produces Equation \ref{eq:on_cone}, we verify below that the uncertainty principle holds, for $\Delta\theta \to 0$.  JCH+AC provide the Gaussian width of the radiation in the Fourier domain: $\sigma_{\rm \nu}$, where $\nu$ represents the frequency in Hz.  Generally speaking, Fourier transform pairs must obey $\sigma_\nu \sigma_t \geq 1/(2\pi)$.  The following procedure is used to compute the width $\sigma_{\rm t}$ of the on-cone field.  First, the $t_{\rm r} < 0$ and $t_{\rm r} > 0$ cases are each treated as probability distributions and normalized.  Next, the average positive and negative retarded times, $\bar{t}_{\rm r,+}$ and $\bar{t}_{\rm r,-}$, are computed.  Finally, subtracting the two averages yields $\sigma_{\rm t}$:

\begin{equation}
\sigma_{\rm t} = \bar{t}_{\rm r,+} - \bar{t}_{\rm r,-} = \frac{\epsilon+2}{\omega_{\rm 0}} = \frac{1}{\omega_{\rm C}} + \frac{2}{\omega_{\rm 0}} \label{eq:width_oncone}
\end{equation}

The result has the correct units and the limiting cases are sensible.  Suppose $\epsilon \to 0$ ($\omega_{\rm C} \gg \omega_{\rm 0}$), then $\sigma_{\rm t} \to 2/\omega_{\rm 0}$, which is expected from observing Equation \ref{eq:on_cone} if the $\omega_{\rm C}$ exponential disappears.  If $\epsilon = 1$ ($\omega_{\rm C} = \omega_{\rm 0}$), then $\sigma_{\rm t} = 3/\omega_{\rm 0}$.  That is, the pulse is wider if there is more than one relevant cutoff frequency.

The expression for $\sigma_{\rm \nu}$ is given by Equation 36 of JCH+AC \cite{10.1016/j.astropartphys.2017.03.008}:

\begin{equation}
\sigma_{\rm \nu} = \frac{c}{2\pi a \Delta\cos\theta}\left(1+\eta^2\right)^{1/2} \label{eq:sigmanu}
\end{equation}

Expanding to first order in $\Delta\cos(\theta) = \cos(\theta) - \cos(\theta_{\rm C})$,

\begin{equation}
\sigma_{\rm \nu} \approx \frac{c}{2\pi a \sin(\theta_{\rm C}) \Delta\theta}\left(1+\eta^2\right)^{1/2} \label{eq:sigmanu2}
\end{equation}

From Table \ref{tab:features_1}: $\omega_{\rm C}^{-1} = n a^2 \sin^2(\theta_{\rm C})/(rc)$, and $\omega_{\rm 0}^{-1} = n l \sin(\theta_{\rm C})/c$, with $l = \sqrt{3/2}/(\sqrt{2\pi} \rho_{\rm 0})$.  (Recall that $\rho_{\rm 0}$ is a parameter discussed in \cite{10.1016/j.astropartphys.2017.03.008}). Multiplying $\sigma_{\rm t}$ and $\sigma_{\rm \nu}$ with the far-field limit ($\eta < 1$) gives the inequality

\begin{equation}
\sigma_{\rm \nu} \sigma_{\rm t} \geq \frac{n}{2\pi} \left( \left(\frac{a}{r}\right) \frac{\sin(\theta_{\rm C})}{\Delta \theta} + 2 \left( \frac{l}{a} \right) \frac{1}{\Delta \theta} \right)
\end{equation}

Therefore, in order to satisfy $\sigma_{\rm \nu} \sigma_{\rm t} > 1/(2\pi)$,

\begin{equation}
n\left(\frac{a}{r}\right)\sin(\theta_{\rm C}) + 2 n \left( \frac{l}{a} \right) > \Delta\theta \label{eq:siglimit}
\end{equation}

Although $a/r \ll 1$ and $l/a \ll 1$, as long as these expressions do not approach zero as fast as $\Delta\theta \to 0$ in Equation \ref{eq:siglimit}, the uncertainty principle holds.  Yet these are exactly the conditions of the problem: a displacement $r$ in the far-field (but not infinitely far away) and a longitudinal length $a$ much larger (but not infinitely larger) than the lateral ICD width $l$.  Thus, $\sigma_{\rm \nu} \sigma_{\rm t} > 1/(2\pi)$ holds.

\section{Off-Cone Field Equations}
\label{sec:ofc}

Turning to the case for which $\theta \neq \theta_{\rm C}$, the $\hat{\theta}$-component of the electromagnetic field  will now be built in the time-domain.  The RB field equations for the $\hat{\theta}$ and $\hat{r}$ components are summarized in both RB and JCH+AC \cite{10.1103/physrevd.65.016003,10.1016/j.astropartphys.2017.03.008}, and Appendix \ref{app:a}.  Recall the general form of the electromagnetic field, given in Equation \ref{eq:main}:

\begin{equation}
r\vec{E}(\omega,\theta) = E_{\rm 0} \left( \frac{\omega}{2\pi} \right) \psi \vec{\mathcal{E}}(\omega,\theta) \widetilde{F}(\omega,\theta) \label{eq:main2}
\end{equation}

The first task is to simplify $\vec{\mathcal{E}}(\omega,\theta)$ before taking the inverse Fourier transform. The simplification inolves expanding $\vec{\mathcal{E}}(\omega,\theta)$ in a Taylor series such that $u = 1-i\eta \approx 1$, restricting $\eta < 1$ (far-field).  Once $\vec{\mathcal{E}}(\omega,\theta)$ is simplified, the inverse Fourier transform of Equation \ref{eq:main2} may be evaluated to produce the result.  Table \ref{tab:features_2} contains useful variable definitions, Table \ref{tab:features_3} contains useful function definitions, and Table \ref{tab:features_4} contains special cases of the functions in Table \ref{tab:features_3}.

\begin{table}
\renewcommand{\arraystretch}{1.5}
\begin{tabular}{| c | c |} \hline
Variable & Definition \\ \hline
$u$ & $1-i\eta$ \\
$x$ & $\cos(\theta)$ \\
$x_{\rm C}$ & $\cos(\theta_C)$ \\
$q$ & $(x x_C - x^2_C)/(1-x^2)$ \\
$y$ & $\left( \frac{1}{2} \right) (ka)^2 (\cos\theta - \cos\theta_C)^2$ \\
$p$ & $\frac{1}{2}\left(\frac{a}{c}\right)^2 \left(\cos\theta - \cos\theta_C\right)^2$ \\ \hline
\end{tabular}
\caption{\label{tab:features_2} Useful variables for the derivation of the off-cone Askaryan electromagnetic field.}
\end{table}

\begin{table}
\renewcommand{\arraystretch}{2}
\begin{tabular}{| c | c |} \hline
Function & Definition \\ \hline
$f(u,x)$ & $\left( u + 3 \frac{(1-u)^2}{u} \frac{x^2 - x x_C}{1-x^2} \right)^{-1/2}$ \\
$g(u,x)$ & $\exp \left( -\frac{1}{2}(ka)^2 (x - x_C)^2 u^{-1} \right)$ \\
$h(u,x)$ & $\left( \frac{1-u}{u} \right) q$ \\
$\vec{\mathcal{E}}(u,x) \cdot \hat{\theta}$ & $f(u,x) g(u,x)(1 - h(u,x))$ \\ \hline
\end{tabular}
\caption{\label{tab:features_3} Useful functions for the derivation of the off-cone Askaryan electromagnetic field.  The last row contains the vector structure of the $\hat{\theta}$-component of the field.}
\end{table}

\begin{table}
\begin{tabular}{| c | c |} \hline
Function ($u = 1$) & Result \\ \hline
$f(x,1)$ & $1$ \\
$\dot{f}|_{u = 1}$ & $-\frac{1}{2}$ \\
$g(x,1)$ & $\exp(-y)$ \\
$\dot{g}|_{u = 1}$ & $y \exp(-y)$ \\
$h(x,1)$ & $0$ \\
$\dot{h}|_{u = 1}$ & $-q$ \\ \hline
\end{tabular}
\caption{\label{tab:features_4} Special cases of the functions defined in Table \ref{tab:features_3}, when $u = 1$.}
\end{table}

The original form of $\vec{\mathcal{E}}(\eta,\theta)$ is shown in Appendix \ref{app:a}.  Changing variables to $u$ and $x$ (Tab. \ref{tab:features_2}) and using the function definitions and values in Tabs. \ref{tab:features_3}-\ref{tab:features_4}, $\vec{\mathcal{E}}(u,x) \cdot \hat{\theta} = \mathcal{E}(u,x)$ becomes

\begin{equation}
\mathcal{E}(u,x) = f(u,x) g(u,x) (1 - h(u,x))
\end{equation}

Expanding $\mathcal{E}(u,x)$ near $u = 1$ gives

\begin{equation}
\mathcal{E}(u,x) = \mathcal{E}(x,1) + (u-1) \dot{\mathcal{E}}(x,1) + \mathcal{O}(u-1)^2 \label{eq:expand}
\end{equation}

The details of the expansion are shown in Appendix \ref{app:b}.  The result is

\begin{equation}
\mathcal{E}(x,u) = e^{-y} \left( 1 - \frac{1}{2} j\eta\left( 2y + 2q - 1 \right) \right) \label{eq:expand_rev_1}
\end{equation}

The inverse Fourier transform of the $\hat{\theta}$-component gives the time-domain results, after including the expanded $\mathcal{E}(u,x)$:

\begin{equation}
r E(t,\theta) = \mathcal{F}^{-1} \left\lbrace E_0 \left(\frac{\omega}{2\pi}\right) \widetilde{F} \psi \mathcal{E} \right\rbrace
\end{equation}

Intriguingly, the result is proportional to the \textit{line-broadening function, $H$} (DLMF 7.19, \cite{NIST:DLMF}) common to spectroscopy applications.  There are three terms in Equation \ref{eq:expand_rev_1}.  Two terms ultimately vanish, being integrals over odd integrands (see Appendix \ref{app:b}).  The integral that remains contains $H$, with $\omega_{\rm 1} = t_{\rm r}/(2p)$:

\begin{equation}
I_{\rm 0} = 2 \pi i \left( \frac{\omega_{\rm C}}{\omega_{\rm 0}} \right) e^{- \frac{t_{\rm r}^2}{4p}} H(\sqrt{p}\omega_0,i\omega_1\sqrt{p})
\end{equation}

The line-broadening function is similar to a convolution between a Gaussian function and a Lorentzian function, and cannot be expressed analytically, though there are examples of polynomial expansions \cite{10.1111/j.1365-2966.2006.10450.x}.  Note that, for situations relevant to the current problem, $\omega > \omega_{\rm 1}$.  Requiring that $\omega > \omega_{\rm 1}$ amounts to a restriction between $\Delta\theta$ and $|t_{\rm r}|$:

\begin{equation}
|t_{\rm r}| < | 2 p \omega | \label{eq:restrict}
\end{equation}

It is shown in the next section that $\sqrt{2p}$ is the pulse width $\sigma_{\rm t}$, so $| 2 p \omega |$ has units of time.  Using the results of Sec. \ref{sec:sigma} below, the restriction on the retarded time may be written $|t_{\rm r}|/\sigma_{\rm t} < \omega \sigma_{\rm t} = 2\pi (\sigma_{\rm t}/T)$.  That is, the accuracy of the waveform should be trusted within a number of pulse widths that is less than $2\pi$ times the ratio of the pulse width to the period of the lowest frequency.  This is not a strong requirement, since the field quickly approaches zero after several pulse widths.  Hereafter, this step will be called the \textit{symmetric approximation}, because the result for $r \vec{E}(t_{\rm r},\theta)$ in Equation \ref{eq:off_cone} has equal positive and negative amplitude.  Evaluating the line-broadening function numerically would account for amplitude asymmetry.  The restriction on $\Delta\theta$ is formalized in Sec. \ref{sec:usage}.  Solving $I_{\rm 0}$ using the symmetric approximation clears the way for the final result (see Appendix \ref{app:b}):

\begin{equation}
\boxed{
r E(t,\theta) = -\frac{E_0 \omega_0 \sin(\theta)}{8 \pi p} t_r e^{-\frac{t_r^2}{4p} + p \omega_0^2}\erfc(\sqrt{p}\omega_0)
} \label{eq:off_cone}
\end{equation}

\begin{figure}
\centering
\includegraphics[width=0.45\textwidth,trim=0cm 6cm 0cm 7cm,clip=true]{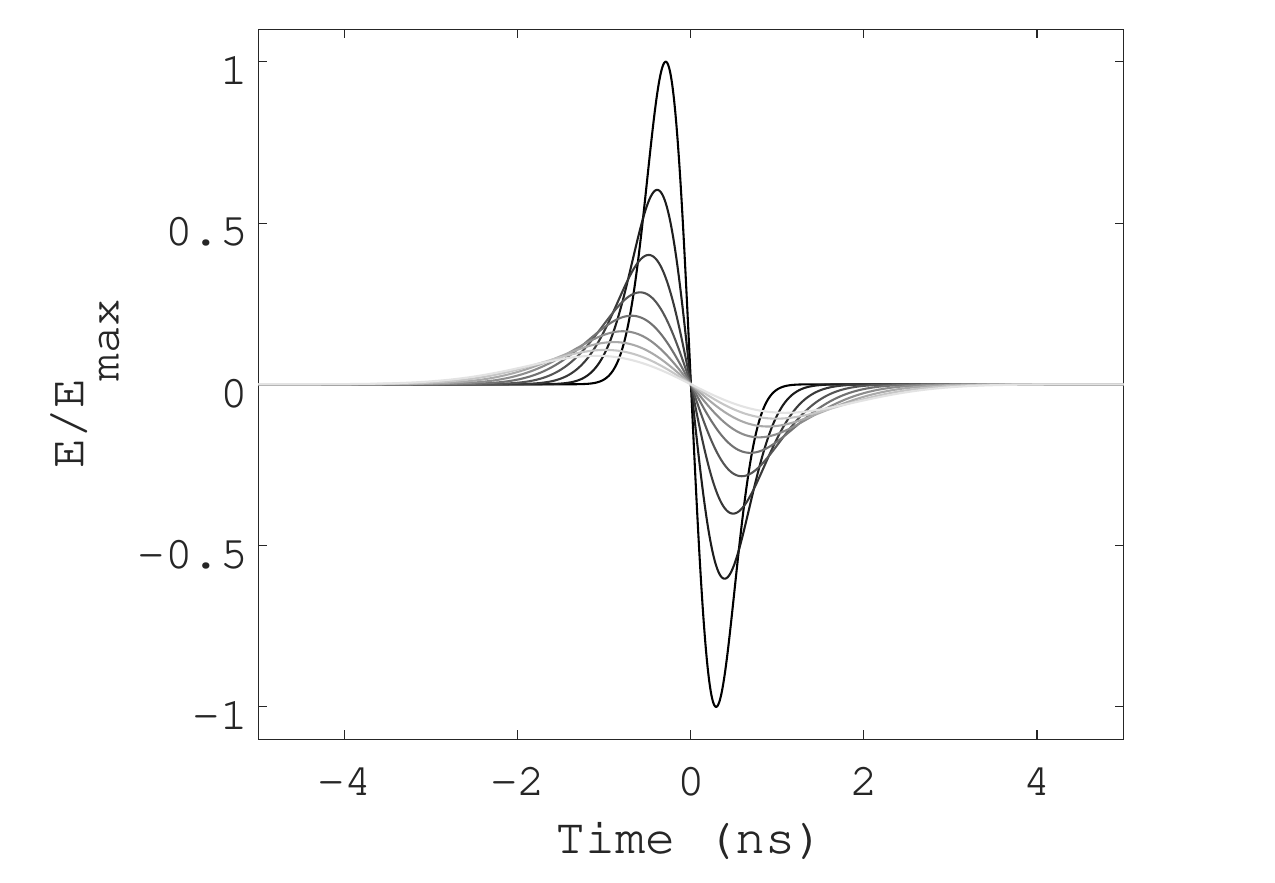}
\includegraphics[width=0.45\textwidth,trim=0cm 6cm 0cm 7cm,clip=true]{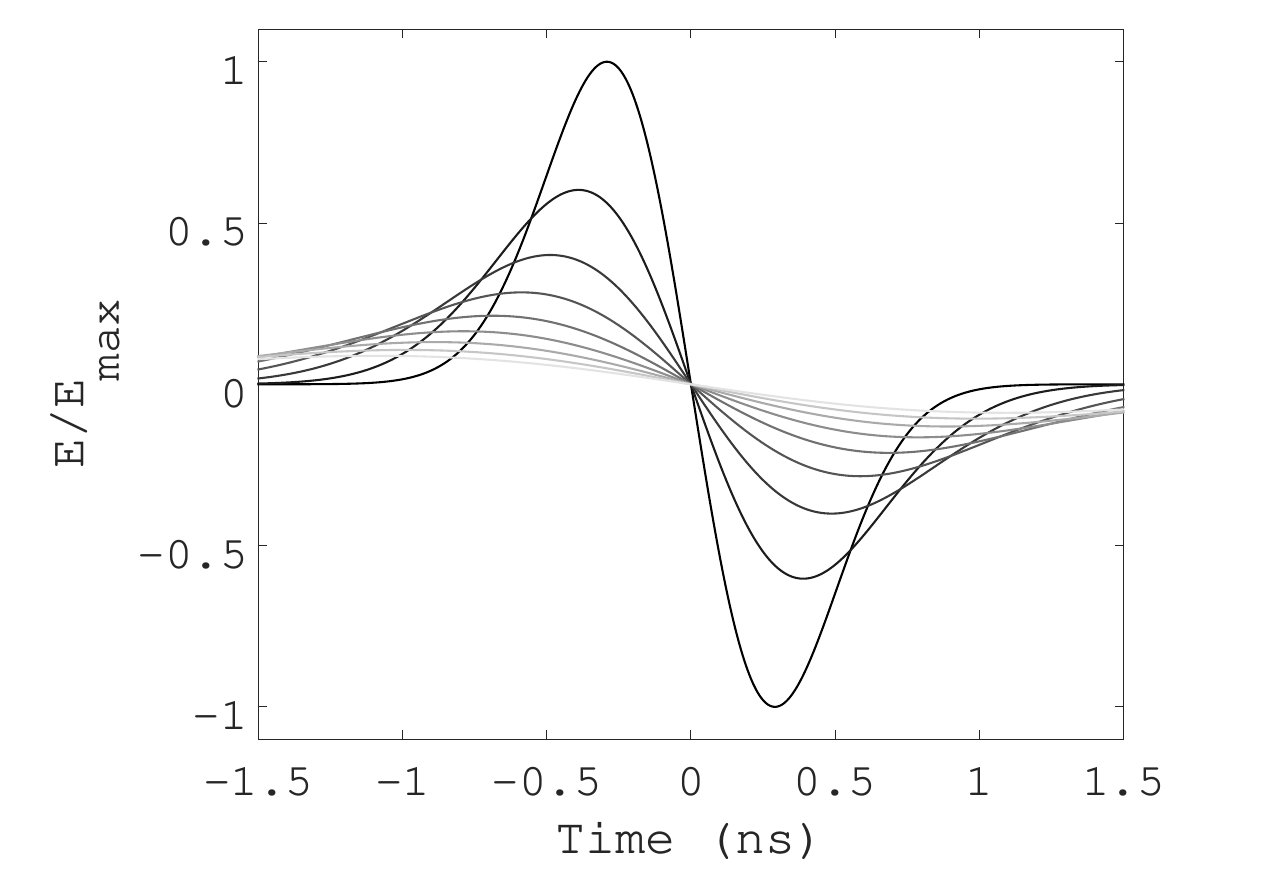}
\caption{\label{fig:off_cone} $E(t,\theta)$ vs. $t_{\rm r}$ (Equation \ref{eq:off_cone}), normalized.  The viewing anlge $\theta$ is varied from $\theta_{\rm C}+1.5^{\circ}$ to $\theta_{\rm C}+5.5^{\circ}$ in steps of $0.5^{\circ}$.  Top: $\omega_{\rm 0}/(2\pi) = 1.0$ GHz.  Bottom: Same as top, zoomed in on central region.}
\end{figure}

\begin{figure}
\centering
\includegraphics[width=0.45\textwidth,trim=0cm 6cm 0cm 6cm,clip=true]{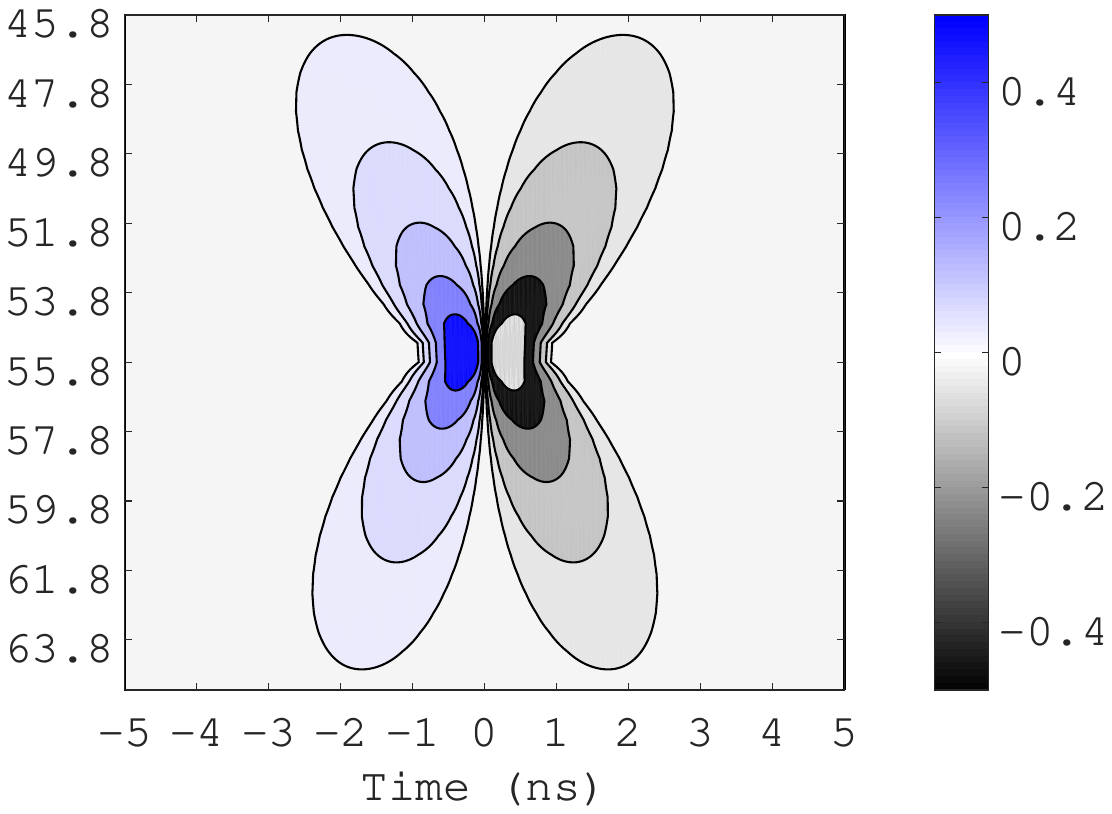}
\caption{\label{fig:off_cone2} Contours of $E(t,\theta)$ vs. $\theta$ vs. $t_{\rm r}$ (Equation \ref{eq:off_cone}), normalized.  The normalization is the same as Fig. \ref{fig:off_cone}.  Although the contour lines extend into the region near $\theta_{\rm C}$, Equation \ref{fig:off_cone2} is only being evaluated at $\Delta \theta > 1.5^{\circ}$ (see text for details).}
\end{figure}

Equation \ref{eq:off_cone} represents the time-domain solution for the off-cone $\hat{\theta}$-component of the Askaryan electric field.  Equation \ref{eq:off_cone} is graphed in Figs. \ref{fig:off_cone} and \ref{fig:off_cone2}. In Fig. \ref{fig:off_cone} (top),  $E(t,\theta)$ is shown normalized to the maximum value for the angular range displayed, $[\theta_{\rm C}+1.5^{\circ}, \theta_{\rm C}+5.5^{\circ}]$, from $t_{\rm} = [-5,5]$ ns.  Pulses with viewing angles closer to $\theta_{\rm C}$ have larger relative amplitudes and shorter pulse widths.  Figure \ref{fig:off_cone} (bottom) contains the same results, but for $t_{\rm} = [-1.5,1.5]$ ns.  The pulses are symmetric and all zero crossings are at $t_{\rm r} = 0$ ns as a result of the symmetric approximation.  Figure \ref{fig:off_cone2} contains contours of the same results as in Fig. \ref{fig:off_cone}.

As in the on-cone result, the overall field amplitude scales with energy ($E_{\rm 0} \sim n_{\rm max} a$).  However, the amplitude scales also with $\omega_{\rm 0}/p$.  The argument of the complementary error function, $\sqrt{p}\omega_0$, is unitless.  This factor is strictly positive, so the range of the complementary error function is $(0,1)$.  The factor $\sqrt{p}\omega_0$ cannot be zero without setting $\theta = \theta_{\rm C}$, or setting $\omega_{\rm CF} = 0$.  Both cases are not allowed.  Equation \ref{eq:off_cone} represents the \textit{off-cone} ($\theta \neq \theta_{\rm C}$) solution, so $p \neq 0$.  Setting $\omega_{\rm CF} = 0$ is not physical, for this implies infinite lateral width ($l$) and cascade particles have finite transverse momentum.  Another possibility is that $p = 0$ if $a = 0$, but this implies $E_{\rm 0}=0$.  Therefore, $0 < \erfc(\sqrt{p}\omega_0) < 1$.

\subsection{Verification of the Uncertainty Principle}
\label{sec:sigma}

As in Section \ref{sec:sigma0}, the uncertainty principle should be checked.  Equation \ref{eq:off_cone} is an anti-symmetric Gaussian function with pulse width $\sigma_{\rm t} = \sqrt{2p}$.  Let $\Delta\cos\theta = (\cos\theta - \cos\theta_{\rm C})$.  Using Table \ref{tab:features_2}, the expression $\sqrt{2p}$ evaluates to

\begin{equation}
\sigma_{\rm t} = \sqrt{2p} = \left(\frac{a}{c}\right)(\Delta \cos\theta) \label{eq:a_versus_theta}
\end{equation}

Recall that $\sigma_{\rm \nu}$ is given by

\begin{equation}
\sigma_{\rm \nu} = \frac{c}{2\pi a \Delta\cos\theta}\left(1+\eta^2\right)^{1/2} \label{eq:sigmanu3}
\end{equation}

The uncertainty product is

\begin{equation}
\sigma_{\rm t} \sigma_{\rm \nu} = \frac{1}{2 \pi} \left(1+\eta^2\right)^{1/2} \label{eq:sigma_result}
\end{equation}

In the far-field, $\eta < 1$, so $\sigma_{\rm t} \sigma_{\rm \nu} \geq 1/(2\pi)$ holds: 

\subsection{Usage of the On-Cone versus Off-Cone Fields}
\label{sec:usage}

The form of Equation \ref{eq:off_cone}, and the restriction between $\Delta\theta$ and $|t_{\rm r}|$ from the symmetric approximation suggests the limit $\Delta\theta \to 0$ must be examined carefully.  Since $p \propto (\cos\theta - \cos\theta_{\rm C})^2$, probing the model near $\theta = \theta_{\rm C}$ is equivalent to taking the limit that $p \to 0$.  Intriguingly, the $p^{-1}$-dependence in the field does not lead to a divergence.  As the field grows in amplitude from $p^{-1}$ as $p \to 0$, the field \textit{width}, $\sqrt{2p}$, approaches zero.

Equations \ref{eq:width_oncone} and \ref{eq:a_versus_theta} contain the pulse widths of the on-cone and off-cone fields, respectively.  Power in the off-cone case is limited by the pulse width $\sqrt{2p}$, and the observed power increases as $\Delta\theta$ and $\sqrt{2p}$ both decrease.  Thus, a reasonable constraint on when $\Delta\theta_{\rm min}$ is large enough to use Equation \ref{eq:off_cone} is given by setting the off-cone pulse width equal to the on-cone pulse width:

\begin{equation}
\frac{1}{\omega_{\rm C}} + \frac{2}{\omega_{\rm 0}} = \sqrt{2p}
\end{equation}

Expanding the expression for $p$ near $\theta = \theta_{\rm C}$, and evaluating the square root leads to

\begin{equation}
\frac{1}{\omega_{\rm C}} + \frac{2}{\omega_{\rm 0}} = \frac{a}{c}\sin\theta_{\rm C} \Delta\theta_{\rm min}
\end{equation}

Using $\epsilon = \omega_{\rm 0}/\omega_{\rm C}$, and letting $k_{\rm 0} = \omega_{\rm 0}/c$, the formula may be rearranged:

\begin{equation}
\epsilon + 2 = a k_{\rm 0} \sin\theta_{\rm 0}\Delta\theta_{\rm min}
\end{equation}

Squaring both sides, and then dividing both sides by $r$ yields

\begin{equation}
\frac{(\epsilon+2)^2}{r} = k_{\rm 0} \left(\frac{k_{\rm 0} (a\sin\theta_{\rm C})^2}{r}\right)\Delta\theta^2_{\rm min}
\end{equation}

The quantity in parentheses on the right-hand side is $\eta$, with $\omega = \omega_{\rm 0}$.  Setting $\omega = \omega_{\rm 0}$ means $\eta = \epsilon$.  Solving for $\Delta\theta_{\rm min}$ gives

\begin{equation}
\Delta\theta_{\rm min} = \frac{\epsilon+2}{\sqrt{\epsilon k_{\rm 0} r}}
\end{equation}

Assuming $\epsilon \approx 1$, $f_{\rm 0} \approx 1$ GHz, $n = 1.78$ for solid ice, and $c = 0.3$ m ns$^{-1}$ (see Sec. \ref{sec:fit:on}), $k_{\rm 0} \approx 35$ m$^{-1}$.  Taking $r = 1000$ m, $\Delta\theta_{\rm min} \approx 1^{\circ}$.  Simple rules-of-thumb for the application of Equation \ref{eq:off_cone} field are:

\begin{align}
\Delta\theta_{\rm min} &\geq 1^{\circ} \\
\Delta\theta_{\rm min} &\propto \frac{1}{\sqrt{kr}}
\end{align}

\section{Comparison to Semi-Analytic Parameterizations}
\label{sec:fit}

The fully analytic model will now be compared to the ARVZ semi-analytic parameterization used in NuRadioMC to predict signals in IceCube-Gen2 Radio \cite{10.1140/epjc/s10052-020-7612-8}.  Specifically, the comparison is between Equations \ref{eq:on_cone} and \ref{eq:off_cone} and the NuRadioMC implementation of the semi-analytic parameterization given in \cite{PhysRevD.101.083005}.  To provide concrete comparisons, a small set of waveforms was generated with NuRadioMC, for both electromagnetic and hadronic cascades, on and off-cone.  The electromagnetic cascades have $E_{\rm C} = 10^{16}$ eV, while the hadronic cascades have $E_{\rm C} = 10^{17}$ eV.  These choices minimize the impact of the LPM effect, though the LPM effect was activated in the NuRadioMC code.

The comparison involves three stages.  First, waveforms and $a$-values are generated for each cascade type, energy, and angle: $\theta = \theta_{\rm C} + 3.0^{\circ}$, and $\theta = \theta_{\rm C}$.  Second, Equations \ref{eq:on_cone} and \ref{eq:off_cone} are tuned to match the waveforms.  In each fit, the Pearson correlation coefficient ($\rho$) is maximized, and the sum-squared of amplitude differences ($(\Delta E)^2$) is minimized.  Finally, best-fit parameters are tabulated. 

Two remarks are important regarding the fit criteria.  First, the Pearson correlation coefficient is not sensitive to changes in amplitude because it is normalized:

\begin{equation}
\rho = \frac{\cov(f_{\rm data},f_{\rm model})}{\sigma_{\rm data} \sigma_{\rm model}}
\end{equation}

Parameters that affect $\rho$ are those that scale $t_{\rm r}$.  Second, parameters that control $(\Delta E)^2$ are those that scale the waveform amplitude.  If $E_i$ represent the samples of the models, then

\begin{equation}
(\Delta E)^2 = \sum_{i = 1}^{N} (E_{i,data} - E_{i,model})^2
\end{equation}

\subsection{Waveform Comparison: $\theta = \theta_{\rm C}$}
\label{sec:fit:on}

\begin{figure}
\centering
\includegraphics[width=0.42\textwidth,trim=0cm 8.5cm 7cm 3.5cm,clip=true]{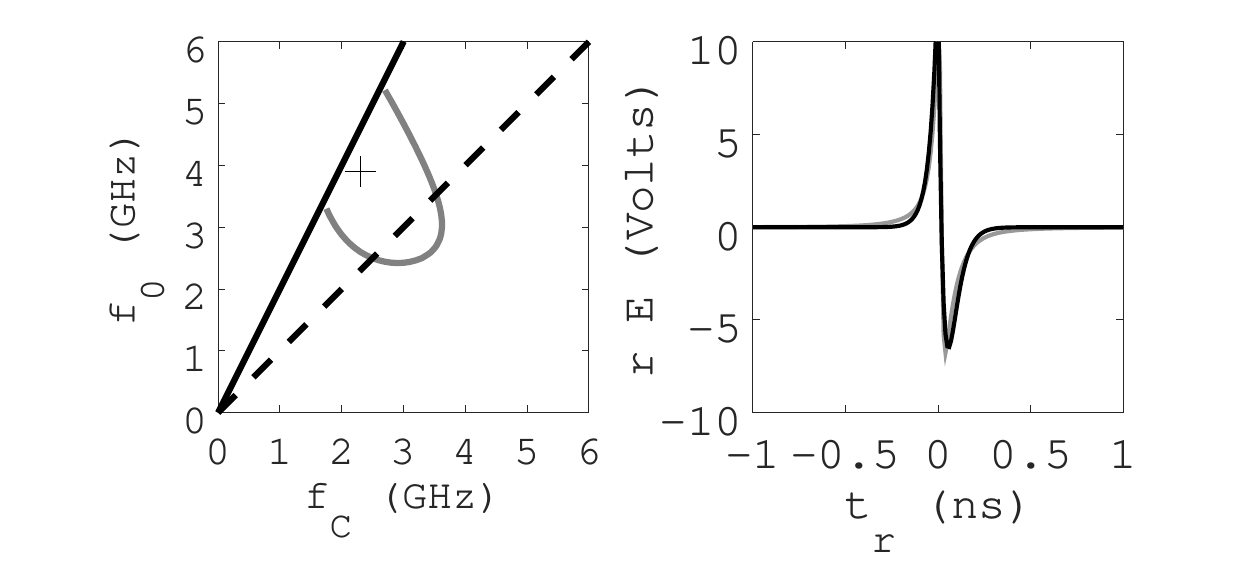}
\includegraphics[width=0.42\textwidth,trim=0cm 8.5cm 7cm 3.5cm,clip=true]{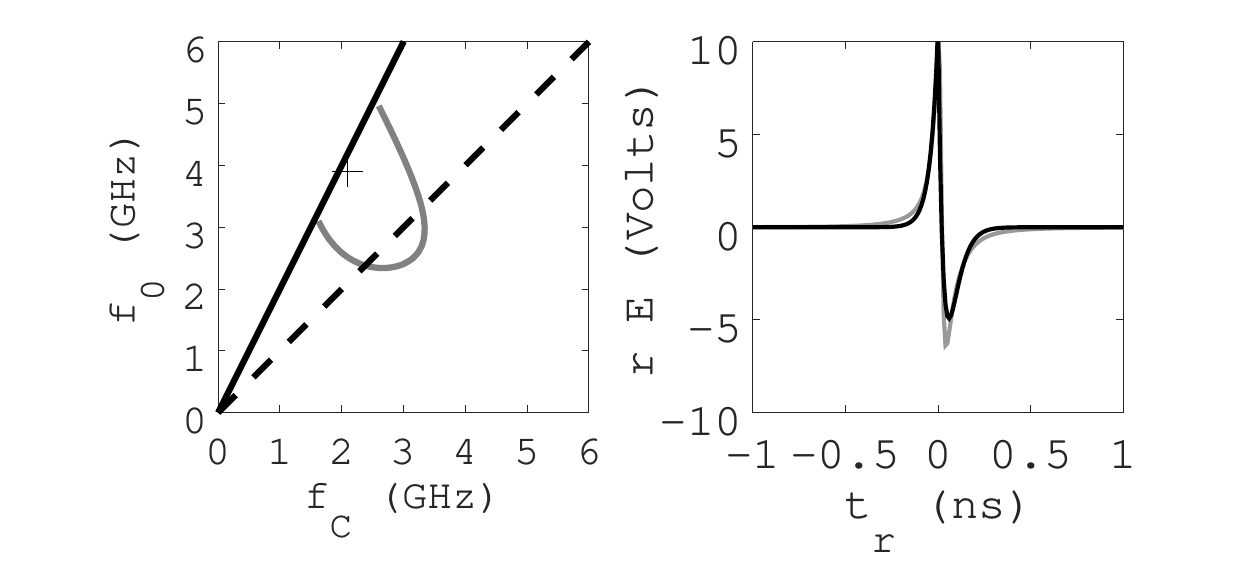}
\includegraphics[width=0.42\textwidth,trim=0cm 8.5cm 7cm 3.5cm,clip=true]{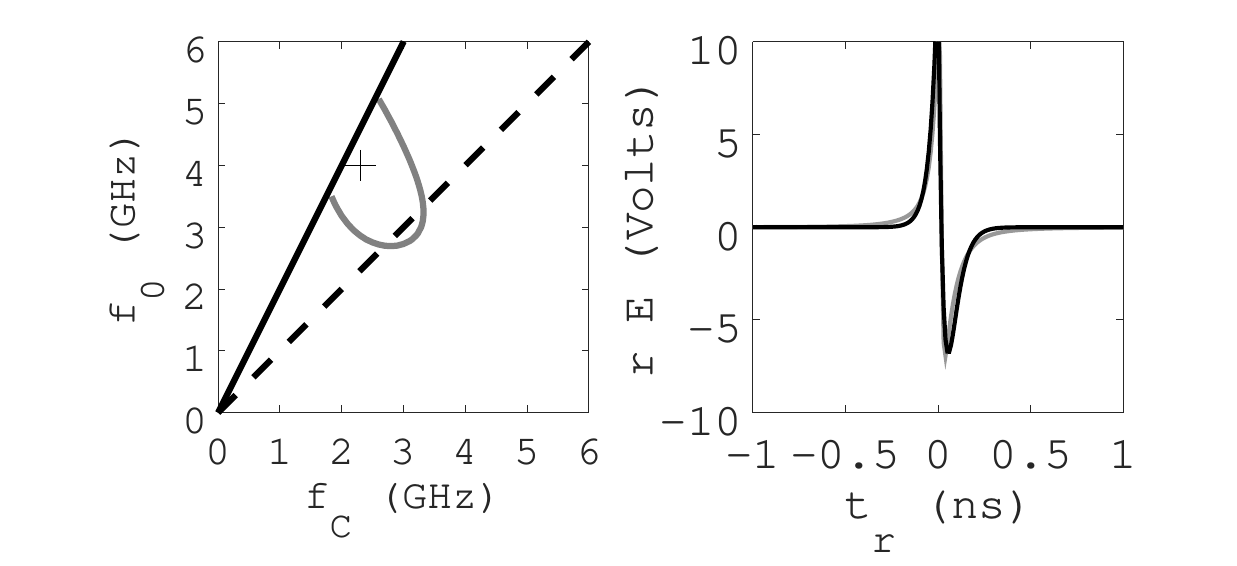}
\includegraphics[width=0.42\textwidth,trim=0cm 8.5cm 7cm 3.5cm,clip=true]{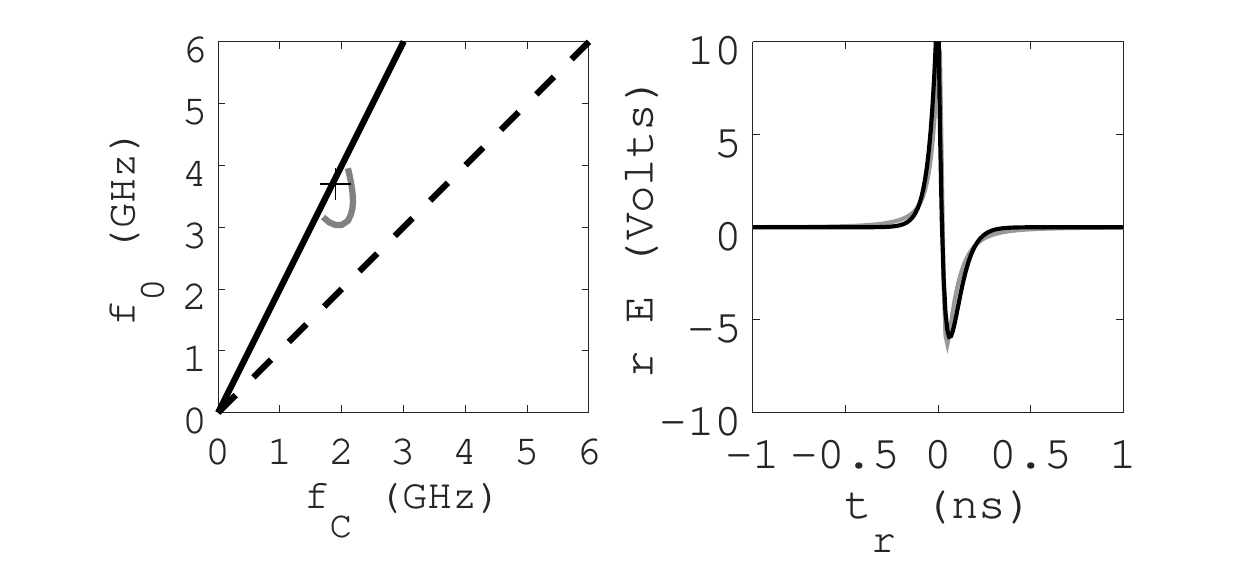}
\includegraphics[width=0.42\textwidth,trim=0cm 8.5cm 7cm 3.5cm,clip=true]{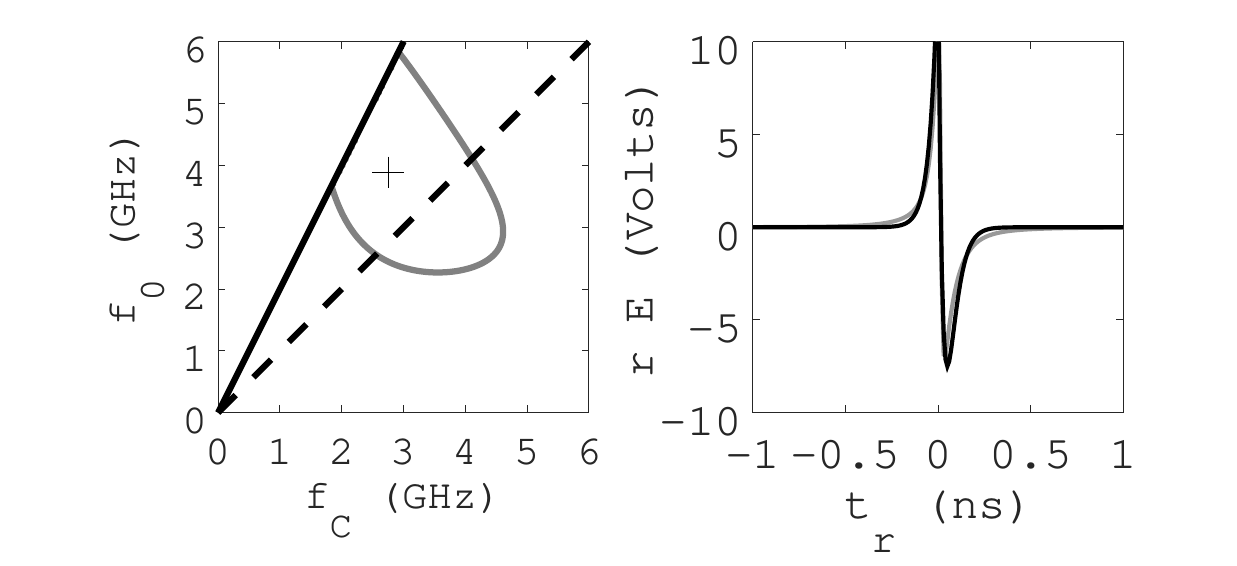}
\includegraphics[width=0.42\textwidth,trim=0cm 8.5cm 7cm 3.5cm,clip=true]{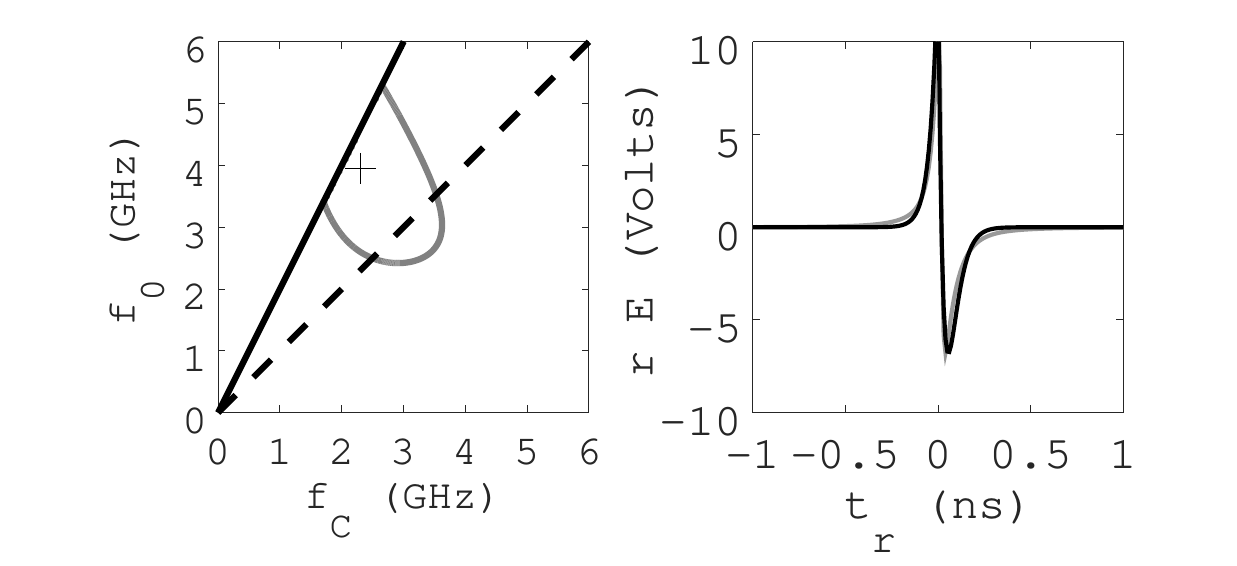}
\caption{\label{fig:fit:oncone} \textbf{Fit results: electromagnetic case, $\theta = \theta_{\rm C}$, $E_{\rm C} = 10$} PeV.  The rows correspond to NuRadioMC waveforms 1-6, 10 PeV electromagnetic cascades.  (Left column) The best-fits for $f_{\rm 0}$ and $f_{\rm C}$.  Dashed line: $\epsilon = 1$.  Solid line: $\epsilon = 2$.  Gray contour: $\rho > 0.95$.  Black cross: best-fit.  (Right column) Best-fit waveforms.  Gray: semi-analytic parameterizations from \cite{10.1140/epjc/s10052-020-7612-8}.  Black: Equation \ref{eq:on_cone}.}
\end{figure}

\textbf{Electromagnetic case.}  Six different electromagnetic cascades and the corresponding Askaryan fields were generated using the ARZ2019 model from NuRadioMC \cite{10.1140/epjc/s10052-020-7612-8} \cite{PhysRevD.101.083005} for comparison to Equation \ref{eq:on_cone}.  The cascades have $E_{\rm C} = 10$ PeV, and $r = 1000$ meters.  The LPM effect is activated in NuRadioMC for all comparisons in this work.  The units of $\vec{E}(t_{\rm r},\theta_{\rm C})$ are mV/m versus nanoseconds, so the units of $r\vec{E}$ are Volts.  The sampling rate of the digitized semi-analytic parameterizations was $100$ GHz, with $N = 2048$ samples.  Let $f_{\rm C} = \omega_{\rm C}/(2\pi)$ and $f_{\rm 0} = \omega_{\rm 0}/(2\pi)$.  The frequencies $f_{\rm C}$ and $f_{\rm 0}$ were varied from [0.6 - 6.0] GHz.  The parameter $E_{\rm 0}$ was varied from [0.05 - 5.0] V GHz$^{-2}$.  In a simple 2-level \verb+for+-loop, the Pearson correlation coefficient $\rho$ was maximized by varying $f_{\rm 0}$ and $f_{\rm C}$.  Next, the sum of the squared amplitude differences $(\Delta E)^2$ was minimized by varying $E_{\rm 0}$, while holding $f_{\rm 0}$ and $f_{\rm C}$ fixed.  Several other schemes were studied, including a 3-level \verb+for+-loop, but the two-stage process produced the best results. The results are shown in Fig. \ref{fig:fit:oncone}.

\begin{figure}
\centering
\includegraphics[width=0.45\textwidth,trim=0cm 6cm 0cm 6cm,clip=true]{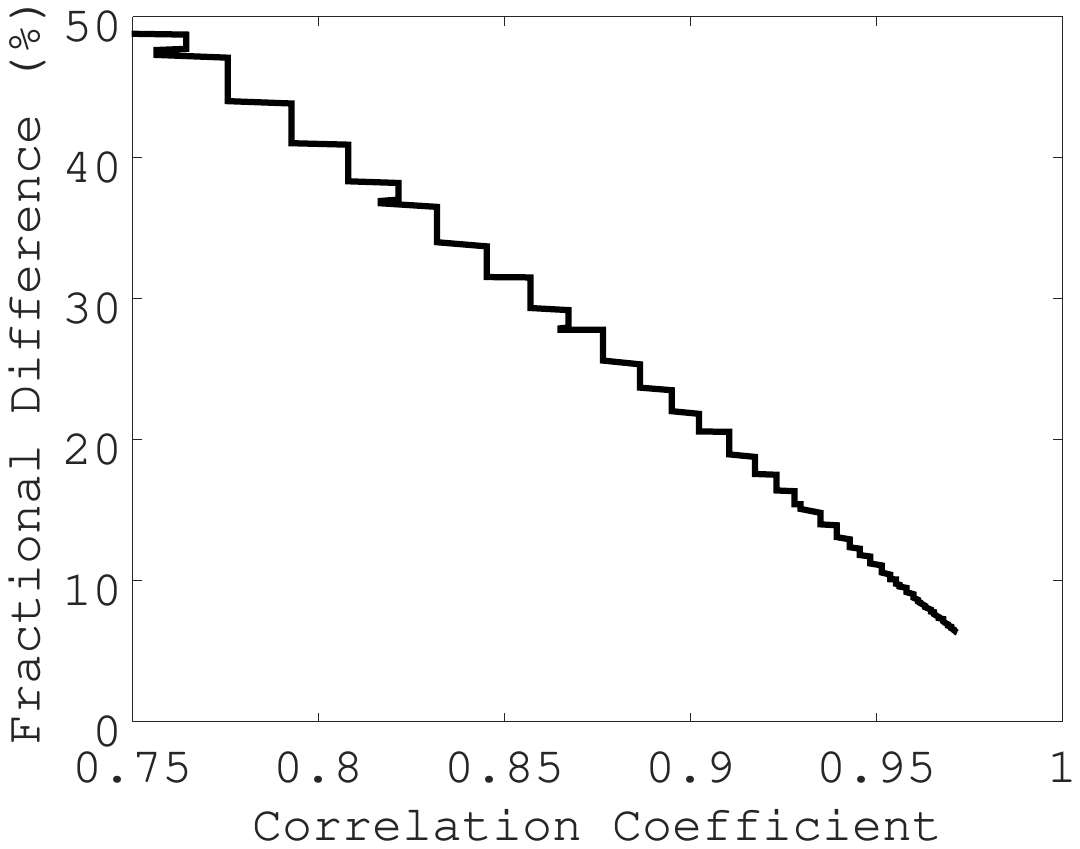}
\caption{\label{fig:f_vs_rho}  The fractional difference in the sum of amplitude differences squared ($(\Delta E)^2$) versus correlation coefficient ($\rho$) for waveform 1 at $E_{\rm C} = 10$ PeV, electromagnetic case.}
\end{figure}

Maximizing $\rho$ corresponds to minimizing $(\Delta E)^2$.  In Fig. \ref{fig:f_vs_rho}, $(\Delta E)^2$ is graphed versus $\rho$ for one event.  Best-fit $\rho$-values are $\approx 0.97$ for this set, corresponding to best-fit $(\Delta E)^2$ values of $\approx 7\%$.  Contours of $\rho>0.95$ for $f_{\rm 0}$ versus $f_{\rm C}$ are shown in Fig. \ref{fig:fit:oncone} (left column). The crosses represent the best-fit location.  The dashed gray line at $y=x$ corresponds to $f_{\rm 0}/f_{\rm C} = \epsilon = 1$.  Though Equation \ref{eq:on_cone} contains an expansion to first order in $\epsilon$, making it resemble the derivative of the vector potential from the ARVZ semi-analytic parameterization \cite{PhysRevD.101.083005}, the expansion is optional.  There is a restriction that $\epsilon \neq 2$ (see Equation \ref{app:eq:yes_eps} of Appendix \ref{app:a}).  Thus, the best-fit $\epsilon$-values avoid the solid black lines ($\epsilon = 2$) in Fig. \ref{fig:fit:oncone}, but are large enough to account for pulse asymmetry.  The best-fit waveforms are shown in Fig. \ref{fig:fit:oncone} (right column).  The gray curves correspond to the semi-analytic parameterization, and the black curves represent Equation \ref{eq:on_cone}.

\begin{figure}
\centering
\includegraphics[width=0.42\textwidth,trim=0cm 8.5cm 7cm 3.5cm,clip=true]{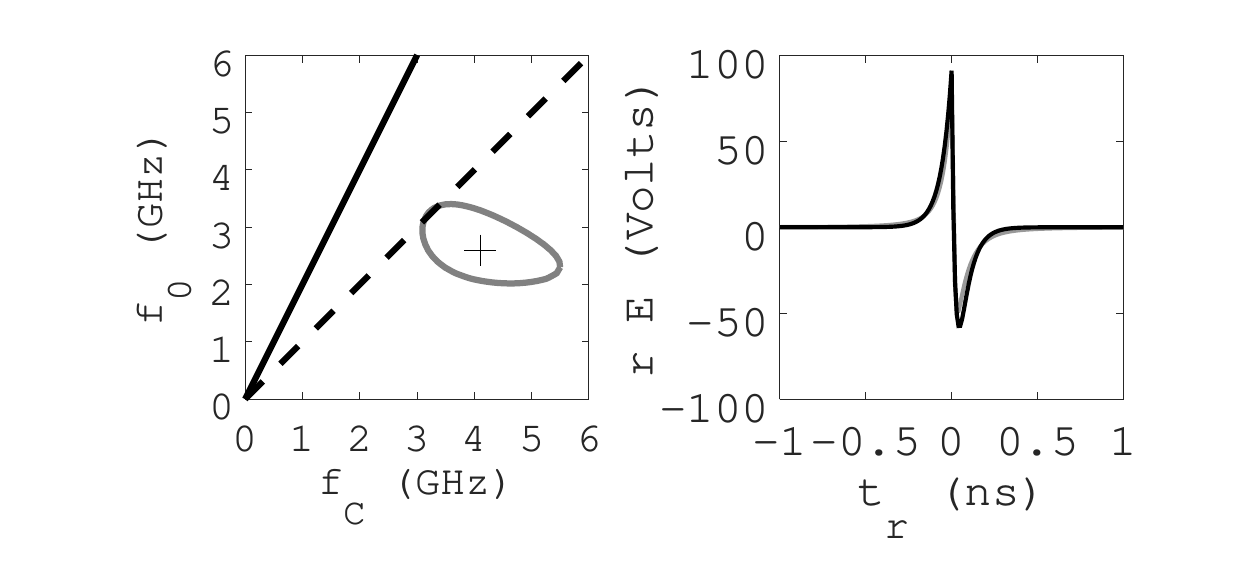}
\includegraphics[width=0.42\textwidth,trim=0cm 8.5cm 7cm 3.5cm,clip=true]{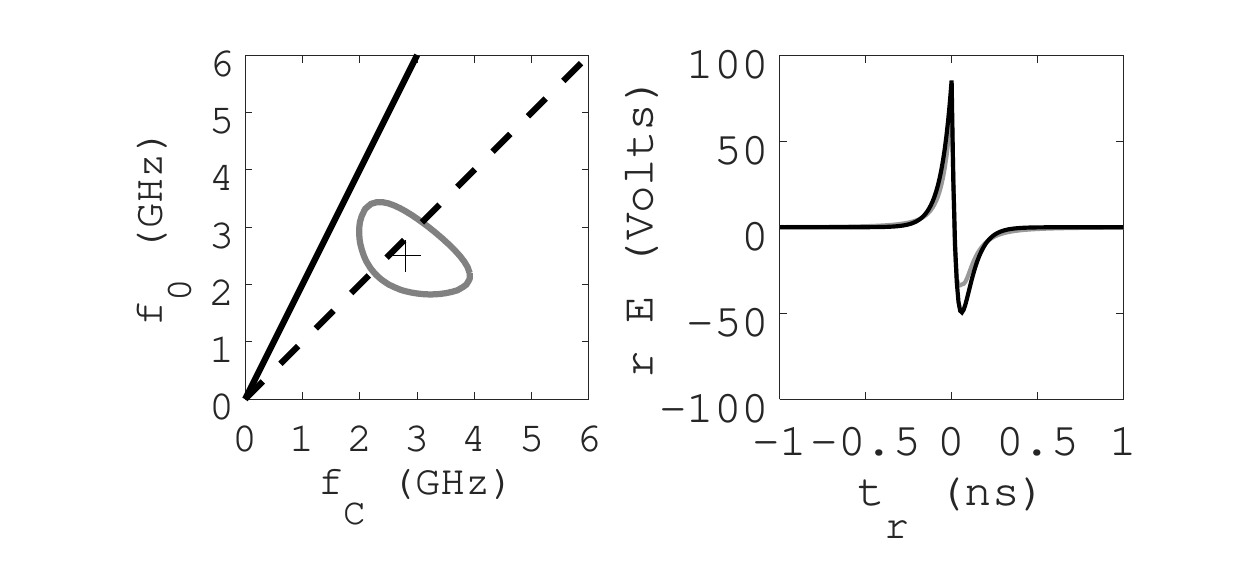}
\includegraphics[width=0.42\textwidth,trim=0cm 8.5cm 7cm 3.5cm,clip=true]{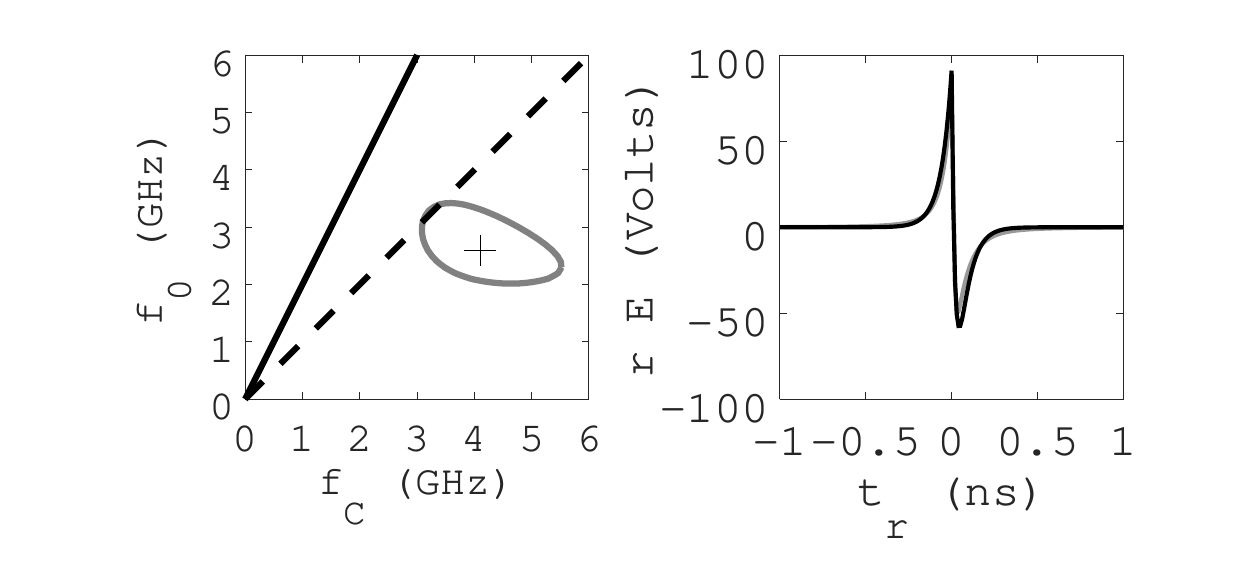}
\includegraphics[width=0.42\textwidth,trim=0cm 8.5cm 7cm 3.5cm,clip=true]{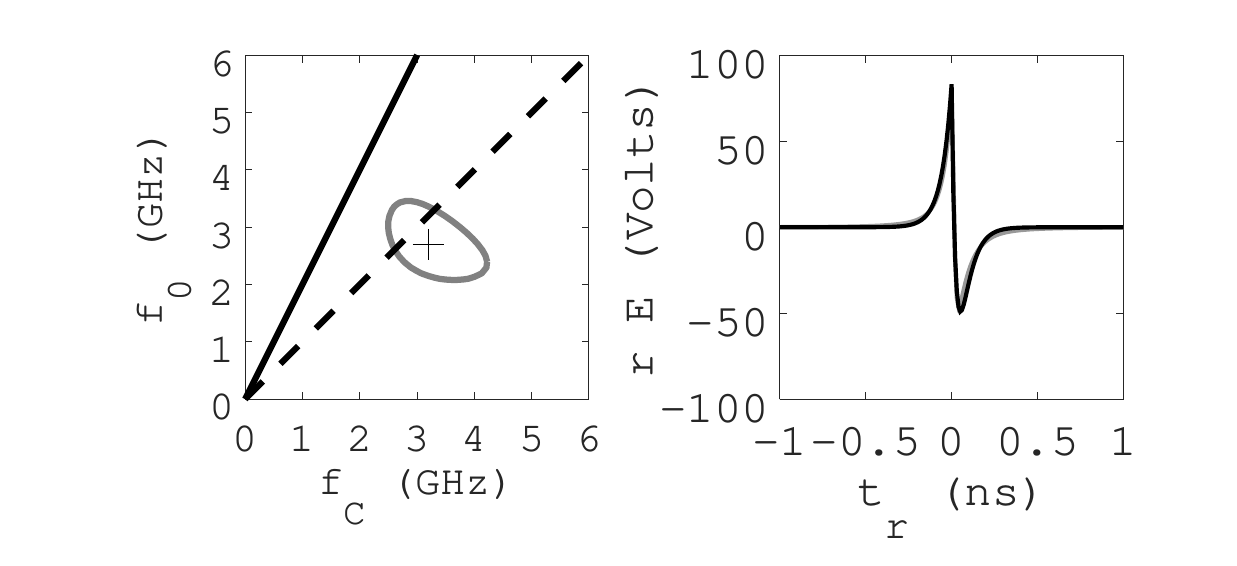}
\includegraphics[width=0.42\textwidth,trim=0cm 8.5cm 7cm 3.5cm,clip=true]{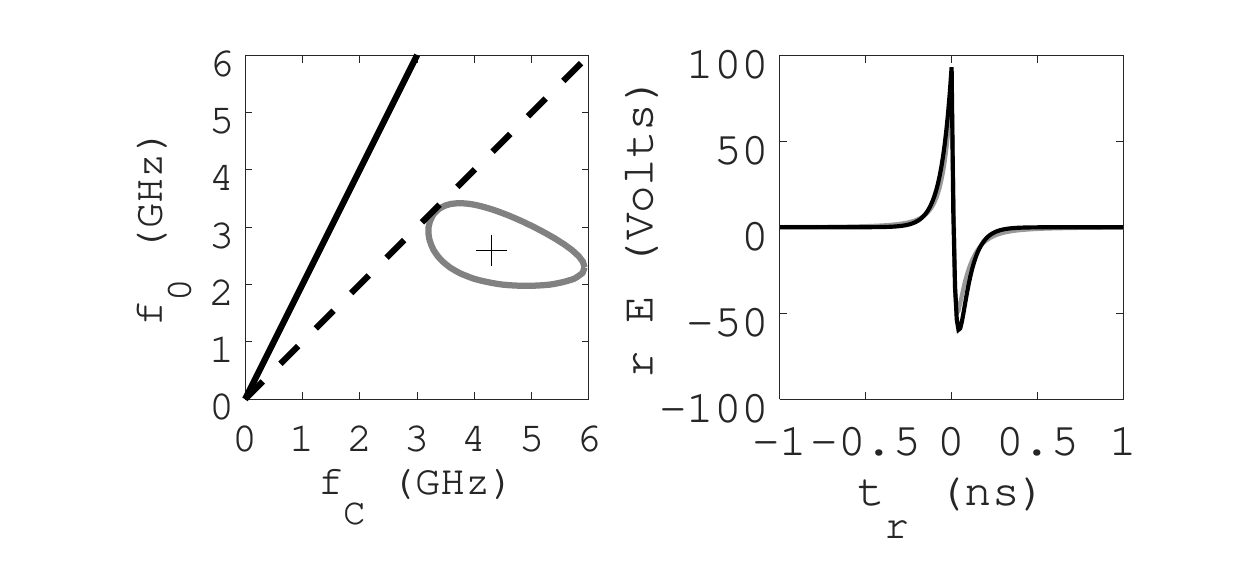}
\includegraphics[width=0.42\textwidth,trim=0cm 8.5cm 7cm 3.5cm,clip=true]{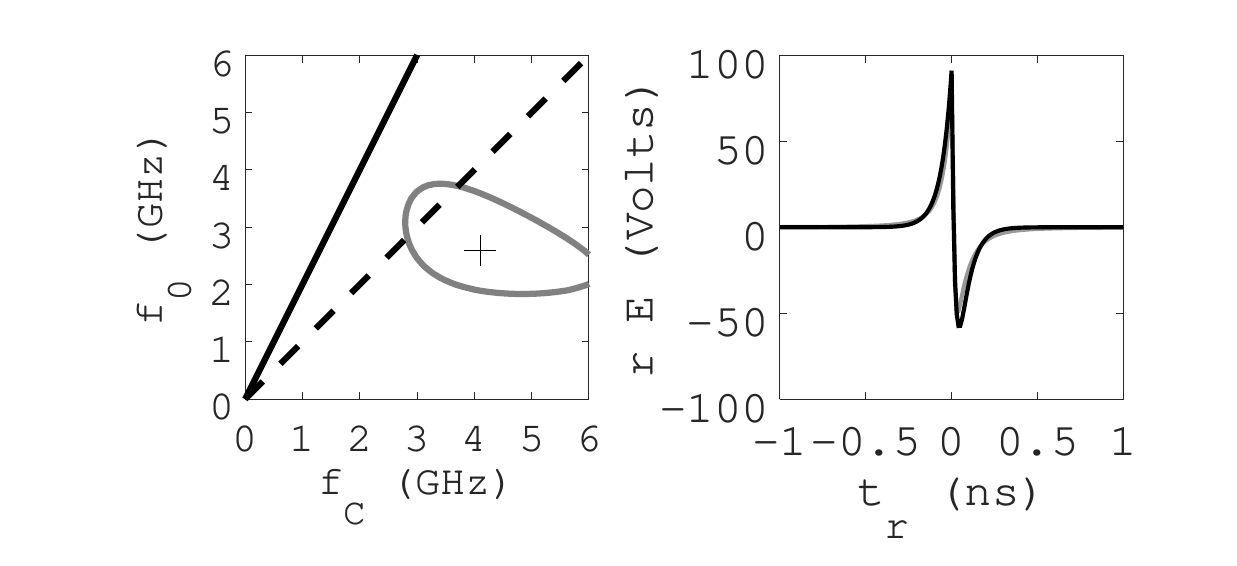}
\caption{\label{fig:fit:oncone2} \textbf{Fit results: hadronic case, $\theta = \theta_{\rm C}$, $E_{\rm C} = 100$} PeV.  The six rows (from top to bottom) correspond to NuRadioMC waveforms 1-6, 100 PeV hadronic cascades.  (Left column) The best-fits for $f_{\rm 0}$ and $f_{\rm C}$.  Dashed line: $\epsilon = 1$.  Solid line: $\epsilon = 2$.  Gray contour: $\rho > 0.9$.  Black cross: best-fit.  (Right column) The best-fit waveforms.  Gray: semi-analytic parameterizations from \cite{10.1140/epjc/s10052-020-7612-8}.  Black: Equation \ref{eq:on_cone}.}
\end{figure}

Table \ref{fit:tab:oncone_em} contains the best-fit results for the Equation \ref{eq:on_cone} parameters, along with best-fit $\rho$-values and $(\Delta E)^2$-values.  The horizontal and vertical distances from the crosses to the $\rho > 0.95$ contour are used as error estimates for $f_{\rm 0}$ and $f_{\rm C}$ in Tab. \ref{fit:tab:oncone_em}.  The $a$-errors typically encompass the $a$-values from NuRadioMC.  The full region in $[f_{\rm 0},f_{\rm C}]$ space for which UHE-$\nu$ signals are expected for IceCube-Gen2 radio will be the topic of future studies, along with the apparent difference in $\epsilon$-value depending on the electromagnetic or hadronic classification of the cascade (see Figure \ref{fig:fit:oncone2}).

\begingroup
\squeezetable
\begin{table}
\centering
\renewcommand{\arraystretch}{2}
\begin{tabular}{|c|c|c|c|c|c|c|}
\hline
\# & \thead{$f_{\rm 0}$ \\ (GHz)} & \thead{$f_{\rm C}$ \\ (GHz)} & \thead{$E_{\rm 0}$ \\ (V \\ GHz$^{-2}$)} & $a_{\rm wave}$ (m), $a_{\rm MC}$ (m) & $\rho$ & \thead{$(\Delta E)^2$ \\ (\%)} \\ \hline
1 & $3.9^{+0.2}_{-1.9}$ & $2.3^{+1.3}_{-0.3}$ & $0.3$ & $4.1^{+1.2}_{-0.3}$, $4.85$ & $0.97$ & $6.5$ \\
2 & $3.9^{+0.3}_{-1.5}$ & $2.1^{0.9}_{-0.1}$ & $0.5$ & $4.3^{+1.8}_{-0.2}$, $6.35$ & $0.97$ & $10.9$ \\
3 & $4.0^{+1.2}_{-1.0}$ & $2.3^{+0.8}_{-0.4}$ & $0.35$ & $4.1^{+0.7}_{-0.4}$, $4.48$ & $0.96$ & $7.5$ \\
4 & $3.7^{+0.1}_{-0.5}$ & $1.9^{+0.5}_{-0.1}$ & $1.85$ & $4.5^{+1.1}_{-0.3}$, $5.6$ & $0.955$ & $8.9$ \\
5 & $3.9^{+1.4}_{-0.9}$ & $2.7^{+1.4}_{-0.8}$ & $0.18$ & $4.0^{+2.0}_{-1.2}$, $4.48$ & $0.97$ & $5.7$  \\
6 & $3.9^{+1.3}_{-1.9}$ & $2.3^{+1.3}_{-0.3}$ & $0.31$ & $4.1^{+2.0}_{-0.5}$, $4.85$ & $0.97$ & $6.4$ \\ \hline
Ave. & $3.88$ & $2.3$ & $0.6$ & $4.18$ & $0.966$ & $7.7$ \\
Err. & $3.08$ & $0.1$ & $0.3$ & $0.07$ & $0.003$ & $0.8$ \\
\hline
\end{tabular}
\caption{\label{fit:tab:oncone_em} \textbf{Fit results: electromagnetic case, $\theta = \theta_{\rm C}$, $E_{\rm C} = 10$} PeV.  The six rows (from top to bottom) correspond to NuRadioMC waveforms 1-6, 10 PeV electromagnetic cascades.  From left to right, the form-factor cutoff-frequency, coherence cuoff-frequency, energy-scaling normalization, longitudinal length parameter, the best-fit correlation coefficient, and the relative power difference between NuRadioMC semi-analytic parameterization and the fully analytic model.  The parameter means and errors in the mean are quoted in the bottom two rows.}
\end{table}
\endgroup

\textbf{Hadronic case.}  Using the same procedure as the electromagnetic case, NuRadioMC was used to generate six hadronic cascades at 100 PeV for comparison to Equation \ref{eq:on_cone}.  The energy was increased to show that the model describes a range of energies, so the waveform amplitudes are larger by a factor of 10 relative to the 10 PeV case.  The LPM effect is activated in NuRadioMC for all comparisons in this work.  The main results are shown in Figure \ref{fig:fit:oncone2}, and the correlation contours represent $\rho = 0.985$.  

The results shown in Figure \ref{fig:fit:oncone2} demonstrate that modeling hadronic cascades at $\theta = \theta_{\rm C}$ is similar to the electromagnetic case, with one interesting difference.  The contours enclose best-fit $\epsilon$-values below the dashed line, whereas the fits to the electromagnetic cases were above the dashed line.  This could indicate a potential discriminator for cascade classification.  Another difference between the electromagnetic and hadronic cases is that the gray contours in Fig. \ref{fig:fit:oncone2} correspond to $\rho = 0.985$, as opposed to $\rho = 0.95$ in the electromagnetic case.

Table \ref{fit:tab:oncone_had} contains the best-fit parameters corresponding to Figure \ref{fig:fit:oncone2}.  The typical power difference $(\Delta E)^2$ has decreased with respect to the electromagnetic case.  The $\rho$-values all exceed 0.985, and the $(\Delta E)^2$ results are typically below 2 percent.  Intriguingly, $\epsilon < 1$ means higher $f_{\rm C}$ values, which in turn yields systematically low $a$-values relative to those generated in NuRadioMC, despite the increased energy.  Reconstructed $a$-values are still within a factor of 2 of the MC-true values.  Despite the systematic offset, the best-fit $a$ and the NuRadioMC $a$-values are tightly correlated (see Fig. \ref{fit:fig:a_vs_a} below).

\begingroup
\squeezetable
\begin{table}
\centering
\renewcommand{\arraystretch}{2}
\begin{tabular}{|c|c|c|c|c|c|c|}
\hline
\# & \thead{$f_{\rm 0}$ \\ (GHz)} & \thead{$f_{\rm C}$ \\ (GHz)} & \thead{$E_{\rm 0}$ \\ (V \\ GHz$^{-2}$)} & $a_{\rm wave}$ (m), $a_{\rm MC}$ (m) & $\rho$ & \thead{$(\Delta E)^2$ \\ (\%)} \\ \hline
1 & $2.6^{+0.6}_{-0.6}$ & $4.1^{+1.1}_{-1.0}$ & $1.0$ & $3.1^{+0.8}_{-0.8}$, $5.23$ & $0.99$ & $1.86$ \\
2 & $2.5^{+0.7}_{-0.6}$ & $2.8^{+0.9}_{-0.8}$ & $1.25$ & $3.75^{+1.2}_{-1.1}$, $6.35$ & $0.99$ & $1.83$ \\
3 & $2.6^{+0.7}_{-0.6}$ & $4.1^{+1.2}_{-0.9}$ & $1.0$ & $3.1^{+0.9}_{-0.7}$, $5.23$ & $0.99$ & $1.83$ \\
4 & $2.7^{+0.6}_{-0.5}$ & $3.2^{+0.8}_{-0.6}$ & $1.0$ & $3.5^{+0.9}_{-0.7}$, $6.35$ & $0.99$ & $2.5$ \\
5 & $2.6^{+0.7}_{-0.6}$ & $4.3^{+1.4}_{-1.1}$ & $1.0$ & $3.0^{+1.0}_{-0.75}$, $4.85$ & $0.99$ & $1.755$ \\
6 & $2.6^{+1.4}_{-0.7}$ & $4.1^{+1.9}_{-1.2}$ & $1.0$ & $3.1^{+1.4}_{-0.9}$, $5.23$ & $0.99$ & $1.86$ \\ \hline
Ave. & $2.60$ & $3.75$ & $1.04$ & $3.3$ & $0.99$ & $1.9$\\
Err. & $0.03$ & $0.25$ & $0.04$ & $0.1$ & $0.0$ & $0.1$ \\
\hline
\end{tabular}
\caption{\label{fit:tab:oncone_had} \textbf{Fit results: hadronic case, $\theta = \theta_{\rm C}$, $E_{\rm C} = 100$} PeV.  The six rows (from top to bottom) correspond to NuRadioMC waveforms 1-6, 100 PeV hadronic cascades.  From left to right, the form-factor cutoff-frequency, coherence cuoff-frequency, energy-scaling normalization, longitudinal length parameter, the best-fit correlation coefficient, and the relative power difference between NuRadioMC semi-analytic parameterization and the fully analytic model.  The parameter means and errors in the mean are quoted in the bottom two rows.}
\end{table}
\endgroup

\subsection{Waveform Comparison: $\theta \neq \theta_{\rm C}$}
\label{sec:fit:off}

\textbf{Electromagnetic case.}  The general comparison procedure of Section \ref{sec:fit:on} was repeated with the same semi-analytic parameterization from NuRadioMC, but with twelve new events each viewed at $\theta = \theta_{\rm C} + 3.0^{\circ}$ (six electromagnetic cascades, six hadronic).  One difference is that $\omega_{\rm 0}$ only changes the waveform amplitude, along with $E_{\rm 0}$.  The pulse width $\sigma_{\rm t} = \sqrt{2p}$ connects the longitudinal length $a$ and the viewing angle with respect to the Cherenkov angle.

\begin{figure}
\centering
\includegraphics[width=0.45\textwidth,trim=0cm 8.5cm 1cm 10.2cm,clip=true]{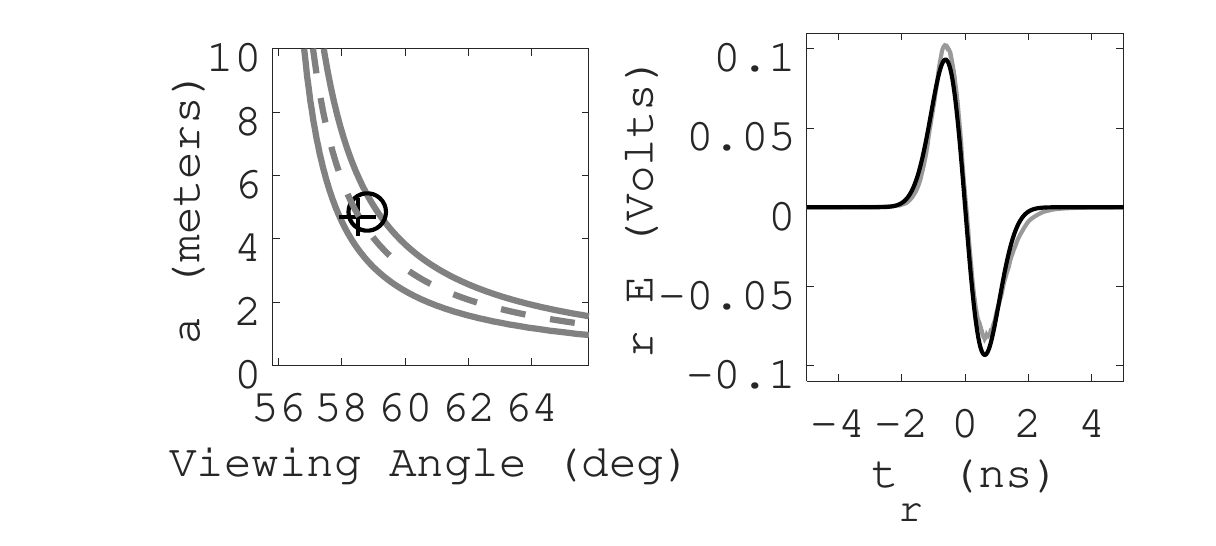}
\includegraphics[width=0.45\textwidth,trim=0cm 8.5cm 1cm 10.2cm,clip=true]{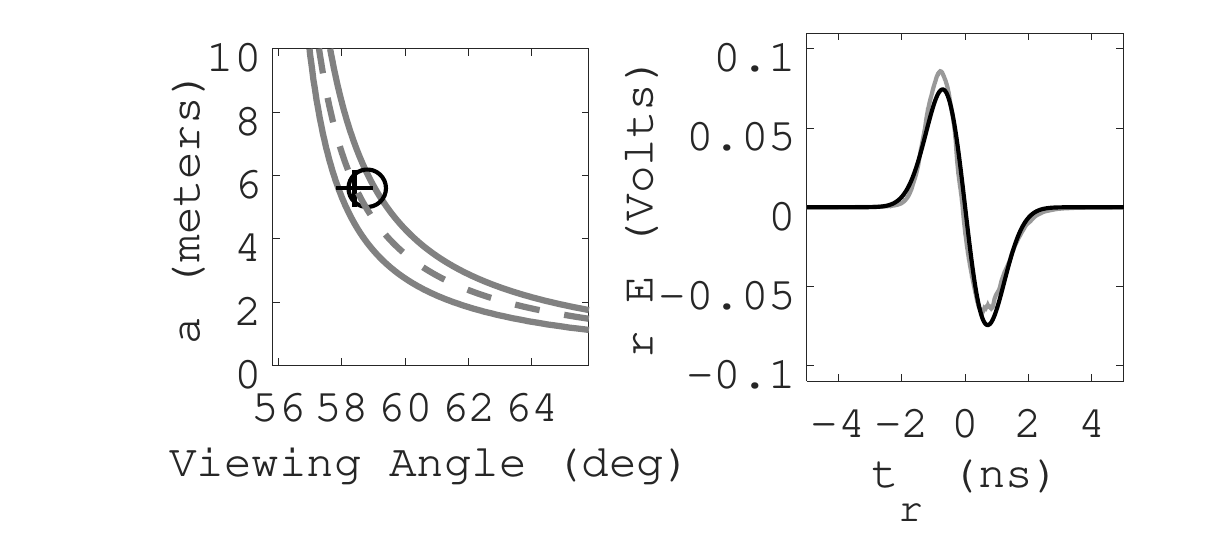}
\includegraphics[width=0.45\textwidth,trim=0cm 8.5cm 1cm 10.2cm,clip=true]{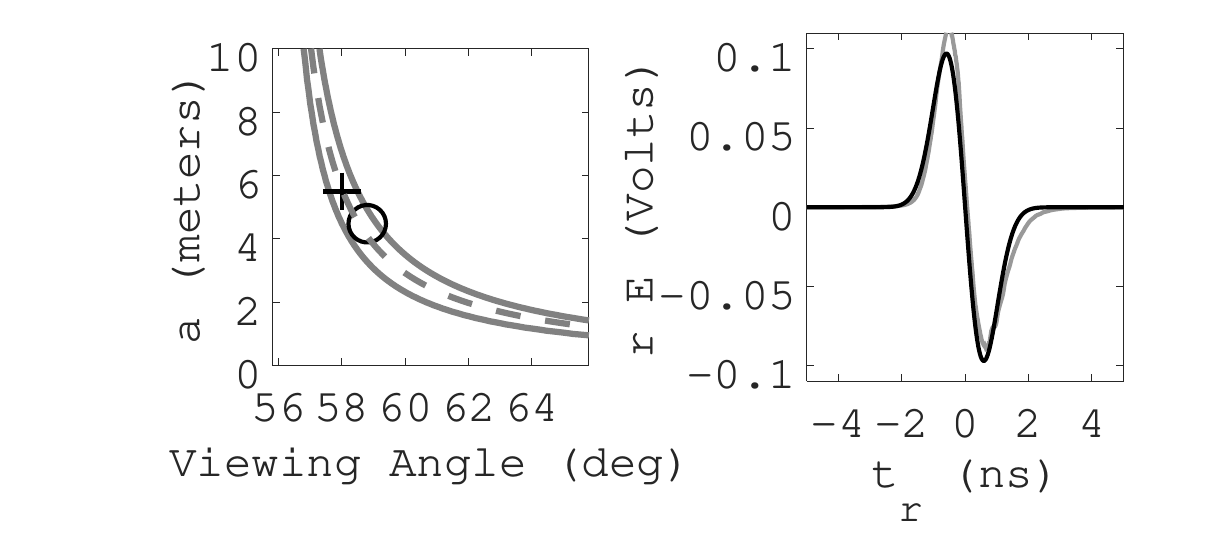}
\includegraphics[width=0.45\textwidth,trim=0cm 8.5cm 1cm 10.2cm,clip=true]{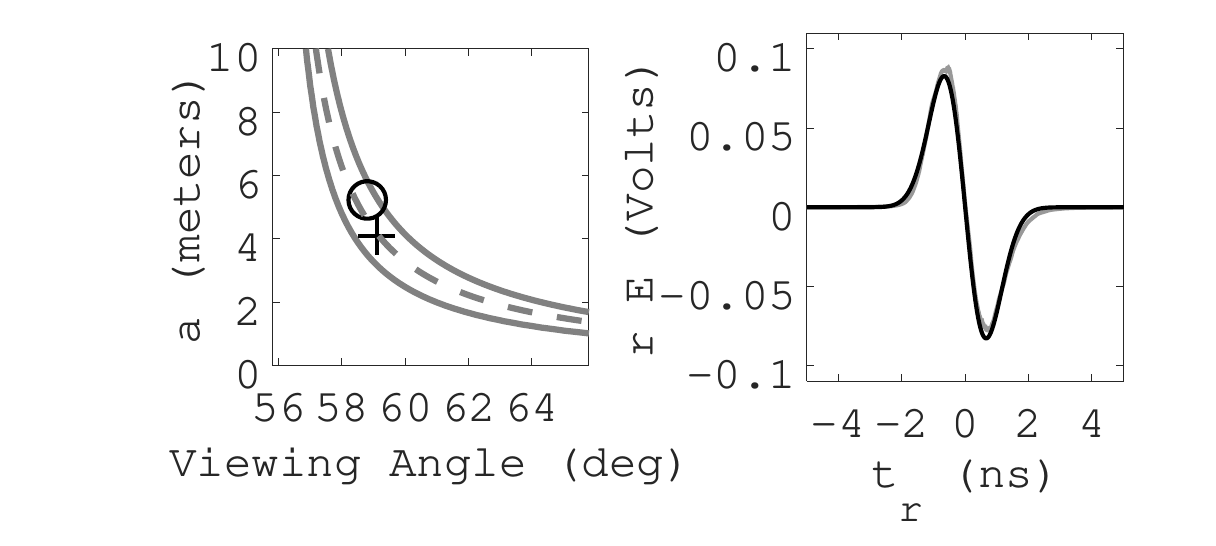}
\includegraphics[width=0.45\textwidth,trim=0cm 8.5cm 1cm 10.2cm,clip=true]{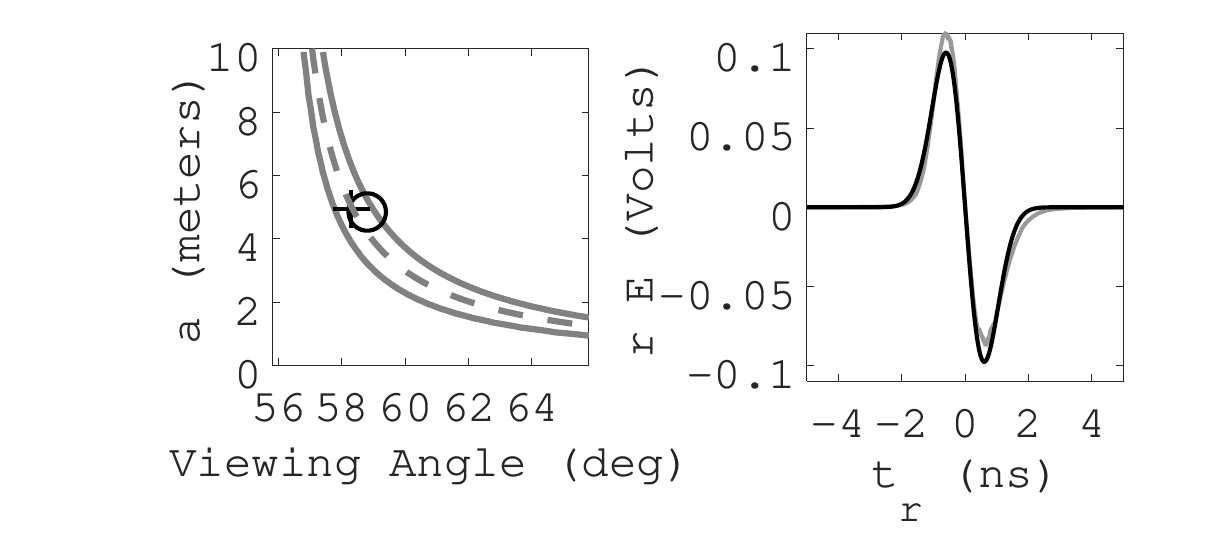}
\includegraphics[width=0.45\textwidth,trim=0cm 8.5cm 1cm 10.2cm,clip=true]{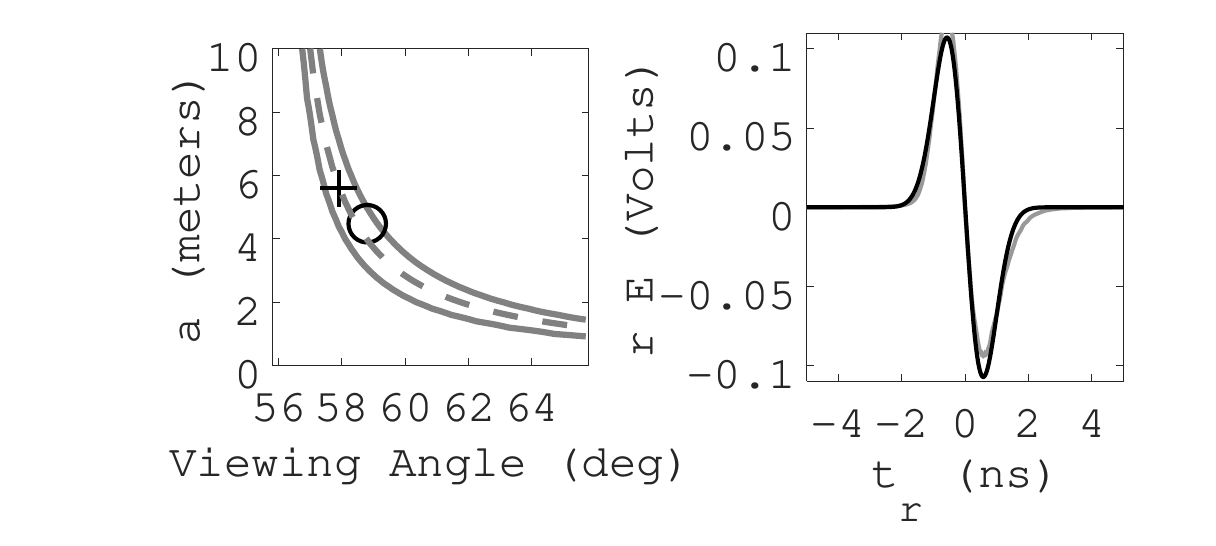}
\caption{\label{fig:fit:offcone} \textbf{Fit results: electromagnetic case, $\theta \neq \theta_{\rm C}$, $E_{\rm C} = 10$} PeV.  The six rows (from top to bottom) correspond to NuRadioMC waveforms 1-6, 10 PeV electromagnetic cascades.  (Left column) Best-fit $\theta$ and $a$-values. Crosses: best-fits.  Circles: MC true values.  Gray contour: $\rho > 0.95$.  Dashed line: $a$ versus $\theta$ from Equation \ref{eq:a_versus_theta}.  (Right column) The best-fit waveforms.  Gray: semi-analytic parameterizations from \cite{10.1140/epjc/s10052-020-7612-8}.  Black: Equation \ref{eq:off_cone}.}
\end{figure}

The fit procedure was performed in two stages.  First, $\theta$-values and $a$-values were scanned from $[\theta_{\rm C} + 1.5^{\circ},\theta_{\rm C} + 10.0^{\circ}]$ and $[0.1,10]$ meters, respectively, to maximize $\rho$.  Once the best-fit values for $a$ and $\theta$ were determined, $(\Delta E)^2$ was minimized by varying $f_{\rm 0} = \omega_{\rm 0}/(2\pi)$ and $E_{\rm 0}$ from $[0.3,3.0]$ GHz and $[0.1,2.0]$ V GHz$^{-2}$, respectively.  The $(\theta,a)$ scan and the $(f_{\rm 0},E_{\rm 0})$ scan were each separate 2-level \verb+for+ loops.  The results are shown in Figure \ref{fig:fit:offcone}.

In Figure \ref{fig:fit:offcone} (left column), the best-fit $a$-values and $\theta$-values are marked with a cross.  The circles represent the MC-true values.  Circles and crosses lie on the dashed lines, because an uncertainty principle connects $a$-values to $\theta$-values (see Section \ref{sec:sigma}).  Specifically, Equation \ref{eq:a_versus_theta} may be used to show, to first-order in $\Delta \theta = \theta - \theta_{\rm C}$:

\begin{equation}
a \Delta\theta = \frac{c\sqrt{2p}}{\sin\theta_{\rm C}} = constant \label{eq:uncertainty}
\end{equation}

The pulse width $\sigma_{\rm t} = \sqrt{2 p}$ is a constant derived from the waveform, implying that the product of $a$ and $\Delta \theta$ is constant.  The parameters $a$ and $\Delta\theta$ are therefore inversely proportional: $a \propto \Delta\theta^{-1}$.  The shape of the $\rho>0.95$ contour follows this inverse proportionality.  The dashed lines represent Equation \ref{eq:uncertainty}.  These results suggest that a measurement of the Askaryan pulse width would constrain the cascade shape and geometry.  The best-fit waveforms are shown in Figure \ref{fig:fit:offcone} (right column).  Typical correlation coefficients exceed $\rho = 0.98$.  Table \ref{fit:tab:offcone_em} contains the fit results.  The fit results include estimates of the lateral width parameter, $l$, derived from $f_{\rm 0}$ (see Section \ref{sec:ff1}).  Despite making the symmetric approximation to arrive at Equation \ref{eq:off_cone}, the fits include fractional power differences of $\approx$ 3\%.

\begingroup
\squeezetable
\begin{table}
\centering
\renewcommand{\arraystretch}{2}
\begin{tabular}{|c|c|c|c|c|c|c|c|}
\hline
\# & \thead{$\theta_{\rm wave}$ \\ (deg), \\ $\theta_{\rm MC}$ \\ (deg)} & \thead{$a_{\rm wave}$ \\ (m), \\ $a_{\rm MC}$ \\ (m)} & \thead{$f_{\rm 0}$ \\ (GHz)} & \thead{$E_{\rm 0}$ \\ (V \\ GHz$^{-2}$)} & \thead{$l$ \\ (cm)} & $\rho$ & \thead{$(\Delta E)^2$ \\ (\%)} \\ \hline
1 & \thead{$58.5^{+0.7}_{-0.6}$, \\ $58.8$} & \thead{$4.7^{+1.3}_{-1.0}$, \\ $4.85$} & $0.75$ & $1.2$ & $3.4^{+0.9}_{-0.7}$ & $0.99$ & $1.93$ \\
2 & \thead{$58.4^{+0.6}_{-0.5}$, \\ $58.8$} & \thead{$5.6^{+1.4}_{-1.1}$, \\ $5.60$} & $1.0$ & $1.2$ & $2.6^{+0.4}_{-0.3}$ & $0.99$ & $2.61$ \\
3 & \thead{$58.0^{+0.5}_{-0.4}$, \\ $58.8$} & \thead{$5.5^{+1.3}_{-1.0}$, \\ $4.48$} & $1.0$ & $1.1$ & $2.6^{+0.3}_{-0.2}$ & $0.98$ & $4.47$ \\
4 & \thead{$59.1^{+0.9}_{-0.7}$, \\ $58.8$} & \thead{$4.1^{+1.2}_{-0.9}$, \\ $5.23$} & $0.75$ & $1.2$ & $3.4^{+0.5}_{-0.5}$ & $0.995$ & $0.80$ \\
5 & \thead{$58.3^{+0.7}_{-0.5}$, \\ $58.8$} & \thead{$4.95^{+1.4}_{-1.1}$, \\ $4.85$} & $0.75$ & $1.2$ & $3.4^{+0.4}_{-0.3}$ & $0.99$ & $1.8$ \\
6 & \thead{$57.9^{+0.6}_{-0.4}$, \\ $58.8$} & \thead{$5.6^{+1.5}_{-1.2}$, \\ $4.48$} & $0.75$ & $1.2$ & $3.5^{+0.5}_{-0.4}$ & $0.99$ & $1.83$ \\
\hline
Ave. & $58.4$ & $5.1$ & $0.83$ & $1.18$ & $3.2$ & $0.989$ & $2.2$ \\
Err. & $0.2$ & $0.2$ & $0.05$ & $0.02$ & $0.2$ & $0.002$ & $0.5$ \\
\hline
\end{tabular}
\caption{\label{fit:tab:offcone_em} \textbf{Fit results: electromagnetic case, $\theta \neq \theta_{\rm C}$, $E_{\rm C} = 10$} PeV.  The six rows (from top to bottom) correspond to NuRadioMC waveforms 1-6, 10 PeV electromagnetic cascades.  From left to right, the viewing angle, longitudinal length parameter, form-factor cutoff frequency, the energy-scaling normalization, the lateral width of the cascade, the best-fit correlation coefficient, and the relative power difference between NuRadioMC semi-analytic parameterization and the fully analytic model.  The parameter means and errors in the mean are quoted in the bottom two rows.}
\end{table}
\endgroup

\begin{figure}
\centering
\includegraphics[width=0.45\textwidth,trim=0cm 8.5cm 1cm 10.2cm,clip=true]{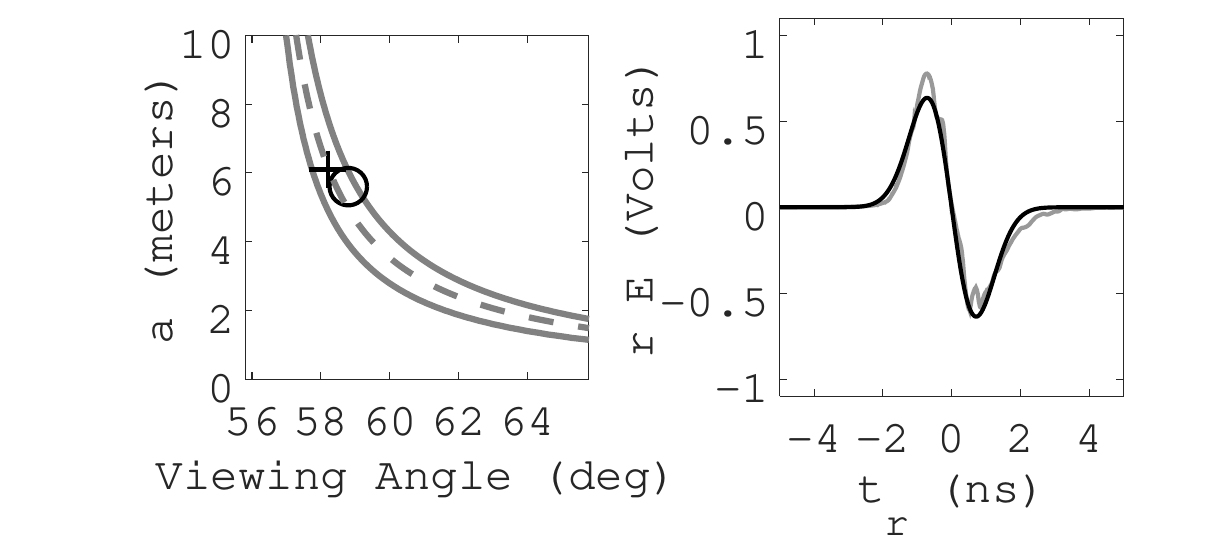}
\includegraphics[width=0.45\textwidth,trim=0cm 8.5cm 1cm 10.2cm,clip=true]{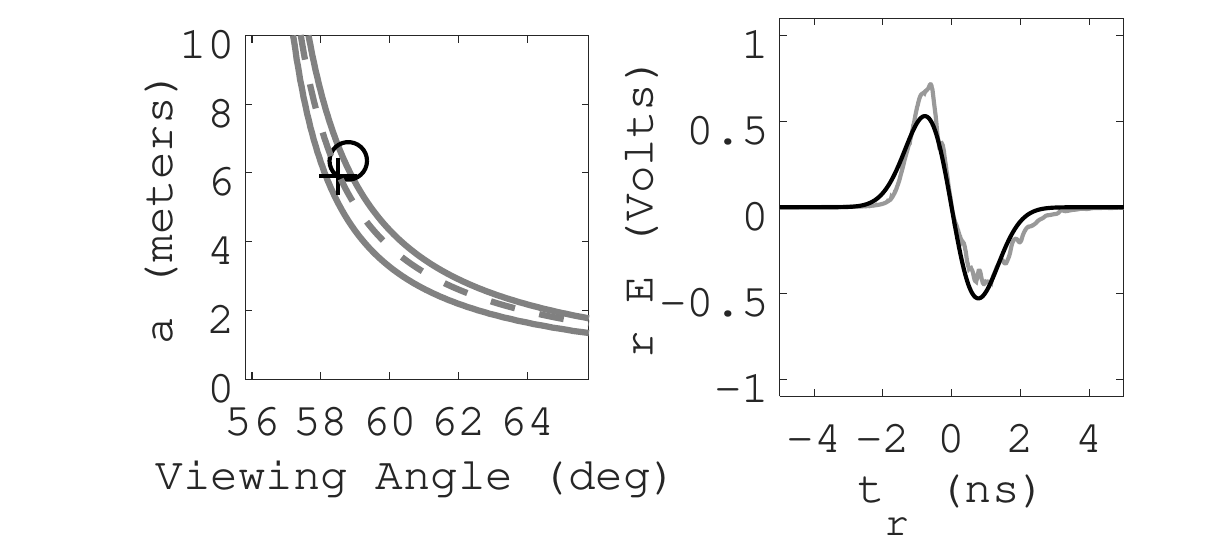}
\includegraphics[width=0.45\textwidth,trim=0cm 8.5cm 1cm 10.2cm,clip=true]{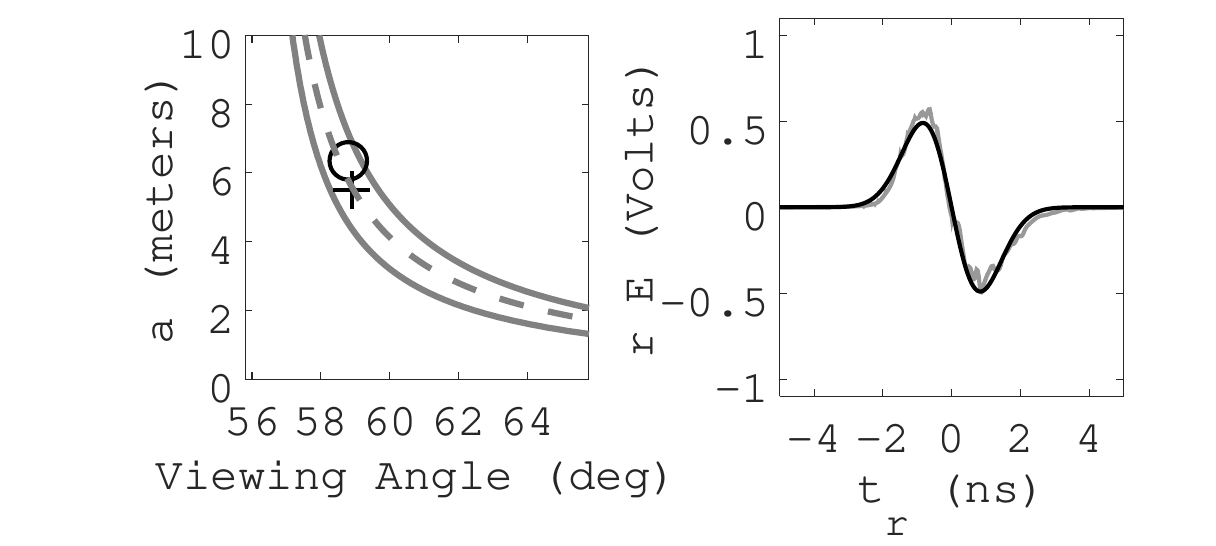}
\includegraphics[width=0.45\textwidth,trim=0cm 8.5cm 1cm 10.2cm,clip=true]{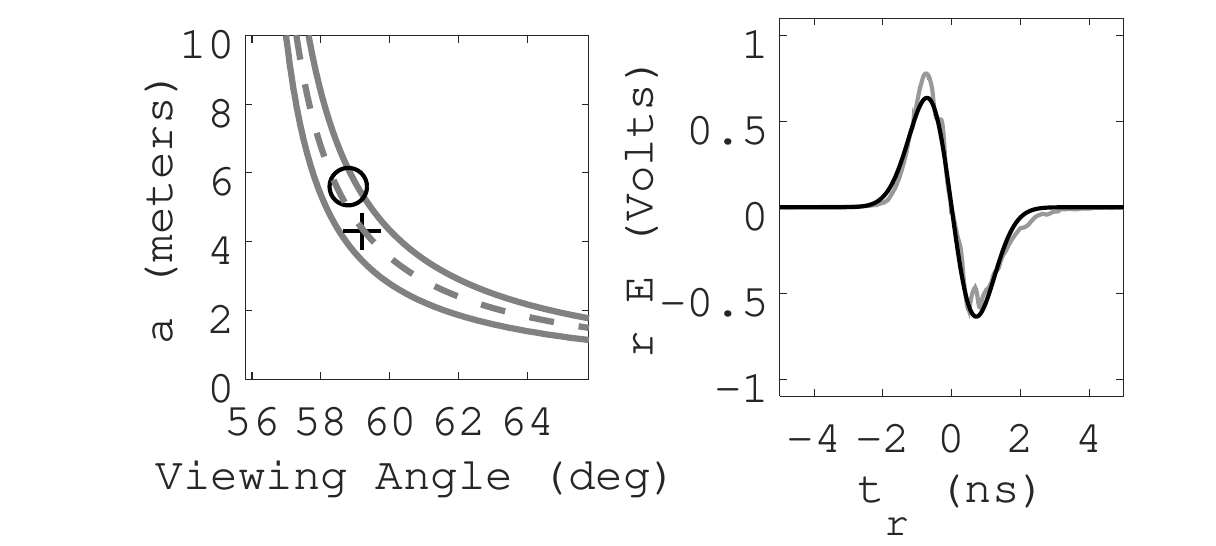}
\includegraphics[width=0.45\textwidth,trim=0cm 8.5cm 1cm 10.2cm,clip=true]{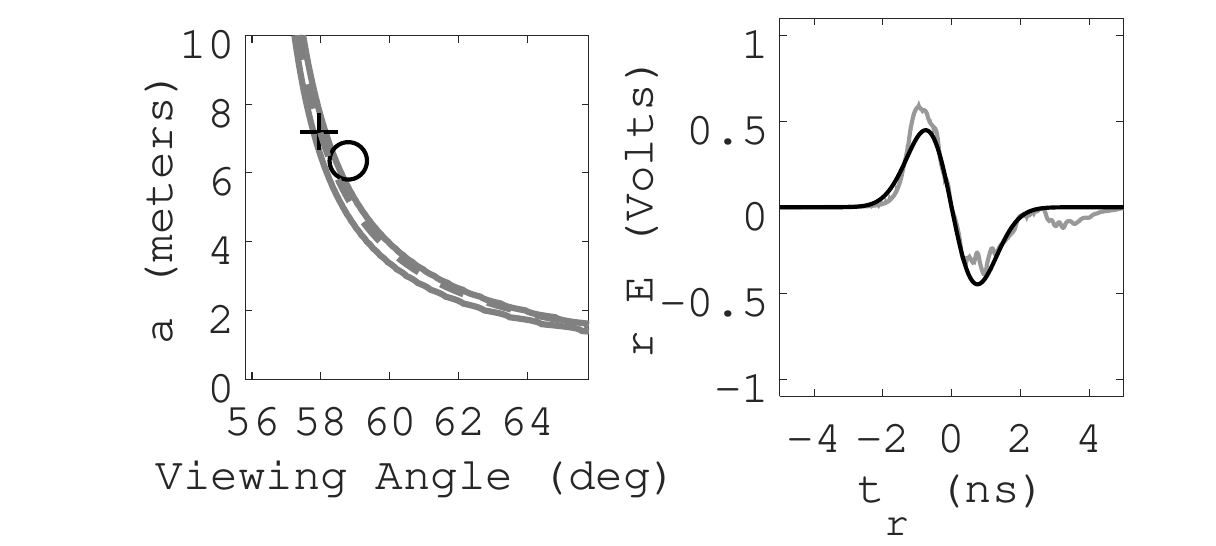}
\includegraphics[width=0.45\textwidth,trim=0cm 8.5cm 1cm 10.2cm,clip=true]{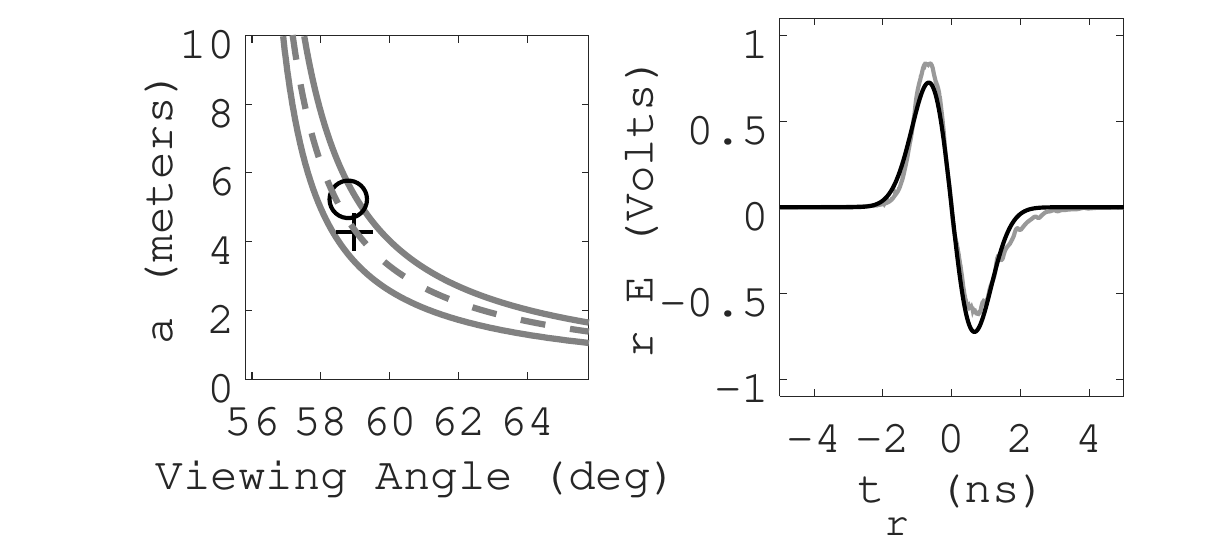}
\caption{\label{fig:fit:offcone2} \textbf{Fit results: hadronic case, $\theta \neq \theta_{\rm C}$, $E_{\rm C} = 100$} PeV. The six rows (from top to bottom) correspond to NuRadioMC waveforms 1-6, 100 PeV hadronic cascades.  (Left column) Best-fit $\theta$ and $a$-values. Crosses: best-fits.  Circles: MC true values.  Gray contour: $\rho > 0.95$.  Dashed line: $a$ versus $\theta$ from Equation \ref{eq:a_versus_theta} (uncertainty principle).  (Right column) The best-fit waveforms.  Gray: semi-analytic parameterizations from \cite{10.1140/epjc/s10052-020-7612-8}.  Black: Equation \ref{eq:off_cone}.}
\end{figure}

\textbf{Hadronic case.}  The fit procedure for the hadronic cascades was the same as the electromagnetic case, except that the range for $E_{\rm 0}$ was expanded to $[1.0,20.0]$ V GHz$^{-2}$.  As in the on-cone procedure, the hadronic cascade energy was $E_{\rm C} = 100$ PeV.  The results are shown in Figure \ref{fig:fit:offcone2}.

As with the electromagnetic case, $\rho$ is maximized and $(\Delta E)^2$ is minimized.  Table \ref{fit:tab:offcone_had} contains the best-fit parameters, along with $\rho$ and $(\Delta E)^2$.  Solutions with $\rho \approx 0.98$ and $(\Delta E)^2 \approx 5$ \% were found.  Similar to the results shown in Table \ref{fit:tab:offcone_em}, the results in Table \ref{fit:tab:offcone_had} are in agreement with the MC values from NuRadioMC.  The $E_{\rm 0}$-values match expectations for 100 PeV cascacdes, because they are a factor of 10 higher than those of the 10 PeV electromagnetic case.  The results for $a$, $l$, and $f_{\rm 0}$, however, are not statistically different between Tables \ref{fit:tab:offcone_em} and \ref{fit:tab:offcone_had}.  Future studies will require computing the probability distributions of these parameters from large numbers of UHE-$\nu$ cascades.

\begingroup
\squeezetable
\begin{table}
\centering
\renewcommand{\arraystretch}{2}
\begin{tabular}{|c|c|c|c|c|c|c|c|}
\hline
\# & \thead{$\theta_{\rm wave}$ \\ (deg), \\ $\theta_{\rm MC}$ \\ (deg)} & \thead{$a_{\rm wave}$ \\ (m), \\ $a_{\rm MC}$ \\ (m)} & \thead{$f_{\rm 0}$ \\ (GHz)} & \thead{$E_{\rm 0}$ \\ (V \\ GHz$^{-2}$)} & \thead{$l$ \\ (cm)} & $\rho$ & \thead{$(\Delta E)^2$ \\ (\%)} \\ \hline
1 & \thead{$58.2^{+0.6}_{-0.4}$, \\ $58.8$} & \thead{$6.1^{+1.5}_{-1.2}$, \\ $5.6$} & $0.8$ & $10.6$ & $3.2^{+0.5}_{-0.5}$ & $0.98$ & $3.55$ \\
2 & \thead{$58.5^{+0.4}_{-0.3}$, \\ $58.8$} & \thead{$5.9^{+0.9}_{-0.8}$, \\ $6.35$} & $0.85$ & $10.3$ & $3.0^{+0.3}_{-0.2}$ & $0.96$ & $7.1$ \\
3 & \thead{$58.9^{+0.8}_{-0.6}$, \\ $58.8$} & \thead{$5.5^{+1.4}_{-1.1}$, \\ $6.35$} & $0.9$ & $10.8$ & $2.8^{+0.5}_{-0.5}$ & $0.98$ & $2.64$ \\
4 & \thead{$59.2^{+0.8}_{-0.7}$, \\ $58.8$} & \thead{$4.3^{+1.1}_{-0.8}$, \\ $5.6$} & $0.85$ & $10.5$ & $3.0^{+0.5}_{-0.5}$ & $0.98$ & $3.10$ \\
5 & \thead{$58.0^{+0.2}_{-0.2}$, \\ $58.8$} & \thead{$7.2^{+0.6}_{-0.6}$, \\ $6.35$} & $0.9$ & $8.2$ & $2.9^{+0.3}_{-0.3}$ & $0.955$ & $8.76$ \\
6 & \thead{$59.0^{+0.8}_{-0.6}$, \\ $58.8$} & \thead{$4.3^{+1.1}_{-0.9}$, \\ $5.23$} & $0.85$ & $10.4$ & $3.0^{+0.5}_{-0.5}$ & $0.985$ & $3.00$ \\
\hline
Ave. & $58.6$ & $5.5$ & $0.86$ & $10.1$ & $3.2$ & $0.973$ & $5$ \\
Err. & $0.2$ & $0.5$ & $0.015$ & $0.4$ & $0.2$ & $0.005$ & $1$ \\
\hline
\end{tabular}
\caption{\label{fit:tab:offcone_had} \textbf{Fit results: hadronic case, $\theta \neq \theta_{\rm C}$, $E_{\rm C} = 100$} PeV.  The six rows (from top to bottom) correspond to NuRadioMC waveforms 1-6, 10 PeV hadronic cascades.  From left to right, the viewing angle, longitudinal length parameter, form-factor cutoff frequency, the energy-scaling normalization, the lateral width of the cascade, the best-fit correlation coefficient, and the relative power difference between NuRadioMC semi-analytic parameterization and the fully analytic model.  The parameter means and errors in the mean are quoted in the bottom two rows.}
\end{table}
\endgroup

\begin{figure}
\centering
\includegraphics[width=0.5\textwidth]{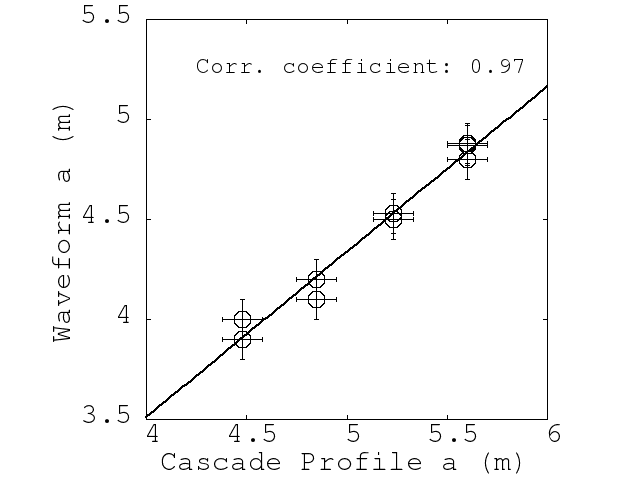}
\caption{\label{fit:fig:a_vs_a} The longitudinal length parameter $a$ derived from the Equation \ref{eq:off_cone} best-fit verus the $a$-value derived from the cascade profile in NuRadioMC.  A linear fit and correlation coefficient are shown (slope: $0.83\pm 0.05$, intercept: $0.2\pm 0.2$ (m), correlation coefficient $= 0.97$).}
\end{figure}

As a first exercise for statistical energy reconstruction from waveform parameters, assume that $\theta = \theta_{\rm C} + 3.0^{\circ}$ is already measured.  For example, $\theta$ could be determined by measuring the cutoff-frequency in the Fourier domain below 1 GHz (see Fig. 5 of \cite{10.1016/j.astropartphys.2017.03.008}, for example).  Scanning Equation \ref{eq:off_cone} over all NuRadioMC waveforms at fixed $\theta=\theta_{\rm C} + 3.0^{\circ}$ yields Figure \ref{fit:fig:a_vs_a}, in which the fitted $a$-value from each waveform is graphed versus the MC-true $a$-value.  The $a$-errors in all cases are taken to be $\pm 10$ cm ($\pm$ two $\Delta$a step-sizes).  A least-squares linear fit was applied to the data.  The linear function fits the data, and the correlation coefficient is 0.97.  The results in Figure \ref{fit:fig:a_vs_a} imply an energy reconstruction technique using the formulas found in Section \ref{sec:ff2}.  Consider the relationship between $a$ and $\ln(E_{\rm C}/E_{\rm crit})$: $a = c_1 \sqrt{\ln(E_{\rm C}/E_{\rm crit})}$.  The fractional error in $\ln(E_{\rm C}/E_{\rm crit})$ is proportional to the fractional error in $a$:

\begin{equation}
\frac{\sigma_{\ln(E_{\rm C}/E_{\rm crit})}}{\ln(E_{\rm C}/E_{\rm crit})} = 2 c_1 \left( \frac{\sigma_a}{a}\right) \label{eq:energy}
\end{equation}

If a reliable fit for the $a$-parameter is obtained from observed Askaryan waveforms, Equation \ref{eq:energy} shows that the logarithm of the energy can be constrained.

\section{Conclusion}
\label{sec:conc}

We have presented a fully analytic Askaryan model in the time-domain, and we have shown that it matches results generated with semi-analytic parameterizations used in NuRadioMC.  Pearson correlation coefficients between the fully analytic and semi-analytic paremeterizations were found to be greater than $0.95$, and typical fractional differences in total power were found to be $\approx 5$\%.  New results and potential applications are summarized in the following sections.

\subsection{Summary of New Results}

\begingroup
\squeezetable
\begin{table}
\renewcommand{\arraystretch}{1.75}
\centering
\begin{tabular}{|c|c|}
\hline
Result & Location \\ \hline
$r\vec{E}(t_{\rm r},\theta_{\rm C})$, on-cone field ($\hat{\theta}$) & Eq. \ref{eq:on_cone}, Sec. \ref{sec:onc} \\
$\sigma_{t} \sigma_{\rm \nu} \geq 1/(2\pi)$, on-cone & Eq. \ref{eq:siglimit}, Sec. \ref{sec:sigma0} \\
$r\vec{E}(t_{\rm r},\theta)$, off-cone field ($\hat{\theta}$) & Eq. \ref{eq:off_cone}, Sec. \ref{sec:ofc} \\
$\sigma_{t} \sigma_{\rm \nu} \geq 1/(2\pi)$, off-cone & Eq. \ref{eq:sigma_result}, Sec. \ref{sec:sigma} \\
On-cone EM comparison to \cite{PhysRevD.101.083005} & Fig. \ref{fig:fit:oncone}, Tab. \ref{fit:tab:oncone_em} \\ 
On-cone HAD comparison to \cite{PhysRevD.101.083005} & Fig. \ref{fig:fit:oncone2}, Tab. \ref{fit:tab:oncone_had} \\ 
Off-cone EM comparison to \cite{PhysRevD.101.083005} & Fig. \ref{fig:fit:offcone}, Tab. \ref{fit:tab:offcone_em} \\ 
Off-cone HAD comparison to \cite{PhysRevD.101.083005} & Fig. \ref{fig:fit:offcone2}, Tab. \ref{fit:tab:offcone_had} \\ 
\hline
\end{tabular}
\caption{\label{tab:conc} A summary of results in this work.}
\end{table}
\endgroup

The main results are summarized in Table \ref{tab:conc}.  This work represents the first time the two distinct pole frequencies $f_{\rm 0}$ and $f_{\rm C}$ have been used to characterize the time-domain field equations of the Askaryan effect for both $\theta = \theta_{\rm C}$ and $\theta \neq \theta_{\rm C}$.  The uncertainty principle was verified on-cone ($\theta = \theta_{\rm C}$), serving as a check on the model.  By fitting on-cone cascade parameters, we have shown that an analytic model matches semi-analytic predictions.  The $\epsilon$ parameter reveals a potential cascade classification scheme.  Next, the off-cone ($\theta \neq \theta_{\rm C}$) field equations were derived, and again the uncertainty principle was verified.  Off-cone cascade parameters were fit, and the results are in excellent agreement with semi-analytic results.  Fitting $a$-values has revealed a potential energy reconstruction.

To obtain the fields on and off-cone, $\eta < 1$ was assumed.  The restriction $\eta < 1$ means that Eqs. \ref{eq:on_cone} and \ref{eq:off_cone} must be applied to the far-field.  Given that $a$ and $\theta_{\rm C}$ are fixed by cascade physics and ice density, and that the relevant Askaryan bandwidth for ice is $[0.1-1]$ GHz, the parameter most easily varied within $\eta$ is the observer distance $r$.  Taking $\nu = 0.5$ GHz, $n = 1.78$, $c = 0.3$ m GHz, $\theta = \theta_{\rm C}$, and $a = 5$ m, requiring that $\eta = 1$ gives $r \geq 0.4$ km.  Scaling to $\nu = 0.25$ GHz gives $r \geq 0.2$ km.  According to NuRadioMC \cite{10.1140/epjc/s10052-020-7612-8} (Fig. 13), the $r$ corresponding to UHE-$\nu$ at $10^{18}$ eV ranges from 0.7-3.2 km, and 0.2 km is rare.

The ``acceleration argument'' invoked by RB in \cite{10.1103/physrevd.65.016003} states that if $r(t)$ points to the ICD, $r(t)$ must be constant enough to ensure that $\Delta r < \lambda$.  Using the law of cosines, with two sides being $r$ and $r+\Delta r$, and a third being $a$, the criteria that $(a/r)^2 \ll 1$ leads to $|\Delta r| \approx a/n$ which is $\mathcal{O}(2)$ m.  When in doubt about usage and event geometry, determining if $(a/r)^2 \ll 1$ is a good check.  If the UHE-$\nu$ event is a charged-current interaction with an electromagnetic cascade far above the LPM energy for ice, $a$ grows faster than $\sqrt{\ln(E_{\rm C}/E_{\rm crit})}$ \cite{10.1103/physrevd.82.074017}.

\subsection{Utility of the Analytic Model}

There are at least four advantages of fully analytic Askaryan models.  First, when analytic models are matched to observed data, cascade properties may be derived directly from the waveforms.  Second, in large scale simulations, evaluating a fully analytic model technically provides a speed advantage over other approaches.  Third, fully analytic models, combined with RF channel response, can be embedded in firmware to form a \textit{matched filter} that enhances UHE-$\nu$ detection probability.  Fourth, parameters in analytic models may be \textit{scaled} to produce results that apply to media of different density than ice.  This application is useful for understanding potential signals in the Antarctic \textit{firn}, or the upper layer of snow and ice that is of lower density than the solid ice beneath it.

The ability to fit cascade properties from waveforms will be a useful tool for the radio component of IceCube-Gen2.  Examples of current reconstruction techniques include the forward-folding method \cite{10.1088/1748-0221/15/09/p09039} and information field theory (IFT) \cite{IFT}.  In particular, the longitudinal length parameter $a$ leads to a reconstruction of $\ln(E_{\rm C})$, given knowledge of $\Delta \theta$ (Fig. \ref{fit:fig:a_vs_a} and Equation \ref{eq:energy}).  Further, all designs for detector stations in IceCube-Gen2 radio include many distinct RF channels and one phased-array of channels.  Matching our analytic model to each channel waveform will provide a separate measurement of parameters like $a$ and $\theta$ (see gray contours of Figures \ref{fig:off_cone} and \ref{fig:off_cone2}).  The ensuing global fit should constrain the event energy and geometry.

The most intriguing usage for a fully analytic Askaryan model would be to embed the model as a \textit{matched filter} in detector firmware.  Because cascade properties are unknown \textit{a priori}, an array of matched filters could be implemented to form a \textit{matched filter bank.}  One example of this approach was the TARA experiment \cite{ABBASI20171}, which was designed to detect low-SNR cosmic ray radar echoes.  This is similar to the challenge faced by IceCube-Gen2 radio: pushing the limit of low-SNR RF pulse detection in a remote setting.  For example, a matched filter bank could be formed with an array of off-cone field formulas with fixed $a$-value and varying $\theta$-values, which would then be convolved with the RF channel impulse response (see Section 6 of \cite{10.1016/j.astropartphys.2014.09.002}).

Finally, a fully analytic model enhances the ability of IceCube-Gen2 radio to identify signals that originate in the firn.  At the South Pole, the RF index of refraction begins around 1.35 and does not reach the solid ice value of 1.78 until 150-200 meters \cite{Barwick:2018497}.  There are at least two signals that could originate in the firn: UHE-$\nu$ events that create Askaryan radiation, and UHE cosmic ray cascades partially inside or fully inside the firn.  The altitude of the South Pole makes the latter possible.  The Askaryan radiation of the firn UHE-$\nu$ events could be modeled via appropriate density-scaling of the cascade parameters.

\section{Acknowledgements}

We would like to thank our families for their support throughout the COVID-19 pandemic.  We could not have completed this work without their help.  We would also like to thank our colleagues for helpful discussions regarding analysis techniques.  In particular, we want to thank Profs. Steve Barwick, Dave Besson, and Christian Glaser for useful discussions.  Finally, we would like to thank the Whittier College Fellowships Committee, and specifically the Fletcher-Jones Fellowship Program for providing financial support for this work.  This work was partially funded by the Fletcher-Jones Summer Fellowship of 2020, Whittier College Fellowships program.

\appendix

\section{Details of the On-Cone Field Equation Derivation}
\label{app:a}

The original equations for the $\hat{\theta}$-component of $\vec{\mathcal{E}}$ are:

\begin{align}
\mathcal{W}(\eta,\theta) &= \frac{\exp\left( -\frac{1}{2}(ka)^2 \frac{(\cos\theta - \cos\theta_{\rm C})^2}{1-i\eta} \right)}{\left(1-i\eta\left(1-3i\eta \frac{\cos\theta}{\sin^2\theta}\frac{\cos\theta - \cos\theta_{\rm C}}{1-i\eta}\right) \right)^{1/2}} \\
\vec{\mathcal{E}}(\eta,\theta) \cdot \hat{\theta} &= \mathcal{W}(\eta,\theta) \left(1-i\eta \frac{\cos\theta_{\rm C}}{\sin^2\theta} \frac{\cos\theta - \cos\theta_{\rm C}}{1-i\eta} \right) \label{app:eq:big_one_1}
\end{align}

Letting $\theta = \theta_{\rm C}$ yields

\begin{equation}
\vec{\mathcal{E}}(\eta,\theta) \cdot \hat{\theta} = \frac{1}{\sqrt{1-i\eta}}
\end{equation}

The complete field from the original RB model \cite{10.1103/physrevd.65.016003}, including the form factor $\widetilde{F}$, $\psi = -i \exp(ikr) \sin\theta$, and $\vec{\mathcal{E}}$ is 

\begin{equation}
r\vec{E}(\omega,\theta) = E_{\rm 0} \left( \frac{\omega}{2\pi} \right) \psi \vec{\mathcal{E}}(\eta,\theta) \widetilde{F} \label{app:eq:main}
\end{equation}

Let Equation \ref{eq:F_main} for the form factor, with $\sigma = \omega/\omega_{\rm CF}$ and $\eta = \omega/\omega_{\rm CF}$, and letting $E_{\rm 0}$ be proportional to cascade energy $E_{\rm C}$:

\begin{equation}
r\widetilde{E}(\omega,\theta_{\rm C}) = \frac{(-i\omega) E_0 \sin(\theta_{\rm C}) e^{i\omega r/c}}{(1-i \omega/\omega_{\rm C})^{1/2} (1+(\omega/\omega_{\rm CF})^2)^{3/2}} \label{app:eq:re_1}
\end{equation}

Suppose $\omega < \omega_{\rm C}$, and $\omega < \omega_{\rm CF}$, such that the following approximations of the factors in the denominator are valid:

\begin{align}
(1-i \omega / \omega_{\rm C})^{1/2} &\approx 1 - \frac{i}{2} \frac{\omega}{\omega_{\rm C}} \\
(1+(\omega/\omega_{\rm CF})^2)^{3/2} & \approx 1 + \frac{3}{2} \left(\frac{\omega}{\omega_{\rm CF}}\right)^2
\end{align}

Using the approximations introduces simple poles into the complex formula for the frequency-dependent electric field.  Inserting the approximations in the denominator of Equation \ref{app:eq:re_1}, we have

\begin{equation}
r\widetilde{E}(\omega,\theta_{\rm C}) = \frac{(-i\omega) E_0 \sin(\theta_{\rm C}) e^{i\omega R/c}}{\left(1-\frac{i}{2} \omega/\omega_{\rm C}\right) \left(1+\frac{3}{2}(\omega/\omega_{\rm CF})^2 \right) }
\end{equation}

The denominator can be rearranged by factoring the $\omega$ coefficients, and defining $\omega_0 = \sqrt{\frac{2}{3	}}\omega_{\rm CF}$.

\begin{equation}
r\widetilde{E}(\omega,\theta_{\rm C}) = \frac{ 2 i \omega_{\rm C} \omega_{\rm 0}^2 (-i\omega) E_0 \sin(\theta_{\rm C}) e^{i\omega r/c}}{ \left(2 i \omega_{\rm C} + \omega \right) (\omega+i\omega_{\rm 0}) (\omega - i\omega_{\rm 0}) }
\end{equation}

Let $\hat{E}_{\rm 0} = E_{\rm 0} \sin(\theta_{\rm C})$, and let the retarded time be $t_{\rm r} = t - r/c$.  Taking the \textit{inverse} Fourier transform, using the same sign convention as RB \cite{10.1103/physrevd.65.016003} ($f(t) = (2\pi)^{-1} \int_{-\infty}^{\infty} \widetilde{F}(\omega) e^{-i\omega t} d\omega$), converts the field to the time-domain:

\begin{widetext}
\begin{equation}
r E(t,\theta_{\rm C}) = \frac{\hat{E}_0 i \omega_{\rm C} \omega_{\rm 0}^2}{\pi} \frac{d}{dt_{\rm r}}\int_{-\infty}^{\infty} \frac{ e^{-i\omega t_r}}{ \left(2 i \omega_{\rm C} + \omega \right) (\omega+i\omega_{\rm 0}) (\omega - i\omega_{\rm 0}) } d\omega \label{app:eq:part1}
\end{equation}
\end{widetext}

\begin{enumerate}
\item If $t_r > 0$: Consider the contour comprised of the real axis and the clockwise-oriented negative infinite semi-circle.  On the contour, the exponential phase factor in Equation \ref{app:eq:part1} goes as
\begin{equation}
\exp(-i\omega t_r) = \exp(-i(R\cos\phi +iR\sin\phi) t_r)
\end{equation}
For the semi-circle, $\phi \in [\pi,2\pi]$, so $\sin\phi < 0$ and $t_r > 0$.  Exponential decay occurs and the integrand vanishes on the semi-circle for $|\omega| = R \to \infty$.
\item If $t_r < 0$: Consider the contour comprised of the real axis and the counter-clockwise-oriented positive infinite semi-circle.  On the contour, the exponential phase factor in Equation \ref{app:eq:part1} goes again as
\begin{equation}
\exp(-i\omega t_r) = \exp(-i(R\cos\phi +iR\sin\phi) t_r)
\end{equation}
For the semi-circle, $\phi \in [0,\pi]$, so $\sin\phi > 0$ and $t_r < 0$.  Exponential decay occurs and the integrand vanishes on the semi-circle for $|\omega| = R \to \infty$.
\end{enumerate}

Using cases 1 and 2, Equation \ref{app:eq:part1} can be solved using the Cauchy integral formula.  Beginning with $t_r > 0$, two poles are enclosed in the semi-circle: one that originated from the coherence cutoff frequency, and the other that originated from the form factor.  The Cauchy integral formula yields

\begin{widetext}
\begin{equation}
r E(t,\theta_{\rm C}) = 2 \hat{E}_0 \omega_{\rm C} \omega_{\rm 0}^2 \frac{d}{dt_r} \left( \frac{e^{-2\omega_{\rm C} t_r} }{i^2(-2\omega_{\rm C} + \omega_{\rm 0})(-2\omega_{\rm C}-\omega_{\rm 0})} + \frac{e^{-\omega_{\rm 0}t_r}}{i^2(-\omega_{\rm 0} + 2\omega_{\rm C})(-2\omega_{\rm 0})} \right) \label{app:eq:no_eps}
\end{equation}
\end{widetext}

Define the ratio of the cutoff frequencies: $\epsilon = \omega_{\rm 0}/\omega_{\rm C}$.  After evaluating the time derivatives, Equation \ref{app:eq:no_eps} becomes

\begin{equation}
r E(t,\theta_{\rm C}) = \hat{E}_0 \omega_{\rm 0}^2 \left( \frac{ e^{-2\omega_{\rm C} t_r} }{(1 -\frac{\epsilon}{2})(1+\frac{\epsilon}{2})} - \frac{e^{-\omega_{\rm 0}t_r}}{(2)(1-\frac{\epsilon}{2})} \right) \label{app:eq:yes_eps}
\end{equation}

Expanding to linear order in $\epsilon$, assuming $\epsilon < 1$, and recalling that $\omega_{\rm 0}^2 = \frac{2}{3}\omega_{\rm CF}^2$:

\begin{equation}
r E(t,\theta_{\rm C}) \approx \frac{1}{3} \hat{E}_0 \omega_{\rm CF}^2 \left( 2 e^{-2\omega_{\rm C} t_r} - \left(1+\frac{\epsilon}{2} \right) e^{-\omega_{\rm 0}t_r} \right)
\end{equation}

Turning to the case of $t_r < 0$, consider integrating Equation \ref{app:eq:part1} along the contour comprised of the real axis and the counter-clockwise-oriented positive infinite semi-circle.  The contour encloses one pole, and the exponent ensures convergence:

\begin{equation}
r E(t,\theta_{\rm C}) = (2\pi i)\hat{E}_0 (\pi)^{-1} i \omega_{\rm C} \omega_{\rm 0}^2 \frac{d}{dt_r}\left( \frac{ e^{\omega_{\rm 0} t_r}}{ \left(2 i \omega_{\rm C} + i\omega_{\rm 0} \right) (2i\omega_{\rm 0}) } \right)
\end{equation}

After evaluating the derivative, the expression simplifies with $\epsilon = \omega_0/\omega_{\rm C}$:

\begin{equation}
r E(t,\theta_{\rm C}) = \frac{1}{2} \hat{E}_0 \omega_{\rm 0}^2  \left( \frac{ e^{\omega_{\rm 0} t_r} }{1 + \frac{1}{2}\epsilon } \right) \label{app:eq:yes_eps2}
\end{equation}

Finally, using the same first-order approximation in $\epsilon$ as the $t_{\rm r} > 0$ case:

\begin{equation}
r E(t,\theta_{\rm C}) \approx \frac{1}{3} \hat{E}_0 \omega_{\rm CF}^2  \left(1 - \frac{1}{2}\epsilon \right)e^{\omega_{\rm 0} t_r}
\end{equation}

Collecting the $t_{\rm r} > 0$ and $t_{\rm r} < 0$ results together:

\begin{widetext}
\begin{equation}
r E(t,\theta_{\rm C}) = \frac{1}{3} \hat{E}_0 \omega_{\rm CF}^2
\begin{cases}
\left(1 - \frac{1}{2}\epsilon \right)e^{\omega_{\rm 0} t_r} ~~~~~~~~~~~~~~~~~~~~~ t_{\rm r} < 0 \\
\left( 2 e^{-2\omega_{\rm C} t_r} - \left(1+\frac{1}{2}\epsilon \right) e^{-\omega_{\rm 0}t_r} \right) ~~ t_{\rm r} > 0
\end{cases}
\label{app:eq:on_cone}
\end{equation}
\end{widetext}

\section{Details of the Off-Cone Field Equation Derivation}
\label{app:b}

Using Tabs. \ref{tab:features_2}-\ref{tab:features_4}, Equation \ref{app:eq:big_one_1} reduces to

\begin{equation}
\mathcal{E}(u,x) = f(u,x)g(u,x)(1-h(u,x))
\end{equation}

Expanding to first-order with respect to $u$ near $(u = 1)$ gives

\begin{equation}
\mathcal{E}(u,x) = \mathcal{E}(x,1) + (u-1) \dot{\mathcal{E}}(x,1) + \mathcal{O}(u-1)^2 \label{app:eq:expand}
\end{equation}

The first term is $fg(1-h)$ evaluated at $u = 1$: $\exp(-y)$ (Table \ref{tab:features_4}).  The second term requires the first derivative of $\mathcal{E}(u,x)$ with respect to $u$, evaluated at $u = 1$.

\begin{align}
\dot{\mathcal{E}}(u,x) &= f\dot{g} + \dot{f} g - (f g \dot{h} + f\dot{g} h + \dot{f} g h) \\
\dot{\mathcal{E}}(1,x) &= \left(f\dot{g} + \dot{f} g - (f g \dot{h} + f\dot{g} h + \dot{f} g h)\right)|_{u=1}
\end{align}

The first-derivatives of $f$, $g$, and $h$, evaluated at $u = 1$, are given in Tab. \ref{tab:features_4}.  Because $h(x,1) = 0$, terms proportional to $h$ will vanish.  The result is

\begin{equation}
\dot{\mathcal{E}}(1,x) = \frac{1}{2} e^{-y} \left(2 y + 2q - 1\right) \label{app:eq:dot}
\end{equation}

Inserting Equation \ref{app:eq:dot} into Equation \ref{app:eq:expand},

\begin{equation}
\mathcal{E}(u,x) = e^{-y} \left( 1 + \frac{1}{2} (u-1)\left(  2 y + 2q - 1 \right) \right)
\end{equation}

Using the definition of $u$ (Table \ref{tab:features_2}), the result may be written

\begin{equation}
\mathcal{E}(u,x) = e^{-y} \left( 1 - \frac{1}{2} j\eta\left( 2y + 2q - 1 \right) \right)
\end{equation}

Proceding with the inverse Fourier transform of the $\hat{\theta}$-component:

\begin{equation}
r E(t,\theta) = \mathcal{F}^{-1} \left\lbrace E_0 \left(\frac{\omega}{2\pi}\right) \widetilde{F} \psi \mathcal{E} \right\rbrace
\end{equation}

Let $\eta = \omega/\omega_C$, $y = p \omega^2$ (Table \ref{tab:features_2}).  Inserting the Taylor series for $\mathcal{E}$, the form factor $\widetilde{F}$, and $\psi = -i \exp(ikr) \sin\theta$ (Sec. \ref{sec:unit}), and following the same steps as the on-cone case produces

\begin{widetext}
\begin{equation}
2\pi r E(t,\theta) = \frac{E_0 \omega_0^2 \sin(\theta)}{4 \pi i \omega_C} \frac{d}{dt_{r}} \int_{-\infty}^{\infty} \frac{e^{-i \omega t_r - p\omega^2}  \left( 2 i \omega_C + 2 p \omega^3 + (2 q - 1) \omega \right)}{\omega^2+\omega_0^2}d\omega \label{app:eq:intrigue}
\end{equation}
\end{widetext}

Unlike the on-cone case, Equation \ref{app:eq:intrigue} cannot be integrated with infinite semi-circle contours, because the exponential term diverges along the imaginary axis far from the origin.  Let $I_{\rm 0}$ represent the constant term with respect to $\omega$ in the numerator:

\begin{equation}
I_{\rm 0} = \int_{-\infty}^{\infty} \frac{e^{-i \omega t_r - p\omega^2}  \left( 2 i \omega_C \right)}{\omega^2+\omega_0^2}d\omega \label{app:eq:intrigue2}
\end{equation}

Further, let $I_{\rm 1}$ and $I_{\rm 3}$ represent the linear and cubic terms, respectively.  Completing the square in the exponent of $I_{\rm 0}$, with $\omega_{\rm 1} = t_{\rm r}/(2p)$, yields

\begin{equation}
I_{\rm 0} = 2 i \omega_C e^{- \frac{t_{\rm r}^2}{4p}} \int_{-\infty}^{\infty} \frac{e^{-p(\omega + i\omega_{\rm 1})^2}}{\omega^2+\omega_0^2}d\omega \label{app:eq:intrigue3}
\end{equation}

Equation \ref{app:eq:intrigue3} may be re-cast as the \textit{line-broadening function, $H$} (DLMF 7.19, \cite{NIST:DLMF}) common to spectroscopy applications:

\begin{equation}
I_{\rm 0} = 2 \pi i \left( \frac{\omega_{\rm C}}{\omega_{\rm 0}} \right) e^{- \frac{t_{\rm r}^2}{4p}} H(\sqrt{p}\omega_0,i\sqrt{p}\omega_1)
\end{equation}

Assume that $\omega > \omega_{\rm 1}$.  This approximating step will be called the \textit{symmetric approximation.}

\begin{equation}
I_{\rm 0} \approx 2 i \omega_C e^{- \frac{t_{\rm}^2}{2p}} \int_{-\infty}^{\infty} \frac{e^{-p\omega^2}}{\omega^2+\omega_0^2}d\omega \label{app:eq:intrigue4}
\end{equation}

The result for $I_{\rm 0}$ involves the complementary error function (DLMF 7.7.1, \cite{NIST:DLMF}):

\begin{equation}
I_{\rm 0} = 2 i \omega_C e^{- \frac{t_{\rm}^2}{2p}} \pi \omega_{\rm 0}^{-1} e^{p \omega_{\rm 0}^2} \erfc(\sqrt{p} \omega_{\rm 0}) \label{app:eq:intrigue5}
\end{equation}

The integrals $I_1$ and $I_3$ are zero by symmetry, with odd integrands over $(-\infty,\infty)$.  Inserting the result for $I_0$ into Equation \ref{app:eq:intrigue} and evaluating the derivative finishes the problem (see Sec. \ref{sec:ofc}).

\bibliography{apssamp}

\begin{thebibliography}{41}
\expandafter\ifx\csname natexlab\endcsname\relax\def\natexlab#1{#1}\fi
\expandafter\ifx\csname bibnamefont\endcsname\relax
  \def\bibnamefont#1{#1}\fi
\expandafter\ifx\csname bibfnamefont\endcsname\relax
  \def\bibfnamefont#1{#1}\fi
\expandafter\ifx\csname citenamefont\endcsname\relax
  \def\citenamefont#1{#1}\fi
\expandafter\ifx\csname url\endcsname\relax
  \def\url#1{\texttt{#1}}\fi
\expandafter\ifx\csname urlprefix\endcsname\relax\def\urlprefix{URL }\fi
\providecommand{\bibinfo}[2]{#2}
\providecommand{\eprint}[2][]{\url{#2}}

\bibitem[{\citenamefont{{The IceCube
  Collaboration}}(2013)}]{10.1126/science.1242856}
\bibinfo{author}{\bibnamefont{{The IceCube Collaboration}}},
  \bibinfo{journal}{Science} \textbf{\bibinfo{volume}{342}},
  \bibinfo{pages}{1242856} (\bibinfo{year}{2013}), ISSN
  \bibinfo{issn}{0036-8075}, \eprint{1311.5238}.

\bibitem[{\citenamefont{Ahlers et~al.}(2010)\citenamefont{Ahlers, Anchordoqui,
  Gonzalez–Garcia, Halzen, and Sarkar}}]{10.1016/j.astropartphys.2010.06.003}
\bibinfo{author}{\bibfnamefont{M.}~\bibnamefont{Ahlers}},
  \bibinfo{author}{\bibfnamefont{L.}~\bibnamefont{Anchordoqui}},
  \bibinfo{author}{\bibfnamefont{M.}~\bibnamefont{Gonzalez–Garcia}},
  \bibinfo{author}{\bibfnamefont{F.}~\bibnamefont{Halzen}}, \bibnamefont{and}
  \bibinfo{author}{\bibfnamefont{S.}~\bibnamefont{Sarkar}},
  \bibinfo{journal}{Astroparticle Physics} \textbf{\bibinfo{volume}{34}},
  \bibinfo{pages}{106} (\bibinfo{year}{2010}), ISSN \bibinfo{issn}{0927-6505}.

\bibitem[{\citenamefont{Kotera et~al.}(2010)\citenamefont{Kotera, Allard, and
  Olinto}}]{10.1088/1475-7516/2010/10/013}
\bibinfo{author}{\bibfnamefont{K.}~\bibnamefont{Kotera}},
  \bibinfo{author}{\bibfnamefont{D.}~\bibnamefont{Allard}}, \bibnamefont{and}
  \bibinfo{author}{\bibfnamefont{A.}~\bibnamefont{Olinto}},
  \bibinfo{journal}{Journal of Cosmology and Astroparticle Physics}
  \textbf{\bibinfo{volume}{2010}}, \bibinfo{pages}{013} (\bibinfo{year}{2010}),
  ISSN \bibinfo{issn}{1475-7516}, \eprint{1009.1382}.

\bibitem[{\citenamefont{{The IceCube
  Collaboration}}(2018)}]{10.1103/physrevd.98.062003}
\bibinfo{author}{\bibnamefont{{The IceCube Collaboration}}},
  \bibinfo{journal}{Physical Review D} \textbf{\bibinfo{volume}{98}},
  \bibinfo{pages}{062003} (\bibinfo{year}{2018}), ISSN
  \bibinfo{issn}{2470-0010}, \eprint{1807.01820}.

\bibitem[{\citenamefont{{The ARIANNA
  Collaboration}}(2020{\natexlab{a}})}]{10.1088/1475-7516/2020/03/053}
\bibinfo{author}{\bibnamefont{{The ARIANNA Collaboration}}},
  \bibinfo{journal}{Journal of Cosmology and Astroparticle Physics}
  \textbf{\bibinfo{volume}{2020}}, \bibinfo{pages}{053}
  (\bibinfo{year}{2020}{\natexlab{a}}), \eprint{1909.00840}.

\bibitem[{\citenamefont{{The ARA
  Collaboration}}(2020)}]{10.1103/physrevd.102.043021}
\bibinfo{author}{\bibnamefont{{The ARA Collaboration}}},
  \bibinfo{journal}{Physical Review D} \textbf{\bibinfo{volume}{102}},
  \bibinfo{pages}{043021} (\bibinfo{year}{2020}), ISSN
  \bibinfo{issn}{2470-0010}, \eprint{1912.00987}.

\bibitem[{\citenamefont{{M. Ackermann et
  al}}(2019{\natexlab{a}})}]{Ackermann:201946d}
\bibinfo{author}{\bibnamefont{{M. Ackermann et al}}}
  (\bibinfo{year}{2019}{\natexlab{a}}), \eprint{1903.04334}.

\bibitem[{\citenamefont{{M. Ackermann et
  al}}(2019{\natexlab{b}})}]{Ackermann:20195ec}
\bibinfo{author}{\bibnamefont{{M. Ackermann et al}}}
  (\bibinfo{year}{2019}{\natexlab{b}}), \eprint{1903.04333}.

\bibitem[{\citenamefont{{J. C. Hanson et
  al}}(2015{\natexlab{a}})}]{10.3189/2015jog14j214}
\bibinfo{author}{\bibnamefont{{J. C. Hanson et al}}}, \bibinfo{journal}{Journal
  of Glaciology} \textbf{\bibinfo{volume}{61}}, \bibinfo{pages}{438}
  (\bibinfo{year}{2015}{\natexlab{a}}), ISSN \bibinfo{issn}{0022-1430}.

\bibitem[{\citenamefont{Avva et~al.}(2014)\citenamefont{Avva, Kovac, Miki,
  Saltzberg, and Vieregg}}]{10.3189/2015jog15j057}
\bibinfo{author}{\bibfnamefont{J.}~\bibnamefont{Avva}},
  \bibinfo{author}{\bibfnamefont{J.}~\bibnamefont{Kovac}},
  \bibinfo{author}{\bibfnamefont{C.}~\bibnamefont{Miki}},
  \bibinfo{author}{\bibfnamefont{D.}~\bibnamefont{Saltzberg}},
  \bibnamefont{and} \bibinfo{author}{\bibfnamefont{A.}~\bibnamefont{Vieregg}},
  \bibinfo{journal}{Journal of Glaciology}  (\bibinfo{year}{2014}),
  \eprint{1409.5413}.

\bibitem[{\citenamefont{{The ARA
  Collaboration}}(2012)}]{10.1016/j.astropartphys.2011.11.010}
\bibinfo{author}{\bibnamefont{{The ARA Collaboration}}},
  \bibinfo{journal}{Astroparticle Physics} \textbf{\bibinfo{volume}{35}},
  \bibinfo{pages}{457} (\bibinfo{year}{2012}), ISSN \bibinfo{issn}{0927-6505},
  \eprint{1105.2854}.

\bibitem[{\citenamefont{{G. Askaryan}}(1962)}]{askaryan1}
\bibinfo{author}{\bibnamefont{{G. Askaryan}}}, \bibinfo{journal}{Soviet Physics
  JETP} \textbf{\bibinfo{volume}{15}} (\bibinfo{year}{1962}).

\bibitem[{\citenamefont{Zas et~al.}(1992)\citenamefont{Zas, Halzen, and
  Stanev}}]{zhs}
\bibinfo{author}{\bibfnamefont{E.}~\bibnamefont{Zas}},
  \bibinfo{author}{\bibfnamefont{F.}~\bibnamefont{Halzen}}, \bibnamefont{and}
  \bibinfo{author}{\bibfnamefont{T.}~\bibnamefont{Stanev}},
  \bibinfo{journal}{Physical Review D} \textbf{\bibinfo{volume}{45}},
  \bibinfo{pages}{362} (\bibinfo{year}{1992}).

\bibitem[{\citenamefont{{I. Kravchenko et
  al}}(2012)}]{10.1103/PhysRevD.85.062004}
\bibinfo{author}{\bibnamefont{{I. Kravchenko et al}}},
  \bibinfo{journal}{Physical Review D} \textbf{\bibinfo{volume}{85}},
  \bibinfo{pages}{062004} (\bibinfo{year}{2012}), ISSN
  \bibinfo{issn}{2470-0029}, \eprint{1106.1164}.

\bibitem[{\citenamefont{{The ANITA
  Collaboration}}(2019)}]{10.1103/physrevd.99.122001}
\bibinfo{author}{\bibnamefont{{The ANITA Collaboration}}},
  \bibinfo{journal}{Physical Review D} \textbf{\bibinfo{volume}{99}},
  \bibinfo{pages}{122001} (\bibinfo{year}{2019}), ISSN
  \bibinfo{issn}{2470-0010}, \eprint{1902.04005}.

\bibitem[{\citenamefont{Saltzberg et~al.}(2001)\citenamefont{Saltzberg, Gorham,
  Walz, Field, Iverson, Odian, Resch, Schoessow, and Williams}}]{saltzberg}
\bibinfo{author}{\bibfnamefont{D.}~\bibnamefont{Saltzberg}},
  \bibinfo{author}{\bibfnamefont{P.}~\bibnamefont{Gorham}},
  \bibinfo{author}{\bibfnamefont{D.}~\bibnamefont{Walz}},
  \bibinfo{author}{\bibfnamefont{C.}~\bibnamefont{Field}},
  \bibinfo{author}{\bibfnamefont{R.}~\bibnamefont{Iverson}},
  \bibinfo{author}{\bibfnamefont{A.}~\bibnamefont{Odian}},
  \bibinfo{author}{\bibfnamefont{G.}~\bibnamefont{Resch}},
  \bibinfo{author}{\bibfnamefont{P.}~\bibnamefont{Schoessow}},
  \bibnamefont{and} \bibinfo{author}{\bibfnamefont{D.}~\bibnamefont{Williams}},
  \bibinfo{journal}{Physical review letters} \textbf{\bibinfo{volume}{86}},
  \bibinfo{pages}{2802} (\bibinfo{year}{2001}), ISSN \bibinfo{issn}{0031-9007}.

\bibitem[{\citenamefont{Miocinovic et~al.}(2006)\citenamefont{Miocinovic,
  Field, Gorham, Guillian, Milincic, Saltzberg, Walz, and
  Williams}}]{10.1103/PhysRevD.74.043002}
\bibinfo{author}{\bibfnamefont{P.}~\bibnamefont{Miocinovic}},
  \bibinfo{author}{\bibfnamefont{R.}~\bibnamefont{Field}},
  \bibinfo{author}{\bibfnamefont{P.}~\bibnamefont{Gorham}},
  \bibinfo{author}{\bibfnamefont{E.}~\bibnamefont{Guillian}},
  \bibinfo{author}{\bibfnamefont{R.}~\bibnamefont{Milincic}},
  \bibinfo{author}{\bibfnamefont{D.}~\bibnamefont{Saltzberg}},
  \bibinfo{author}{\bibfnamefont{D.}~\bibnamefont{Walz}}, \bibnamefont{and}
  \bibinfo{author}{\bibfnamefont{D.}~\bibnamefont{Williams}},
  \bibinfo{journal}{Physical Review D} \textbf{\bibinfo{volume}{74}},
  \bibinfo{pages}{043002} (\bibinfo{year}{2006}), ISSN
  \bibinfo{issn}{2470-0029}, \eprint{hep-ex/0602043}.

\bibitem[{\citenamefont{Gorham et~al.}(2007)\citenamefont{Gorham, Barwick,
  Beatty, Besson, Binns, Chen, Chen, Clem, Connolly, Dowkontt
  et~al.}}]{ask_ice}
\bibinfo{author}{\bibfnamefont{P.~W.} \bibnamefont{Gorham}},
  \bibinfo{author}{\bibfnamefont{S.~W.} \bibnamefont{Barwick}},
  \bibinfo{author}{\bibfnamefont{J.~J.} \bibnamefont{Beatty}},
  \bibinfo{author}{\bibfnamefont{D.~Z.} \bibnamefont{Besson}},
  \bibinfo{author}{\bibfnamefont{W.~R.} \bibnamefont{Binns}},
  \bibinfo{author}{\bibfnamefont{C.}~\bibnamefont{Chen}},
  \bibinfo{author}{\bibfnamefont{P.}~\bibnamefont{Chen}},
  \bibinfo{author}{\bibfnamefont{J.~M.} \bibnamefont{Clem}},
  \bibinfo{author}{\bibfnamefont{A.}~\bibnamefont{Connolly}},
  \bibinfo{author}{\bibfnamefont{P.~F.} \bibnamefont{Dowkontt}},
  \bibnamefont{et~al.} (\bibinfo{collaboration}{ANITA Collaboration}),
  \bibinfo{journal}{Phys. Rev. Lett.} \textbf{\bibinfo{volume}{99}},
  \bibinfo{pages}{171101} (\bibinfo{year}{2007}),
  \urlprefix\url{https://link.aps.org/doi/10.1103/PhysRevLett.99.171101}.

\bibitem[{\citenamefont{Alvarez-Muñiz
  et~al.}(2009)\citenamefont{Alvarez-Muñiz, James, Protheroe, and
  Zas}}]{10.1016/j.astropartphys.2009.06.005}
\bibinfo{author}{\bibfnamefont{J.}~\bibnamefont{Alvarez-Muñiz}},
  \bibinfo{author}{\bibfnamefont{C.}~\bibnamefont{James}},
  \bibinfo{author}{\bibfnamefont{R.}~\bibnamefont{Protheroe}},
  \bibnamefont{and} \bibinfo{author}{\bibfnamefont{E.}~\bibnamefont{Zas}},
  \bibinfo{journal}{Astroparticle Physics} \textbf{\bibinfo{volume}{32}},
  \bibinfo{pages}{100} (\bibinfo{year}{2009}), ISSN \bibinfo{issn}{0927-6505}.

\bibitem[{\citenamefont{Gerhardt and Klein}(2010)}]{10.1103/physrevd.82.074017}
\bibinfo{author}{\bibfnamefont{L.}~\bibnamefont{Gerhardt}} \bibnamefont{and}
  \bibinfo{author}{\bibfnamefont{S.~R.} \bibnamefont{Klein}},
  \bibinfo{journal}{Physical Review D} \textbf{\bibinfo{volume}{82}}
  (\bibinfo{year}{2010}), ISSN \bibinfo{issn}{1550-7998}.

\bibitem[{\citenamefont{Dookayka}(2011)}]{dookayka2011characterizing}
\bibinfo{author}{\bibfnamefont{K.}~\bibnamefont{Dookayka}}, Ph.D. thesis,
  \bibinfo{school}{University of California, Irvine} (\bibinfo{year}{2011}).

\bibitem[{\citenamefont{{The ARA Collaboration}}(2015)}]{testbed}
\bibinfo{author}{\bibnamefont{{The ARA Collaboration}}},
  \bibinfo{journal}{Astroparticle Physics} \textbf{\bibinfo{volume}{70}},
  \bibinfo{pages}{62} (\bibinfo{year}{2015}), ISSN \bibinfo{issn}{0927-6505},
  \urlprefix\url{https://www.sciencedirect.com/science/article/pii/S0927650515000687}.

\bibitem[{\citenamefont{{C. Glaser et
  al}}(2020)}]{10.1140/epjc/s10052-020-7612-8}
\bibinfo{author}{\bibnamefont{{C. Glaser et al}}}, \bibinfo{journal}{The
  European Physical Journal C} \textbf{\bibinfo{volume}{80}},
  \bibinfo{pages}{77} (\bibinfo{year}{2020}), ISSN \bibinfo{issn}{1434-6044},
  \eprint{1906.01670}.

\bibitem[{\citenamefont{{C. Glaser et
  al}}(2019)}]{10.1140/epjc/s10052-019-6971-5}
\bibinfo{author}{\bibnamefont{{C. Glaser et al}}}, \bibinfo{journal}{The
  European Physical Journal C} \textbf{\bibinfo{volume}{79}},
  \bibinfo{pages}{464} (\bibinfo{year}{2019}), ISSN \bibinfo{issn}{1434-6044},
  \eprint{1903.07023}.

\bibitem[{\citenamefont{{The ARIANNA
  Collaboration}}(2020{\natexlab{b}})}]{10.1088/1748-0221/15/09/p09039}
\bibinfo{author}{\bibnamefont{{The ARIANNA Collaboration}}},
  \bibinfo{journal}{Journal of Instrumentation} \textbf{\bibinfo{volume}{15}},
  \bibinfo{pages}{P09039} (\bibinfo{year}{2020}{\natexlab{b}}),
  \eprint{2006.03027}.

\bibitem[{\citenamefont{Welling et~al.}(2021)\citenamefont{Welling, Frank,
  Enßlin, and Nelles}}]{IFT}
\bibinfo{author}{\bibfnamefont{C.}~\bibnamefont{Welling}},
  \bibinfo{author}{\bibfnamefont{P.}~\bibnamefont{Frank}},
  \bibinfo{author}{\bibfnamefont{T.~A.} \bibnamefont{Enßlin}},
  \bibnamefont{and} \bibinfo{author}{\bibfnamefont{A.}~\bibnamefont{Nelles}},
  \bibinfo{journal}{arXiv}  (\bibinfo{year}{2021}), \eprint{2102.00258}.

\bibitem[{\citenamefont{{J. C. Hanson et
  al}}(2015{\natexlab{b}})}]{10.1016/j.astropartphys.2014.09.002}
\bibinfo{author}{\bibnamefont{{J. C. Hanson et al}}},
  \bibinfo{journal}{Astroparticle Physics} \textbf{\bibinfo{volume}{62}},
  \bibinfo{pages}{139} (\bibinfo{year}{2015}{\natexlab{b}}), ISSN
  \bibinfo{issn}{0927-6505}, \eprint{1406.0820}.

\bibitem[{\citenamefont{{The ARIANNA Collaboration}}(2018)}]{Barwick:2018497}
\bibinfo{author}{\bibnamefont{{The ARIANNA Collaboration}}},
  \bibinfo{journal}{Journal of Cosmology and Astroparticle Physics}
  \textbf{\bibinfo{volume}{2018}}, \bibinfo{pages}{055} (\bibinfo{year}{2018}).

\bibitem[{\citenamefont{{The ARA Collaboration}}(2019)}]{ALLISON201963}
\bibinfo{author}{\bibnamefont{{The ARA Collaboration}}},
  \bibinfo{journal}{Astroparticle Physics} \textbf{\bibinfo{volume}{108}},
  \bibinfo{pages}{63} (\bibinfo{year}{2019}), ISSN \bibinfo{issn}{0927-6505},
  \urlprefix\url{https://www.sciencedirect.com/science/article/pii/S0927650518301154}.

\bibitem[{\citenamefont{Barwick et~al.}(2015)\citenamefont{Barwick, Berg,
  Besson, Binder, Binns, Boersma, Bose, Braun, Buckley, Bugaev
  et~al.}}]{10.1016/j.astropartphys.2015.04.002}
\bibinfo{author}{\bibfnamefont{S.}~\bibnamefont{Barwick}},
  \bibinfo{author}{\bibfnamefont{E.}~\bibnamefont{Berg}},
  \bibinfo{author}{\bibfnamefont{D.}~\bibnamefont{Besson}},
  \bibinfo{author}{\bibfnamefont{G.}~\bibnamefont{Binder}},
  \bibinfo{author}{\bibfnamefont{W.}~\bibnamefont{Binns}},
  \bibinfo{author}{\bibfnamefont{D.}~\bibnamefont{Boersma}},
  \bibinfo{author}{\bibfnamefont{R.}~\bibnamefont{Bose}},
  \bibinfo{author}{\bibfnamefont{D.}~\bibnamefont{Braun}},
  \bibinfo{author}{\bibfnamefont{J.}~\bibnamefont{Buckley}},
  \bibinfo{author}{\bibfnamefont{V.}~\bibnamefont{Bugaev}},
  \bibnamefont{et~al.}, \bibinfo{journal}{Astroparticle Physics}
  \textbf{\bibinfo{volume}{70}}, \bibinfo{pages}{12} (\bibinfo{year}{2015}),
  ISSN \bibinfo{issn}{0927-6505}, \eprint{1410.7352}.

\bibitem[{\citenamefont{Alvarez-Muniz et~al.}(2011)\citenamefont{Alvarez-Muniz,
  Romero-Wolf, and Zas}}]{10.1103/physrevd.84.103003}
\bibinfo{author}{\bibfnamefont{J.}~\bibnamefont{Alvarez-Muniz}},
  \bibinfo{author}{\bibfnamefont{A.}~\bibnamefont{Romero-Wolf}},
  \bibnamefont{and} \bibinfo{author}{\bibfnamefont{E.}~\bibnamefont{Zas}},
  \bibinfo{journal}{Physical Review D} \textbf{\bibinfo{volume}{84}},
  \bibinfo{pages}{103003} (\bibinfo{year}{2011}), ISSN
  \bibinfo{issn}{2470-0029}, \eprint{1106.6283}.

\bibitem[{\citenamefont{Alvarez-Muniz et~al.}(2020)\citenamefont{Alvarez-Muniz,
  Hansen, Romero-Wolf, and Zas}}]{PhysRevD.101.083005}
\bibinfo{author}{\bibfnamefont{J.}~\bibnamefont{Alvarez-Muniz}},
  \bibinfo{author}{\bibfnamefont{P.~M.} \bibnamefont{Hansen}},
  \bibinfo{author}{\bibfnamefont{A.}~\bibnamefont{Romero-Wolf}},
  \bibnamefont{and} \bibinfo{author}{\bibfnamefont{E.}~\bibnamefont{Zas}},
  \bibinfo{journal}{Phys. Rev. D} \textbf{\bibinfo{volume}{101}},
  \bibinfo{pages}{083005} (\bibinfo{year}{2020}),
  \urlprefix\url{https://link.aps.org/doi/10.1103/PhysRevD.101.083005}.

\bibitem[{\citenamefont{Buniy and Ralston}(2001)}]{10.1103/physrevd.65.016003}
\bibinfo{author}{\bibfnamefont{R.~V.} \bibnamefont{Buniy}} \bibnamefont{and}
  \bibinfo{author}{\bibfnamefont{J.~P.} \bibnamefont{Ralston}},
  \bibinfo{journal}{Physical Review D} \textbf{\bibinfo{volume}{65}}
  (\bibinfo{year}{2001}), ISSN \bibinfo{issn}{2470-0029}.

\bibitem[{\citenamefont{Hanson and
  Connolly}(2017)}]{10.1016/j.astropartphys.2017.03.008}
\bibinfo{author}{\bibfnamefont{J.~C.} \bibnamefont{Hanson}} \bibnamefont{and}
  \bibinfo{author}{\bibfnamefont{A.~L.} \bibnamefont{Connolly}},
  \bibinfo{journal}{Astroparticle Physics} \textbf{\bibinfo{volume}{91}},
  \bibinfo{pages}{75} (\bibinfo{year}{2017}), ISSN \bibinfo{issn}{0927-6505}.

\bibitem[{\citenamefont{Bogorodsky et~al.}(1985)\citenamefont{Bogorodsky,
  Bentley, and Gudmandsen}}]{bog}
\bibinfo{author}{\bibfnamefont{V.}~\bibnamefont{Bogorodsky}},
  \bibinfo{author}{\bibfnamefont{C.}~\bibnamefont{Bentley}}, \bibnamefont{and}
  \bibinfo{author}{\bibfnamefont{P.}~\bibnamefont{Gudmandsen}},
  \emph{\bibinfo{title}{Radioglaciology}} (\bibinfo{publisher}{Springer
  Netherlands}, \bibinfo{year}{1985}).

\bibitem[{\citenamefont{Razzaque et~al.}(2002)\citenamefont{Razzaque,
  Seunarine, Besson, McKay, Ralston, and Seckel}}]{PhysRevD.65.103002}
\bibinfo{author}{\bibfnamefont{S.}~\bibnamefont{Razzaque}},
  \bibinfo{author}{\bibfnamefont{S.}~\bibnamefont{Seunarine}},
  \bibinfo{author}{\bibfnamefont{D.~Z.} \bibnamefont{Besson}},
  \bibinfo{author}{\bibfnamefont{D.~W.} \bibnamefont{McKay}},
  \bibinfo{author}{\bibfnamefont{J.~P.} \bibnamefont{Ralston}},
  \bibnamefont{and} \bibinfo{author}{\bibfnamefont{D.}~\bibnamefont{Seckel}},
  \bibinfo{journal}{Phys. Rev. D} \textbf{\bibinfo{volume}{65}},
  \bibinfo{pages}{103002} (\bibinfo{year}{2002}),
  \urlprefix\url{https://link.aps.org/doi/10.1103/PhysRevD.65.103002}.

\bibitem[{\citenamefont{Andringa et~al.}(2011)\citenamefont{Andringa,
  Conceição, and Pimenta}}]{ANDRINGA2011360}
\bibinfo{author}{\bibfnamefont{S.}~\bibnamefont{Andringa}},
  \bibinfo{author}{\bibfnamefont{R.}~\bibnamefont{Conceição}},
  \bibnamefont{and} \bibinfo{author}{\bibfnamefont{M.}~\bibnamefont{Pimenta}},
  \bibinfo{journal}{Astroparticle Physics} \textbf{\bibinfo{volume}{34}},
  \bibinfo{pages}{360} (\bibinfo{year}{2011}), ISSN \bibinfo{issn}{0927-6505},
  \urlprefix\url{https://www.sciencedirect.com/science/article/pii/S0927650510001830}.

\bibitem[{\citenamefont{Fadhel et~al.}(2021)\citenamefont{Fadhel, Al-Rubaiee,
  Jassim, and Al-Alawy}}]{10.1088/1742-6596/1879/3/032089}
\bibinfo{author}{\bibfnamefont{K.~F.} \bibnamefont{Fadhel}},
  \bibinfo{author}{\bibfnamefont{A.}~\bibnamefont{Al-Rubaiee}},
  \bibinfo{author}{\bibfnamefont{H.~A.} \bibnamefont{Jassim}},
  \bibnamefont{and} \bibinfo{author}{\bibfnamefont{I.~T.}
  \bibnamefont{Al-Alawy}}, \bibinfo{journal}{Journal of Physics: Conference
  Series} \textbf{\bibinfo{volume}{1879}}, \bibinfo{pages}{032089}
  (\bibinfo{year}{2021}), ISSN \bibinfo{issn}{1742-6588}.

\bibitem[{{\relax DLMF}()}]{NIST:DLMF}
{\relax DLMF}, \emph{\bibinfo{title}{{\it NIST Digital Library of Mathematical
  Functions}}}, \bibinfo{howpublished}{http://dlmf.nist.gov/, Release 1.1.1 of
  2021-03-15}, \bibinfo{note}{f.~W.~J. Olver, A.~B. {Olde Daalhuis}, D.~W.
  Lozier, B.~I. Schneider, R.~F. Boisvert, C.~W. Clark, B.~R. Miller, B.~V.
  Saunders, H.~S. Cohl, and M.~A. McClain, eds.},
  \urlprefix\url{http://dlmf.nist.gov/}.

\bibitem[{\citenamefont{García}(2006)}]{10.1111/j.1365-2966.2006.10450.x}
\bibinfo{author}{\bibfnamefont{T.~T.} \bibnamefont{García}},
  \bibinfo{journal}{Monthly Notices of the Royal Astronomical Society}
  \textbf{\bibinfo{volume}{369}}, \bibinfo{pages}{2025} (\bibinfo{year}{2006}),
  ISSN \bibinfo{issn}{0035-8711}.

\bibitem[{\citenamefont{{The Telescope Array
  Collaboration}}(2017)}]{ABBASI20171}
\bibinfo{author}{\bibnamefont{{The Telescope Array Collaboration}}},
  \bibinfo{journal}{Astroparticle Physics} \textbf{\bibinfo{volume}{87}},
  \bibinfo{pages}{1} (\bibinfo{year}{2017}), ISSN \bibinfo{issn}{0927-6505},
  \urlprefix\url{https://www.sciencedirect.com/science/article/pii/S0927650516301682}.

\end{thebibliography}

\end{document}